\documentclass[]{aa}
\usepackage{natbib}
\usepackage{graphicx}
\usepackage{color}
\usepackage{txfonts}

\bibpunct{(}{)}{;}{a}{}{,}

\begin{document}

\title{Magnetised winds in dwarf galaxies}

\author{Yohan Dubois \inst{1, 3} and Romain Teyssier \inst{2, 3}}

\offprints{Y. Dubois}

\institute{Astrophysics, University of Oxford, Denys Wilkinson Building, Keble Road, Oxford, OX13RH, United Kingdom
\and Universit\"at Z\"urich, Institute f\"ur Theoretische Physik, Winterthurerstrasse 190, CH-8057 Z\"urich, Switzerland
\and CEA Saclay, DSM/IRFU/SAP, B\^atiment 709, F-91191 Gif-sur-Yvette, Cedex, France
\\
\email{yohan.dubois@physics.ox.ac.uk}}
\date{Accepted 2010 July 07. Received 2010 June 28; in original form 2009 July 30}

\label{firstpage}

\abstract {The generation and  the amplification of magnetic fields in
  the  current cosmological  paradigm are  still open  questions.  The
  standard theory  is based on  an early field generation  by Biermann
  battery effects, possibly at the Epoch of Reionisation, followed by
  a long period of field amplification by galactic dynamos. The origin
  and the  magnitude of  the inter-galactic magnetic
  field is of  primordial importance in this global  picture, as it is
  considered to  be the missing link between  galactic magnetic fields
  and cluster magnetic fields on much larger scales.}  {We are testing whether dwarf galaxies
  are good candidates to  explain the enrichment of the Inter-Galactic
  Medium (IGM):  after their discs form and  trigger galactic dynamos,
  supernova feedback  will launch strong winds,  expelling magnetic   field  lines  in   the  IGM.}   {We   have  performed
  Magneto-Hydrodynamics  simulations  of  an  isolated  dwarf  galaxy,
  forming self-consistently  inside a cooling halo.   Using the RAMSES
  code,  we have  for  the first  time  simulated the  formation of  a
  magnetised supernova-driven galactic  outflow. This simulation is an
  important  step  towards  a  more realistic  modelling  using  fully
  cosmological  simulations.  }  {Our  simulations reproduce  well the
  observed  properties of  magnetic  fields in  spiral galaxies.   The
  formation and the  evolution of our simulated disc  lead to a strong
  magnetic field  amplification: the magnetic field in  the final wind
  bubble is one order of magnitude larger than the initial value.  The
  magnetic  field  in  the  disc,  essentially  toroidal,  is  growing
  linearly with time as  a consequence of differential rotation.}  {We
  discuss  the consequence  of  this simple  mechanism  on the  cosmic
  evolution of the  magnetic field: we propose a  new scenario for the
  evolution of the  magnetic field, with dwarf galaxies  playing a key
  role  in  amplifying  and  ejecting  magnetic  energy  in  the  IGM,
  resulting in what we call  a ``Cosmic Dynamo'' that could contribute
  to the rather high field  strengths observed in galaxies and clusters
  today.}   

\authorrunning{Dubois and Teyssier} 
  
\titlerunning{Magnetised winds in dwarf galaxies}

\maketitle

\section{Introduction}

The  origin of  magnetic fields  in  the Universe  is a  long-standing
puzzle  in  cosmology. Although  measuring  magnetic  field in  cosmic
structures is very challenging,  we have strong evidence that magnetic
fields exist in many galaxies, with magnitudes between 1 and 10 $\mu$G
at   a   level   close   to   equipartition   (see   the   review   of
\citealp{becketal96}).  Magnetic  fields of the order  of several tens
of  $\mu  G$   are  also  detected  in  large   galaxy  clusters  (see
\citealp{carilli&taylor02}     or    \citealp{govoni&feretti04}    and
references therein).   However, in the very  low density Intergalactic
Medium (IGM),  magnetic fields still remain undetected  on large scale
\citep{blasietal99}.

From  a theoretical  point of  view, small-scale  fluctuating magnetic
fields   could  be   generated   from  quantum   effects  during   the
Big--Bang~(\citealp{turner&widrow88} for  example).  The amplitude and
the correlation length of such  primordial fields should both be quite
small \citep{grasso&rubinstein01}.  Magnetic fields greater than a few
nG would also have been detected as a specific source of anisotropy in
the Cosmic  Microwave Background. More  extreme scenarios can  also be
ruled  out:  for  example,  a  uniform  magnetic  field  greater  than
$10^{-1}\, \mu$G at our epoch would have led to unexpected anistropies
in the  expansion of the  Universe~(\citealp{chengetal94}).  Note also
that the relic of any magnetic field generated  in the early
universe, if  it is  greater than $10^{-7}\,  \mu$G today,  would have
broken the symmetry between left and right neutrinos (spin transition)
during  the   period  of  neutrino   production  \citep{enqvistetal93,
  sciama94}.

When the first structures collapse  ($z   \simeq  10$),  the  Biermann  Battery~(see
\citealp{kulsrudetal97}) is believed to have generated magnetic fields
from  pure collisional microscopic  processes.  These  magnetic fields
appear mainly  at shocks  and ionization fronts,  where the  motions of
electrons and ions are decoupled, creating small microscopic currents.
It was shown that this  battery effect can generate magnetic fields of
the order of $10^{-13}\,  \mu$G field in the IGM \citep{gnedinetal00}.
More recently, shocks around  primordial mini-halos and the associated
first generation  of massive stars  were also considered  as important
sources of magnetic field at high redshift~(\citealp{xuetal08}).

In  order  to  reconcile  the  very small  values  quoted  above  with
observations,  we  need  strong  amplification  processes  up  to  the
observed level  of magnetic field  strength in galaxies  and clusters.
Gravitational contraction of the frozen--in magnetic field will not be
sufficient,  since  we  expect  in   this  case  that  $B$  scales  as
$\rho^{2/3}$.

Using  cosmological simulations  of galaxy  cluster formation,  it was
shown  that  one  could  reproduce  the observed  field  magnitude  by
considering both  gravitational contraction and  turbulent stirring of
the  magnetised gas \citep{roettigeretal99,  dolagetal99, dolagetal02,
  dolagetal05,     sigletal04,    bruggenetal05,    subramanianetal06,
  asaietal07,  dubois&teyssier08cluster}.  In  cooling  flow clusters,
radiative losses  are probabily boosting  this amplification mechanism
even further up \citep{dubois&teyssier08cluster}  as confirmed by
recent observations \citep{carilli&taylor02}.  When an active galactic
nuclei  is releasing energy  into the  intra-cluster medium,  a non-cool
core  develops  and  the magnetic field within the core is amplified by the jet-induced turbulence \citep{duboisetal09}. The conclusion  of this series of  paper is that one  needs an initial
magnetic field of at least $10^{-5}\, \mu$G in the IGM to reach $\mu$G
values  in cluster  cores. Note  that the  finite resolution  of these
numerical  experiments leads  probably  to an  underestimation of  the
turbulent  amplification  of the  magnetic  field. Nevertheless,  this
justifies indirectly rather  low values for the magnetic  field in the
IGM, but significantly larger than Biermann Battery generated fields.

In order to explain the  origin of this diffuse cosmic magnetic field,
\cite{bertoneetal06} proposed galactic winds  as a possible solution to
explain the enrichment of the IGM by metals {\it and} magnetic fields.
In this scenario, magnetic  fields are amplified inside galactic discs
by  the  so--called Galactic  Dynamo~(\citealp{kulsrud99}),  and in  a
second phase, supernova--driven winds  are expelling field lines into
the surrounding IGM.

The   basic   idea   of   the  Galactic   Dynamo   \citep{parker71, ferriere92a,
  brandenburgetal95,  kulsrud99,   shukurov04}  is  that  small--scale
turbulence  (precisely   cyclonic  motions  that   could  result  from
supernova             explosions,            \citealp{balsaraetal04},
\citealp{gissingeretal09})  in the  Interstellar  Medium (ISM)  modify
sufficiently the small--scale magnetic  field to make the large--scale
component  grow  significantly.   An  additional term  $\nabla  \times
(\alpha \vec  B)$ is added  to the induction equation  that represents
this dynamo  amplification of the field,  where $\alpha \propto  ~ < u
. \nabla  \times u >$  tensor stands for  the cyclonic motions  of the
gas.   If  one  can   correctly  approximate  the  $\alpha$  parameter
\citep{ferriere92a,  ferriere92b}, it  is  possible to  show that  the
magnetic  field  can  grow  in  a  few Gyr  up  to  its  equipartition
value~(\citealp{ferriere&schmitt00}).   Note  that  the dynamo  theory
works   only    if   the   galaxy   can    loose   magnetic   helicity
\citep{brandenburg&subramanian05}, even though it is still debated that the small scale saturation of the $\alpha$ effect really occurs \citep{field95}. Numerical simulations of a very small patch (kpc size) of a galaxy from \cite{gresseletal08} demonstrated that such a fast dynamo occurs if supernovae explosions can carry some magnetic helicity from the disc to the hot wind. However one has to reach a tremendous resolution power (parsec scale) in cosmological situations to provide such a long-term dynamo. Galactic   winds    have   been considered as  a nice explanation for suppressing some magnetic helicity from the disc,
but no clear  demonstration of the effect has  been performed in a cosmological context. Another
issue  with galactic  dynamo is  that the  amplification  mechanism is
probably too  slow, especially if  one considers recent  magnetic field
observations     in     high-redshift     galaxies    beyond     $z=1$
\citep{bernetetal08}.    These  measurements  suggest   that  magnetic
amplification must occur very quickly,  in less than 5 Gyr. It is also plausible that a fast growth of the field due to turbulent amplification in the first halos at redshift $z>10$ could provide an equipartition field before the first galaxies form \citep{arshakianetal09}. Of course it has to be proven that small scale intense magnetic fields are able to generate this large-scale magnetic field observed in galaxies, for this purpose, alternative theories must be explored.

\cite{parker92}  first  suggested  that the production of cosmic rays  in
supernova remnants could generate stronger buoyancy in the ISM and amplify the
magnetic field.   Simulations of a cosmic ray  pressurised medium have
shown that  it leads to a rapid  growth of magnetic lines  if they are
strongly  diffused  in  the  ISM~\citep{hanaszetal04, hanaszetal09, hanaszetaldip09}.   \cite{rees87}
considered Biermann battery effects occuring at the surface of massive
stars.  More generally, one  can consider any stellar dynamo mechanism
as    an    efficient    amplification    mechanism    inside    stars
\citep{brunetal04},  followed by  the release  of this  magnetic field
into  the  ISM, thanks  to  stellar  winds  or supernova  explosions.
Although  the  efficiency  of   stellar  dynamo  is  rather  uncertain
\citep{brandenburg&subramanian05},  and the  magnetic  energy released
into supernova remants poorly constrained (\citealp{kennel&coroniti84}
for the Crab nebula  and \citealp{helfandetal01} for the Vela nubula),
this stellar origin represents a very appealing perspective to explain
a fast magnetic field generation in high-redshift galaxies.

Galactic dynamos, or  other  field generation  mechanisms  in the  ISM,
together with galactic winds,  should therefore play an important role
in the generation of magnetic fields in the IGM. In this respect, this
is probably  inside dwarf galaxies that this  double mechanism occurs:
dwarf galaxies are very numerous  in the early universe, and they host
most  of  the cosmic  material  at  early times  \citep{bertoneetal06,
  donnertetal09}. Dwarf  galaxies   are  also  easily  disrupted  by
galactic winds,  in contrast  to Milky Way  like galaxies,  from which
metal     and    magnetic     fields     probably    never     escaped
(\citealp{dubois&teyssier08winds} and references therein). \cite{kronbergetal99} proposed a scenario where primeval galaxies launch strong starbursts that generate a $5.10^{-3}\, \mu$G IG magnetic field. They assume that galactic winds are able to reach Mpc scales and that they substantially amplify the field during the outburst.

Understanding the  evolution of magnetic  fields in dwarf  galaxies is
therefore  an important  goal.  Performing MagnetoHydrodynamics  (MHD)
numerical simulations with  SPH, \cite{kotarbaetal09} studied the
amplification of  the magnetic field  by differential rotation  and by
the spiral pattern of the gaseous disc. Another step was undertaken by
\cite{wang&abel09}, who  performed MHD numerical  simulations with AMR
of a  dwarf galaxy  formed by  the collapse of  a cooling  halo.  They
studied in details the gas fragmentation and the magnetic field energy
amplification  due  to  turbulent,  vortical motions  induced  by  the
gravitational  instability in  the disc.   Starting with  an initially
uniform  field  of  $10^{-3}  \,  \mu$G in  the  halo,  they  reached
equipartition  after only  a few  rotations.  They  haven't simulated,
however, the  effect of supernova feedback in  driving strong outflows
out  of the  galaxy.  \cite{bertoneetal06},  on the  other  hand,
focused their study on the effect of galactic winds and how they might
enrich the  Universe with metals  and magnetic fields.  They  used for
that purpose a semi--analytical model, assuming equipartition magnetic
fields inside  their galactic discs and following  the winds evolution
through cosmic times.

In this paper,  our goal is to model the  evolution of magnetic fields
in a dwarf  galaxy together with the formation of  a galactic wind, in
order to compute  the amount of magnetic energy  that escapes from the
disc. This  outgoing magnetic enegy flux  is the key  quantity that we
need in order to estimate  the efficiency of IGM enrichment.  For that
purpose, we simulate an isolated,  cooling halo, in the same spirit of
\cite{dubois&teyssier08winds} and \cite{wang&abel09}, and form a small
galactic disc in the halo  centre.  We solve numerically the ideal MHD
equations, in the presence of self-gravity, with standard galaxy formation
physics (cooling, heating,  star formation, supernova feedback).  The
main  parameter in  our  study  is the  starting  magnetic field  that
permeates the initial halo. We therefore vary its strength, assuming a
simple but  realistic initial field  geometry.  We will  also consider
the case of zero initial magnetic  field in the halo, in a model where
seed fields are introduced within supernova bubbles, implementing the
stellar feedback scenario suggested  by \cite{rees87} in a way similar
to \cite{hanaszetaldip09}.   In   Sect.~\ref{numerics}  we  present  the
numerical set-up  of our simulations, as  well as some  details in our
implementation of galaxy formation physics, and in Sect.~\ref{ICs}, we
discuss    more    specifically    our   initial    conditions.     In
Sect.~\ref{resultsdisc}   and  \ref{resultsdynamo},  we   present  our
results  on  the magnetic  amplification  in  galactic  discs, and  in
Sect.~\ref{resultswind},  we  show how  the  IGM  can  be enriched  by
magnetised galactic winds. We finally comment the implications of this
work in Sect.~\ref{discussion}.

\section{Numerical methods}
\label{numerics}

\subsection{Gas dynamics with magnetic fields}

We use the Adaptive  Mesh Refinement code RAMSES \citep{teyssier02} to
solve the full set of ideal MHD equations
\begin{eqnarray}
\label{E_massMHD}
\frac{\partial \rho}{\partial{t}}+\nabla \, . \, (\rho \vec{u})&=&0\, ,\\ 
\label{E_momentumMHD}
\frac{\partial \rho \vec{u}}{\partial{t}}+
\nabla \, (\rho \vec{u} \otimes \vec{u} -\vec{B} \otimes \vec{B}  
+P_{\rm tot} \mathbb{I})&=&0\, ,\\
\label{E_energyMHD}
\frac{\partial E}{\partial{t}}+
\nabla\, . \, ((E+P_{\rm tot})\vec{u}-\vec{B}(\vec{B}.\vec{u}))&=&0\, , \\
\frac{\partial \vec{B}}{\partial{t}}-
\nabla\times (\vec{u} \times \vec{B}) &=&0\, ,
\end{eqnarray}
using a  Godunov scheme with Constrained Transport  that preserves the
divergence of the magnetic field \citep{teyssieretal06, fromangetal06}.
In  the   previous  system,  $\rho$  is  the   plasma  density,  $\rho
\vec{u}$  is  the plasma  momentum,  $\vec{B}$  is the  magnetic
field,  $E=0.5 \rho  u^2+E_{\rm th}+B^2/8\pi$  is the  total  energy and
$E_{\rm th}$ is  the thermal energy. This system  of conservation laws
is closed using  the Equation of State (EoS)  for an ideal mono-atomic
gas,    where   the    total    pressure   is    given   by    $P_{\rm
  tot}=(\gamma-1)E_{\rm  th}+B^2/8\pi$,  with $\gamma=5/3$.   The  MHD
solver in the RAMSES code has been tested using idealized cases, as well as
more realistic  astrophysical situations in  \cite{fromangetal06}.  It
has  been used  for  the first  time  in the  cosmological context  by
\cite{dubois&teyssier08cluster}  to study the  evolution of  a cooling
flow cluster,  for which  it produced results  in good  agreement with
previous  studies \citep{dolagetal05,  bruggenetal05}.  Because  it is
based  on  Constrained  Transport,  the  magnetic  divergence  exactly
vanishes in an integral sense: the total magnetic flux across each AMR
cell boundary is zero.  The  electric field is computed at cell edges,
using  a 2D  Riemann  solver,  in order  to  upwind the  edge-centered
electric field with  respect to the 4 neighboring  cell states. The 2D
Riemann solver  is based  on a generalisation  of the 1D  HLLD Riemann
solver  of  \cite{miyoshi05},  assuming  a 5-wave  piecewise  constant
Riemann  solution. We  also account  for self-gravity  in the  gas and
stellar distribution, assuming a  static potential for the dark matter
component.   The Poisson  equation  is solved  with isolated  boundary
conditions, using  Dirichlet boundary  conditions given by  a mutipole
approximation of the mass  distribution in the box.  The gravitational
force is  added as a source term  in both the momentum  and the energy
conservation equations.

\subsection{Cooling, star formation and supernova feedback}

In order to  describe the formation and evolution  of a galactic disc,
we model gas cooling  in the halo as well as in  the star forming disc
using  a  look--up  table  for  the  metallicity--dependent  radiative
cooling function of  \cite{sutherland&dopita93} for a mono--atomic gas
of H  and He with a  standard metal mixture. This  cooling function is
added as a sink  term in the energy equation~(\ref{E_energyMHD}).  The
metal mass  fraction is  advected as a  passive scalar, with  a source
term coming from supernova explosions in the simulated ISM.

Gas   is   not   allowed   to   cool   down   to   arbitrarily   small
temperature.    Following    the    general   ideas    presented    in
\cite{yepesetal97}  and  in   \cite{springel&hernquist03},  we  use  a
density--dependent  temperature floor  to account  for  the unresolved
turbulence in  the ISM, modelling this complex  multiphase medium with
an  effective EoS.  In this  paper, we  use the  following temperature
floor $T_{\rm min}=T_0  (n/n_0)^{2/3}$, where $n_0$ is  the star formation
density threshold  (see below) and  $T_0=10^4\, \rm K$. The  effect of
this temperature floor is to  prevent the gas from fragmenting down to
parsec scales,  well below our  resolution limit. This is  in contrast
with  the  study  presented   in  \cite{wang&abel09},  where  the  gas
temperature was allowed to cool  down to 300~K, triggering strong disc
fragmentation and massive clumps formation.

Star formation is  implemented using a standard approach  based on the
Schmidt law, spawning star particles  as a random Poisson process. The
reader      can     refer     to      \cite{rasera&teyssier06}     and
\cite{dubois&teyssier08winds}  for more details.   We choose  here the
following  star  formation  parameters:  an efficiency  of  $5\%$  per
free--fall--time and a star  formation density threshold of $n_0=0.1\,
\rm  H.cm^{-3}$.  Each  stellar  particle represent  a single  stellar
population  of total  mass $m_*  =n_0m_{\rm H}\Delta x^3  \simeq  10^4\, \rm
M_{\odot}$.  This stellar particle should therefore be considered as a
Stellar Cluster (SC) rather than an individual star.

After  several Myr,  the first  massive stars  explode  in supernova,
giving rise to large expanding bubbles and ultimately to the formation
of a large scale galactic wind. A mass fraction of $\eta_{\rm SN}=10\%$ in
massive  stars $M>8\, \rm M_{\odot}$ (taken from any standard Initial Mass Function, e.g. \citealp{kroupa01}) is considered  for  our  stellar clusters.   Modelling
supernova  feedback  is  a  sensitive numerical  issue  that  depends
crucially  on  the  adopted  spatial  resolution.   It  was  shown  by
\cite{ceverino&klypin09} that modelling  feedback using a pure thermal
dump  of energy  works only  at resolution  smaller than  50 pc  for a
clumpy interstellar  medium. In this  paper, we used for  our smallest
cells a physical  size of 150~pc. We therefore need to  rely on a more
sophisticated   approach   to   model   properly  the   formation   of
supernova--driven  bubbles. We  adopt in  this paper  the  now rather
standard {\it  kinetic} feedback  scheme, proposed by  various authors
\citep{springel&hernquist03,                    dubois&teyssier08winds,
  dallavecchia&schaye08}.  For  each stellar particle  formed, we also
form in the parent gaseous cell an additional collision-less particle,
mimicking the formation of a companion Molecular Cloud (MC).  The mass
of the MC is parametrized  by $m_{\rm MC}=\eta_{\rm w} m_*$.  The free parameter
$\eta_{\rm w}$, similar  to the one  defined in \cite{springel&hernquist03},
\cite{rasera&teyssier06} and \cite{dallavecchia&schaye08}, was choosen
to be $\eta_{\rm w}=2$.

We  assume here  that  after a  delay  of $t_{\rm SN}=10$~Myr,  supernova
feedback destroys the MC, and the corresponding mass, momentum, energy
and   metal   content   are   released  into   the   surrounding   gas
cells. Following  \cite{dubois&teyssier08winds}, we add  to each fluid
element a  Sedov blast wave  profile for density, momentum  and total
energy,  with   a  radius   equal  to  a   fixed  physical   scale  of
$r_{\rm SN}=300$~pc, or  2 AMR  cells. For metal  enrichment, we  assume a
constant yield  of $y_{\rm SN}=10  \%$ for any  exploding star.  A stellar
cluster  with  mass  $m_*=10^4\,   \rm  M_{\odot}$  and  zero  initial
metallicity  will  release  $y_{\rm SN}  \eta_{\rm SN}   m_*=10^2\,  \rm
M_{\odot}$ of metals into the ISM.

When  we form  a stellar  cluster, we  do not  remove any  large scale
magnetic  field component. We  assume that  at small  scales, magnetic
field is  expelled from the  collapsing MC by ambipolar  diffusion, so
that we can neglect the magnetic  field that is trapped into the newly
formed stars.  Similarly, when supernova-driven blast  waves form, we
do not modify  the magnetic field in their  surroundings. The magnetic
field evolution is  therefore driven only by the  large scale velocity
field.  In only one  case do  we consider  supernova seed  fields, as
explained  in the  Appendix, in  order  to test  the stellar  magnetic
feedback scenario proposed by \cite{rees87}.

\section{Initial and boundary conditions}
\label{ICs}

In  this section, we  decribe our  initial conditions,  especially our
initial magnetic topology.  We also  decribe in details how we enforce
the  magnetic  divergence  free  condition  at  the  boundary  of  the
computational domain.

\subsection{Initial hydrostatic halo structure}

\begin{figure*}
\centering{\resizebox*{!}{8cm}{\includegraphics{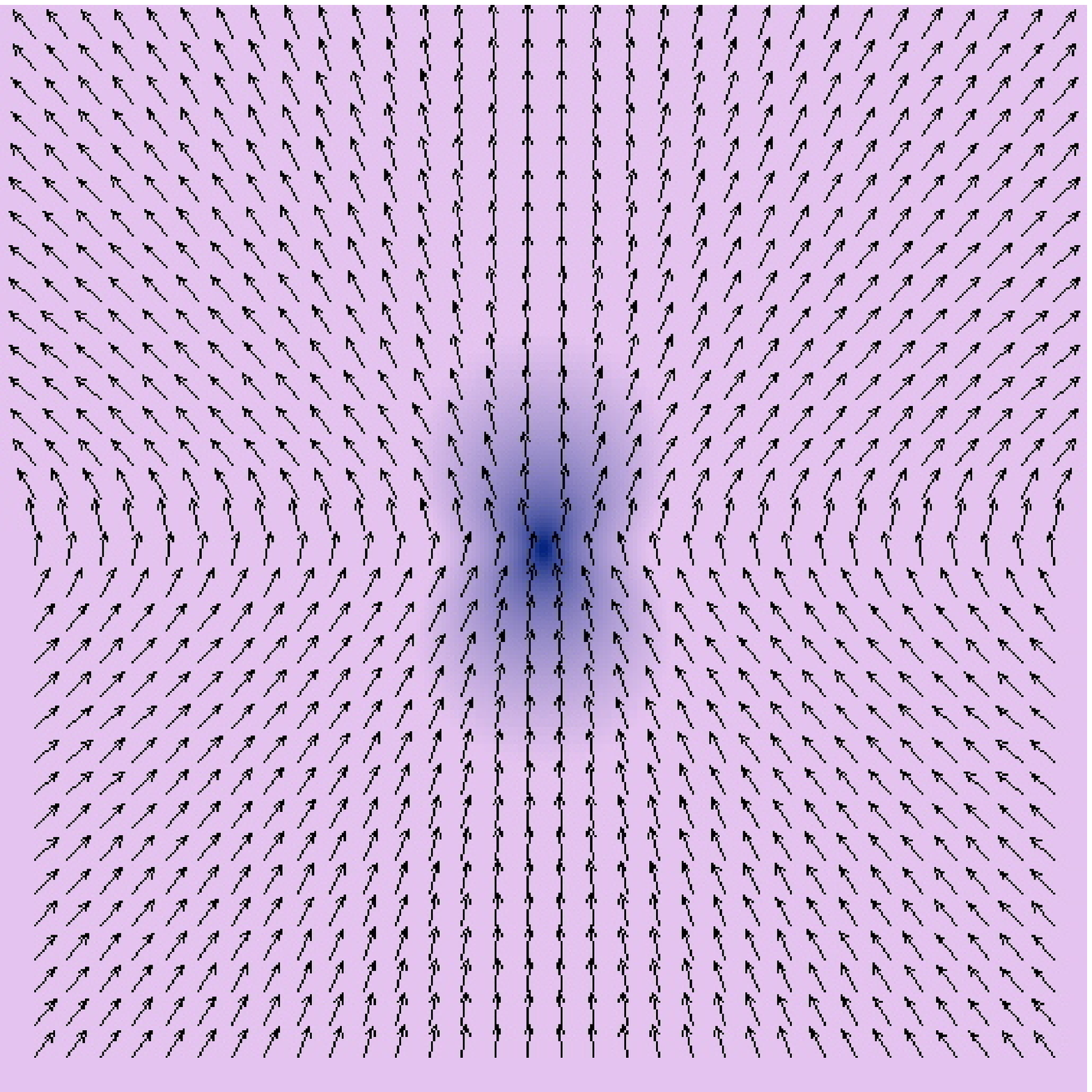}}}
\centering{\resizebox*{!}{8cm}{\includegraphics{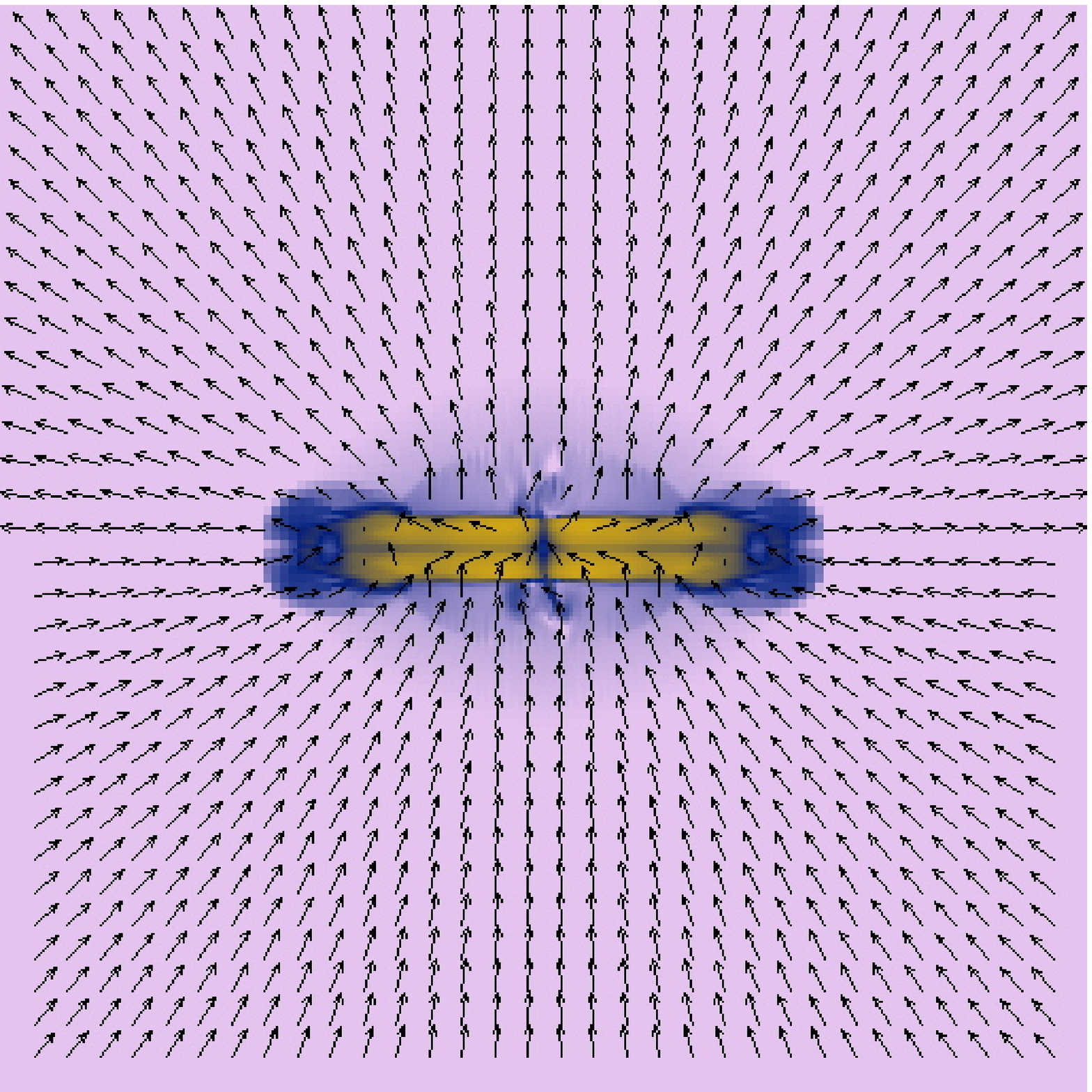}}}
\centering{\resizebox*{!}{8cm}{\includegraphics{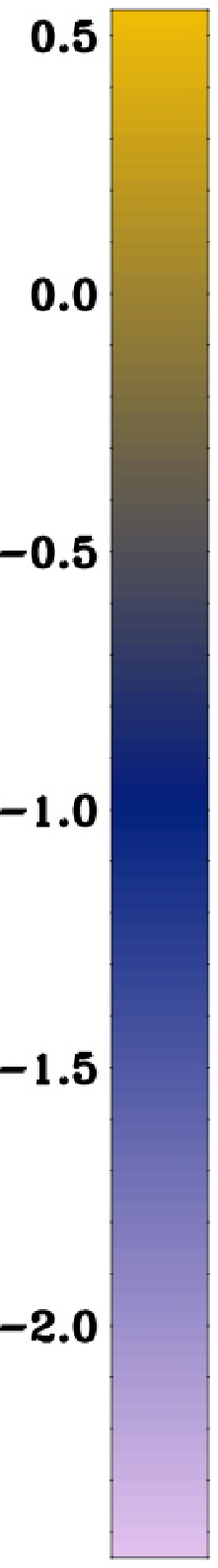}}}
\caption{(Oyz) slice of the  initial magnetic field amplitude in $\log
  \mu$G units  and magnetic  vectors for $B_{\rm  IGM}=10^{-5}\mu$G at
  $t=0$ Gyr (left  pannel) and $t=3$ Gyr (right  pannel). Picture size
  is 40 kpc. }
\label{Bini_snapshot}
\end{figure*}

We use  the same set of  initial conditions as the  one presented in
\cite{dubois&teyssier08winds},  namely an  isolated  dark matter  halo
hosting a  cooling gaseous  component.  This idealized  situation have
been  used many  times to  study various  aspects of  galaxy formation
\citep{springel&hernquist03,  dubois&teyssier08winds, kaufmannetal09}.
Let us  recall here only the  most important aspects.   Dark matter is
modelled  with   a  static   density  distribution  following   a  NFW
(\citealp{navarroetal96}) profile such that
\begin{equation}
\rho={\rho_s \over r/r_s (1+r/r_s)^2 }\, ,
\end{equation}
where $\rho_s$  and $r_s$ are  the characteristic density  and radius.
This profile  can also be parametrized  in terms  of concentration
$c=r_{200}/r_s$ and Virial mass as
\begin{equation}
M_{200}=4\pi \rho_s r_{200}^3 
\left ( \log \left ( 1+c \right ) + {1 \over 1+c} \right) \, , 
\end{equation}
within the virial  radius $r_{200}$.  The virial mass  is defined here
at  an overdensity  equal  to  200 times  the  {\it critical}  density
$\rho_c$. We also assume $H_0= 70 \, \rm km/s/Mpc$.

The gas is initially in  hydrostatic equilibrium with the same profile
than  dark  matter, with  a  total  gas  fraction of  $f_b=15\%$.   We
consider that  the halo is  slowly rotating, with an  angular momentum
distribution  consistent with  the average  specific  angular momentum
profile measured in cosmological simulations \citep{bullocketal01}
\begin{equation}
j(r)=j_{max} {M(<r) \over M_{200}} \, ,
\label{ang_mom}
\end{equation}
with a spin parameter 
\begin{equation}
\lambda={J     \vert     E\vert^{1/2}\over GM^{5/2}_{200}}\, .
\end{equation}

Using   this  particular   setting,    \cite{springel&hernquist03} and
\cite{dubois&teyssier08winds} 
showed that  large-scale outflows are  produced only in  low-mass halos
with $M_{200}\la10^{11}\, \rm M_{\odot}$.  As we are interested in the
role of  galactic winds in the  enrichment of the IGM,  we consider in
this paper only the case  of a small mass halo with $M_{200}=10^{10}\,
\rm  M_{\odot}$.  The  halo  spin  parameter  is  chosen  to  be
$\lambda=0.1$.

The  box  size  is  set  equal   to  3  times  the  Virial  radius  or
$L=150.7$~kpc, and is  covered by a coarse grid  with $64^3$ cells. We
consider  4 additional  levels of  refinement, so  that  our effective
resolution  reaches  $1024^3$ or  $\Delta  x=147$~pc.  Our  refinement
strategy is based on two criteria. First, we refine a cell if its mass
exceeds 8  times our mass  resolution indicator $m_{\rm  res}=2 \times
10^{-6} M_{200}$.  Second, we refine  a cell is the local Jeans length
falls below  4 cells,  so that we  prevent the disc  from artificially
fragmenting \citep{trueloveetal97, bate&burkert97, machaeketal01}. The
initial AMR grid  was built from the coarse  mesh with this refinement
strategy.  We start initially  with a total $\sim  2.10^6$ cells  which number
remains roughly constant  with time. The number  of cells in the
maximum level  of refinement (mainly  located in the gaseous  disc) is
$\sim 4.10 ^4$. After 3 Gyr the total number of stars is $\sim 10^5$.

\subsection{Initial magnetic field}

Although the  choice of the  initial magnetic field is  quite crucial,
the magnetic field topology is usually chosen quite arbitrarily in the
literature.  Using an initial set up very similar to the one presented
here, \cite{wang&abel09} considered  a constant initial magnetic field
$B_z=10^{-9}$~G.   If the  initial halo  field is  the product  of the
collapse of some large  scale cosmological structure, the field should
rather scale as $\rho_{\rm gas}^{2/3}$. This scaling has been observed
in    cluster--size    halo    in   cosmological    MHD    simulations
\citep{sigletal04,             dolagetal05,             bruggenetal05,
  dubois&teyssier08cluster}.   In the  spherically symmetric  set-up we
are considering,  each collapsing gas shell will  compress and amplify
significantly the  frozen-in field, before any  disc-like dynamo comes
into  play.  The  initial magnetic  field amplitude  {\it  profile} is
therefore    quite    important.      In    a    different    context,
\cite{kotarbaetal09}  considered the case  of a  pre-existing galactic
disc.  The initial  magnetic field  was also  constant, but  this time
$B_x=$~constant, the field  lines lying exacty in the  disc plane.  As
argued by these authors,  differential rotation quickly takes over and
create a strong toroidal field  in the star--forming disc, so that the
initial topology is quickly forgotten.

We  nevertheless propose  here to  consider a  more  realistic initial
magnetic  field profile, with  a dipole--like  topology.  In  order to
preserve the  $\nabla . \vec  B=0$ constraint we compute  the magnetic
field from its potential vector $\vec A$
\begin{equation}
\vec{B}= \nabla \times \vec{A}\, .
\label{BrotA}
\end{equation}
$\vec  A$ is  computed assuming  that each  cell in  the computational
domain contain  a small magnetic dipole with  strength proportional to
$\rho_{\rm gas}^{2/3}$. Every dipole have the same direction along the
z axis, so  that they all add up to a  global large scale dipole--like
topology on the halo scale.  The potential vector is written as
\begin{equation}
\vec{A}(x,y,z)=B_{\rm IGM} 
\left[ \frac{\rho_{\rm gas}(r)}{\rho_{\rm IGM}}\right]^{2/3}
\left ( \begin{array} {c}
-y \\
+x \\
0 
\end{array} \right ) \, ,
\end{equation}
\noindent  where $\rho_{\rm  IGM}=  \Omega_{\rm b} \rho_c$, with the baryon density $\Omega_{\rm b}=0.04$ and  $r$ is  the
spherical radius.  Each component  of the potential vector is averaged
over the corresponding  cell edge, and the magnetic  field is computed
using the integral  form of the curl operator on  each cell face.  One
can see  on Figure~\ref{Bini_snapshot}  that the initial  magnetic field
topology is roughly dipolar, preferentially aligned with the $z$ axis,
and that the field strength decreases strongly with radius.

Figure~\ref{Bini_profile} shows the magnetic field profile along the z
axis for  different values  of the magnetic  field in the  IGM $B_{\rm
  IGM}$.  The  magnetic field strength  decreases down to 2  orders of
magnitude   from  the   halo  centre   to  the   outer  parts   at  $2
r_{200}$. Results from cluster  scale cosmological simulations seem to
favor values  of $B_{\rm IGM} \simeq  10^{-5}~{\rm -}~10^{-4}~\mu$G in
order to explain the observed  magnetic field strength inside the core
of     large      X--ray     clusters     (\citealp{carilli&taylor02},
\citealp{govoni&feretti04}  and  references  therein). This  value  is
however to  be considered as  an upper limit since  these cosmological
simulation have  limited resolution and are  believed to underestimate
the magnetic  field amplification  from the initial  IGM value  to the
final cluster's core strength. In  this paper, $B_{\rm IGM}$ is a free
parameter that we  allow to vary between $10^{-7}~\mu$G,  deep into the
pure induction  regime for which the Lorentz  force remains negligible
and  $10^{-4}~\mu$G,  our  maximum value.   \cite{wang&abel09}  initial
magnetic field of $B_z=10^{-3}~\mu$G in the halo corresponds roughly to
our  case  with $B_{\rm  IGM}  \simeq  3\times10^{-5}~\mu$G, using  the
$\rho^{2/3}$ scaling and a mean overdensity of 200 in the halo.

\subsection{Zero--gradient boundary conditions}

A complex numerical  issue for MHD flows is  to design proper boundary
conditions.   We are  dealing  with  an isolated  halo,  for which  we
consider  that the  environment  is a  very  low density,  homogeneous
medium. It  is convenient in this  case to use  {\it outflow} boundary
conditions: gas  is allowed  to flow freely  out of  the computational
domain.   Outflow  boundary  conditions  can be  implemented  using  a
zero-gradient condition at the  domain boundary for all hydro variable
(density, pressure  and the  3 components of  the velocity).   For the
magnetic  field, however,  we  need to  be  more careful  in order  to
enforce  the  divergence free  condition.   We  use the  zero-gradient
condition   only  for   the  transverse   magnetic   field  components
$B_{\perp}$ (perpendicular to the  unit vector normal to the surface),
and  we   use  a  linear   extrapolation  for  the   normal  component
$B_{\parallel}$ (parallel to  the unit vector). We make  sure that the
divergence of the magnetic field takes  the same value on each side of
the domain boundary, namely zero.

This particular set of boundary conditions imposed on the magnetic field do not rigourously ensure that no electromagnetic energy flux crosses the boundaries inwards (non-vanishing Poynting flux).
We have checked that our results are not polluted by any artificial increase of the magnetic energy in the simulation coming from this choice of boundary conditions.
The total magnetic energy increase during the whole simulation course coming from the boundaries is evaluated to be $10^{-6}$ smaller than the magnetic energy increase measured in the galaxy at the end of the simulation.

\section{Disk formation with magnetic field}
\label{resultsdisc}

When radiative cooling  is turned on, our hydrostatic  halo cools down
from  the inside  out. Each  shell looses  pressure support  and sinks
towards the centre  of the halo.  The magnetic  field lines are frozen
in the  free--falling plasma. Although  initially the field  lines are
mainly  vertical  (perpendicular  to  the  future  galaxy),  they  are
entrained and bent by the collapsing gas, so that the radial component
finally dominates.   Figure~\ref{Bini_snapshot} compares  the topology  of the
initial  magnetic  field  and   the  magnetic  field  after  3~Gyr  of
evolution. One clearly sees that the field lines close to the disc are
almost  horizontal  (with  a  strong radial  component).   The  radial
component  changes   sign  above  and  below  the   plane,  a  natural
consequence of the pinching of  the field lines by the collapsing gas.
Because of  angular momentum conservation, each  collapsing shell will
speed up, flatten and form  a centrifugally supported disc. As we will
show  below,  the galactic  velocity  field  will  quickly generate  a
toroidal field component. This toroidal  field is a consequence of the
field lines being wrapped up and amplified by the rotating disc.

One should also note that the halo magnetic field keep its initial magnetic dipole-like structure, even after 3 Gyr evolution. This a direct consequence of the choice of our initial conditions. We will see that when a galactic wind develops, it strongly suppress the memory of the initial configuration of the field.

\begin{figure}
\centering{\resizebox*{!}{7.5cm}{\includegraphics{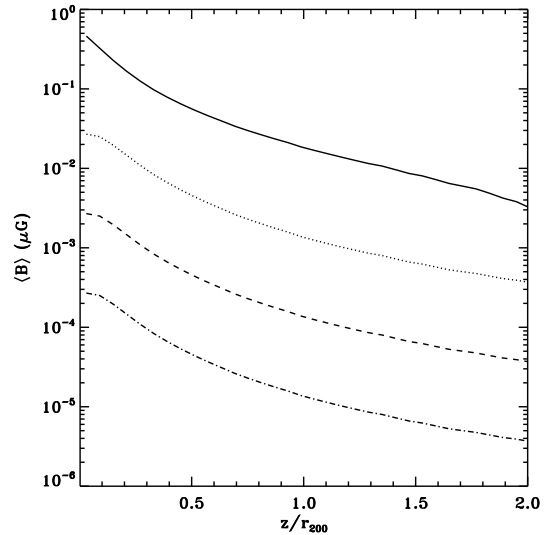}}}
\caption{Magnetic   amplitude   along  the   z   axis  for   different
  normalisations  of the  initial profile:  $B_{\rm  IGM}= 10^{-6}\mu$G
  (dot-dashed),   $B_{\rm  IGM}=10^{-5}\mu$G   (dashed)   and  $B_{\rm
    IGM}=10^{-4}\mu$G  (dotted). The solid  line shows  for comparison
  the magnetic field that  would have been in equipartition throughout
  the halo.}
\label{Bini_profile}
\end{figure}

We  show  in  Figure~\ref{Bini_profile}  the  initial  magnetic  field
amplitude along  the z axis of the  halo. We consider in  this study 3
different  cases:  $B_{\rm IGM}=10^{-6}~\mu$G, $B_{\rm IGM}=10^{-5}~\mu$G  and
$B_{\rm IGM}=10^{-4}~\mu$G. We also  show in Figure~\ref{Bini_profile} for
comparison the  magnetic field  corresponding to equipartition  in the
initial halo.  We define here the standard plasma parameter
\begin{equation}
\beta={8 \pi p \over B^2} \, ,
\end{equation}
which  measure the  ratio of  the  thermal to  the magnetic  pressure.
Equipartition     means     here     $\beta=1$.    We     see     from
figure~\ref{Bini_profile}  that  for  our  3  scenarios,  the  initial
magnetic field is well below equipartition.

\begin{figure*}
\centering{\resizebox*{!}{5.5cm}{\includegraphics{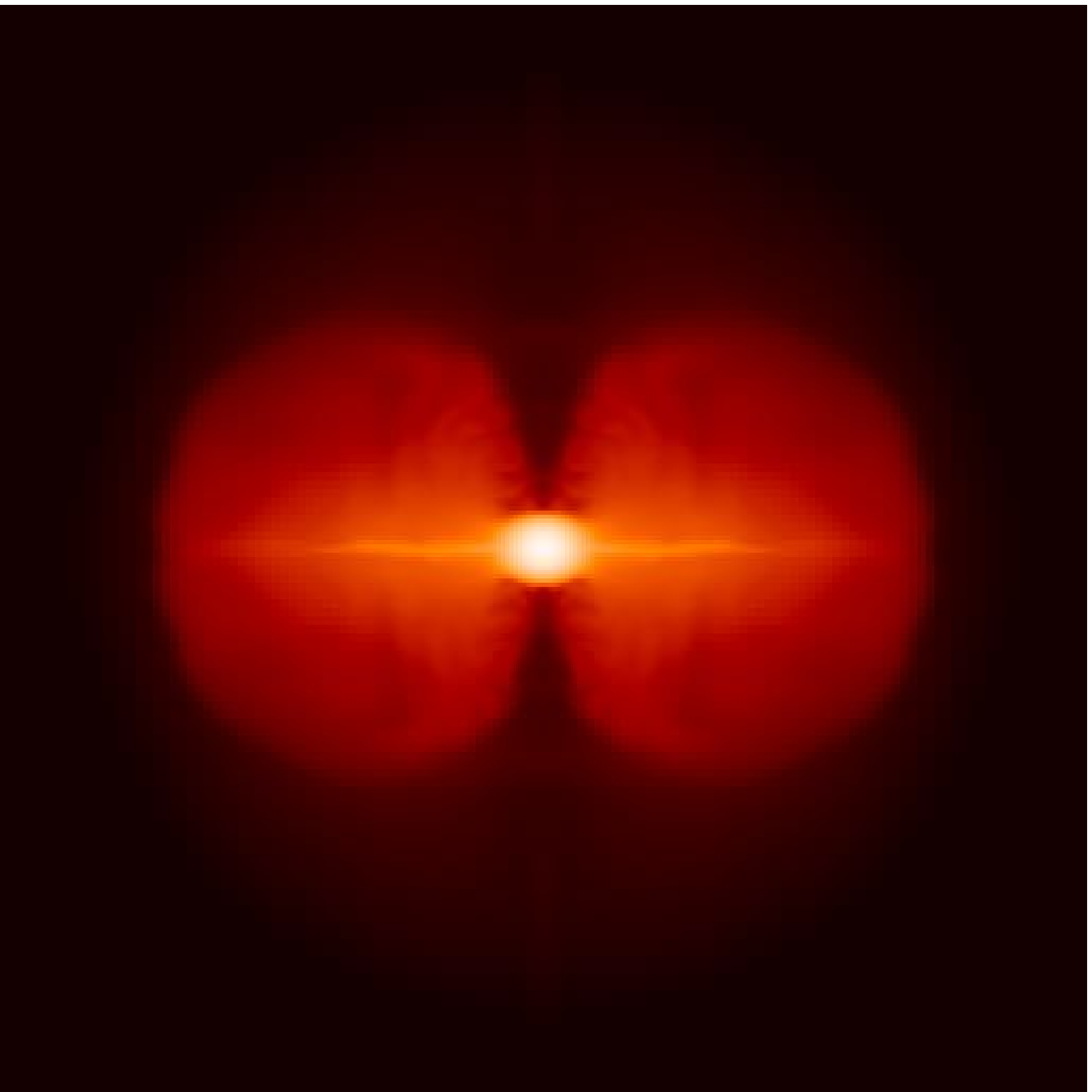}}}
\centering{\resizebox*{!}{5.5cm}{\includegraphics{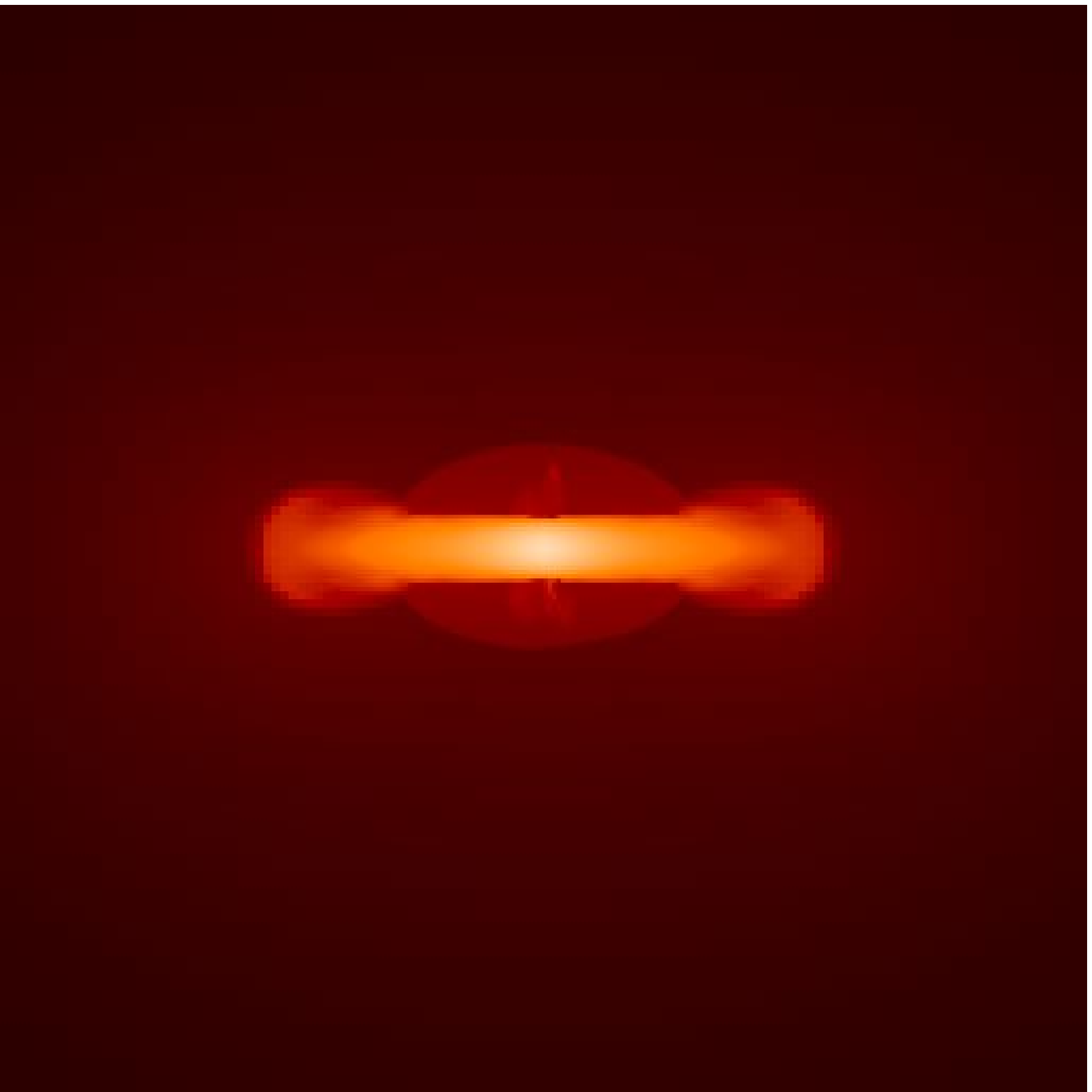}}}
\centering{\resizebox*{!}{5.5cm}{\includegraphics{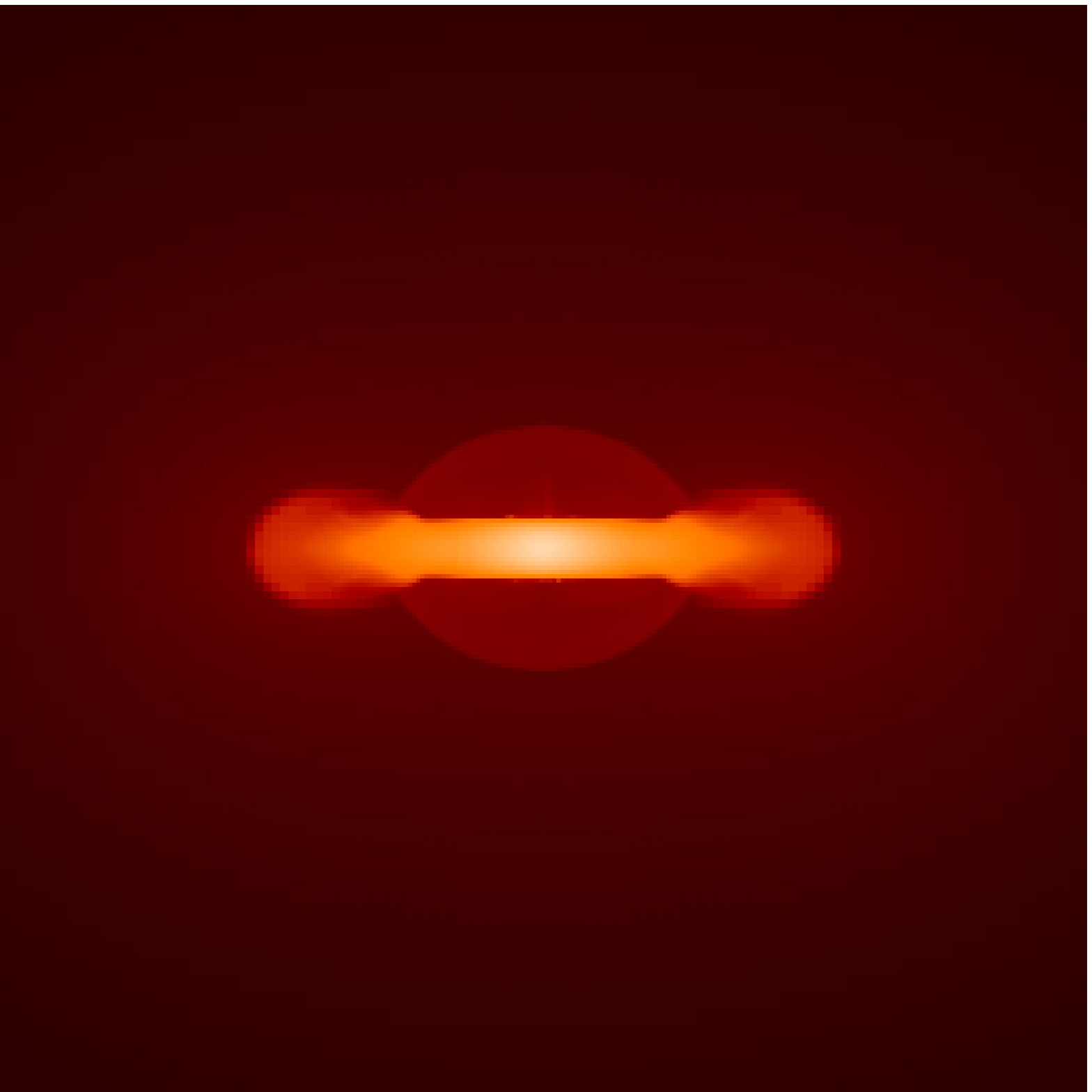}}}
\centering{\resizebox*{!}{5.5cm}{\includegraphics{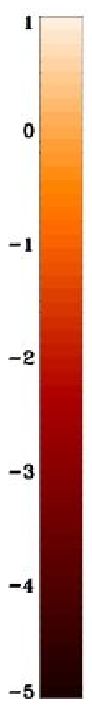}}}\\
\centering{\resizebox*{!}{5.5cm}{\includegraphics{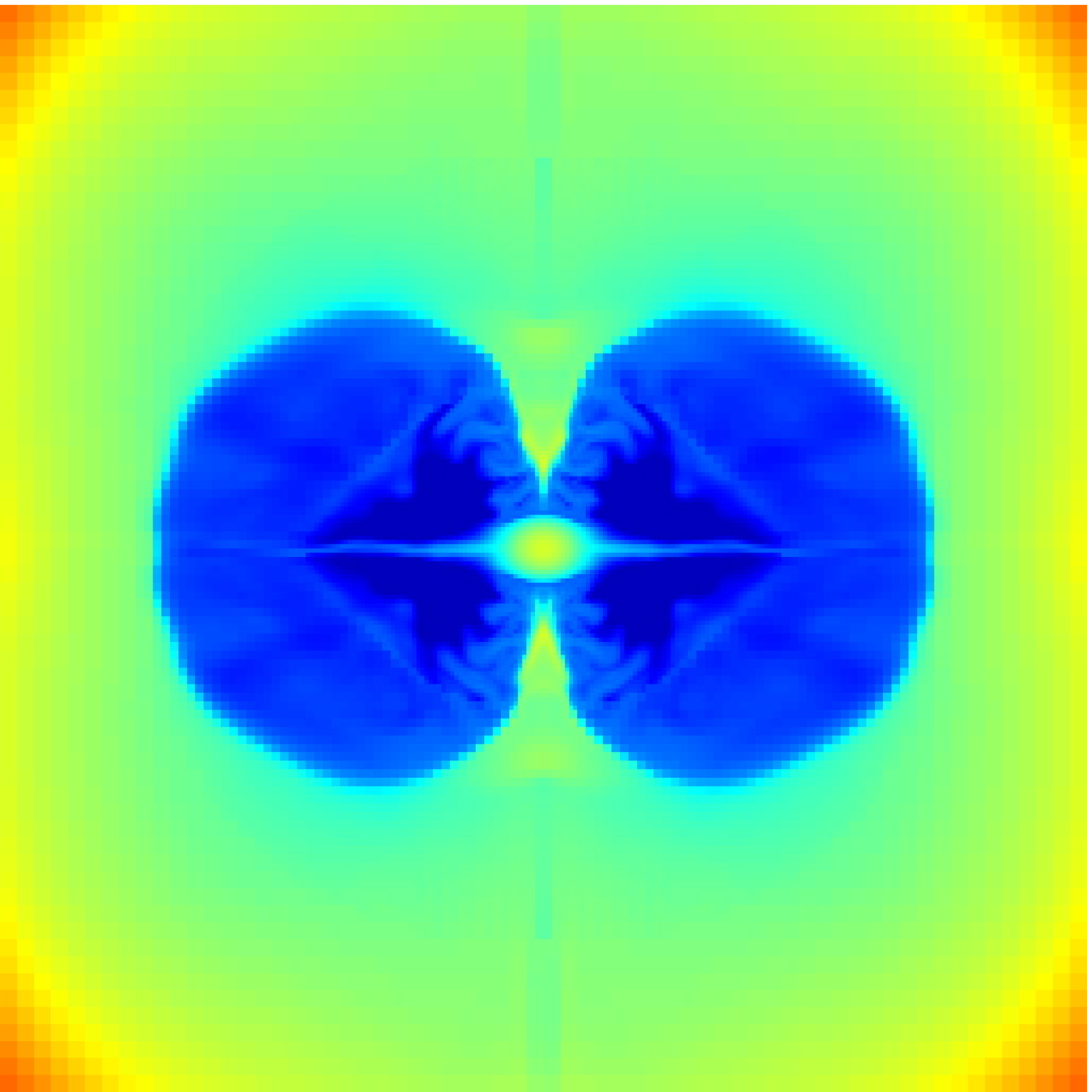}}}
\centering{\resizebox*{!}{5.5cm}{\includegraphics{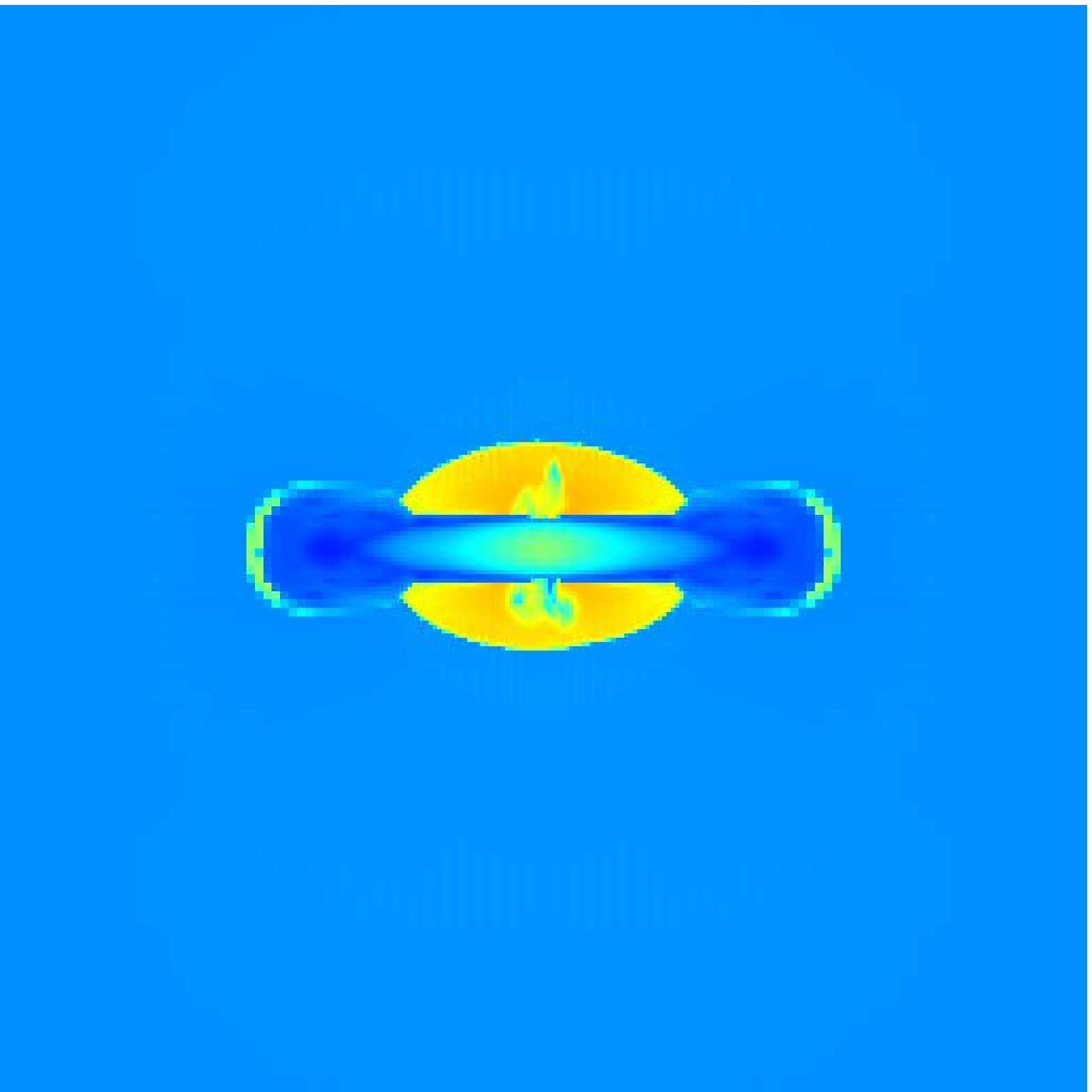}}}
\centering{\resizebox*{!}{5.5cm}{\includegraphics{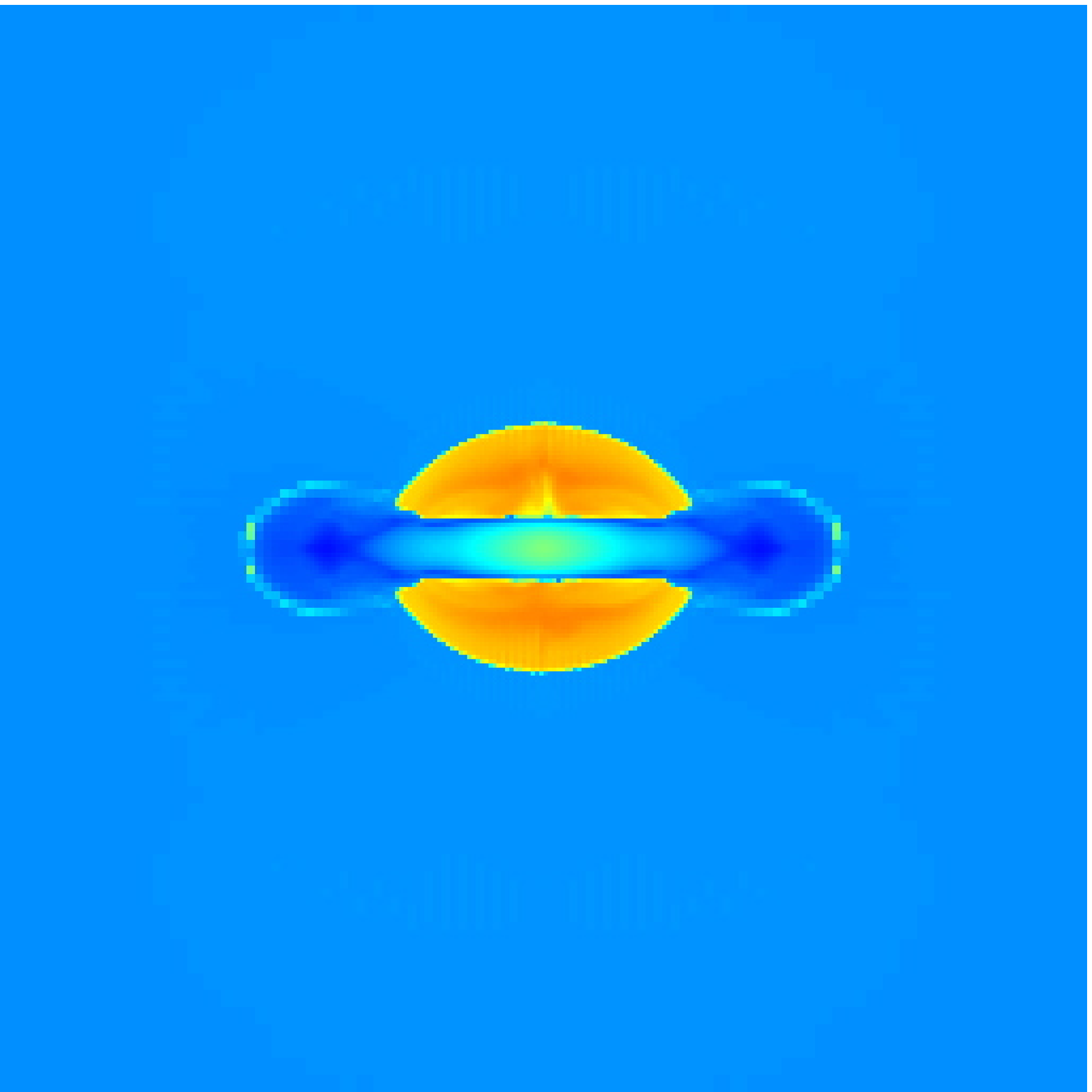}}}
\centering{\resizebox*{!}{5.5cm}{\includegraphics{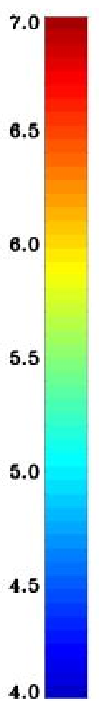}}}\\
\centering{\resizebox*{!}{5.5cm}{\includegraphics{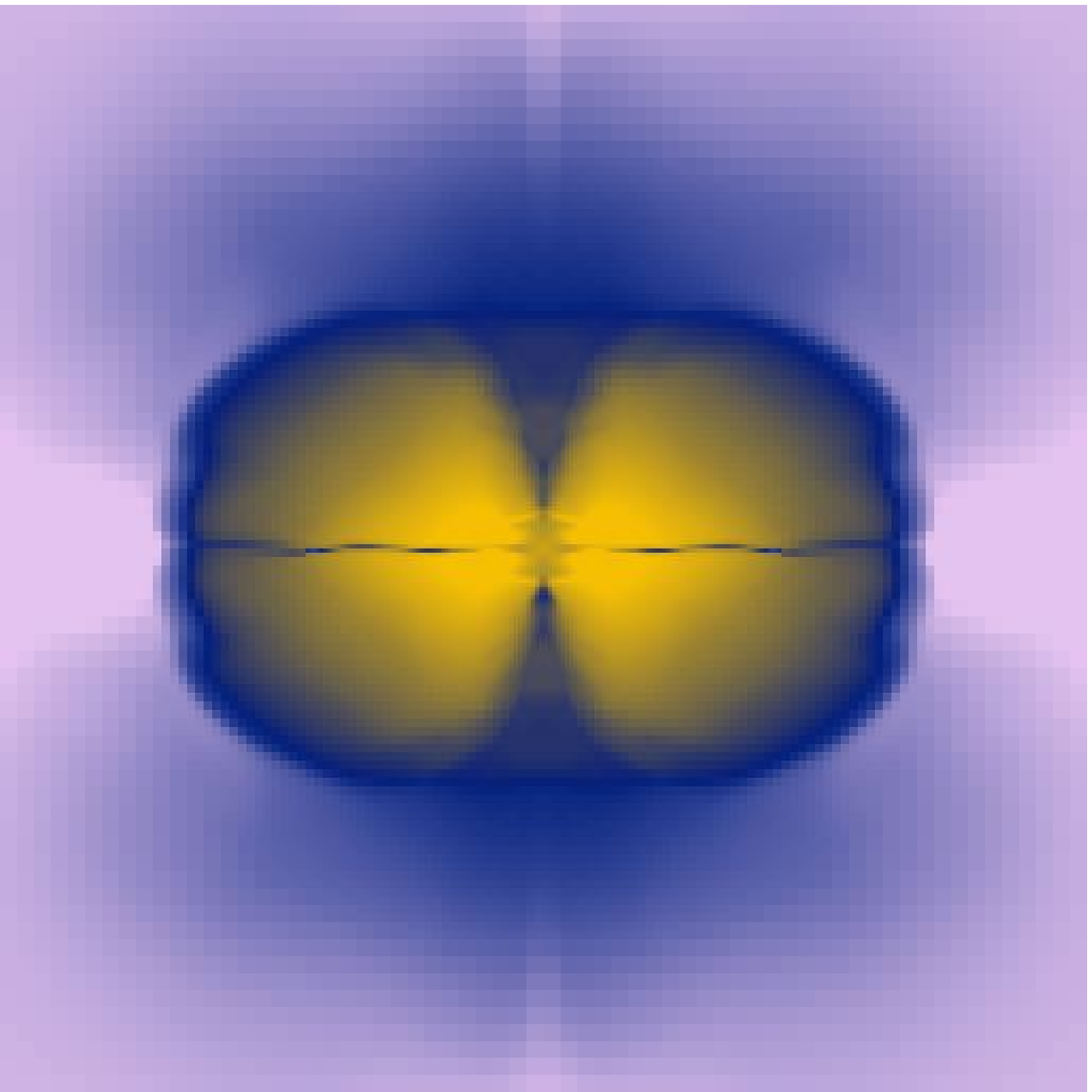}}}
\centering{\resizebox*{!}{5.5cm}{\includegraphics{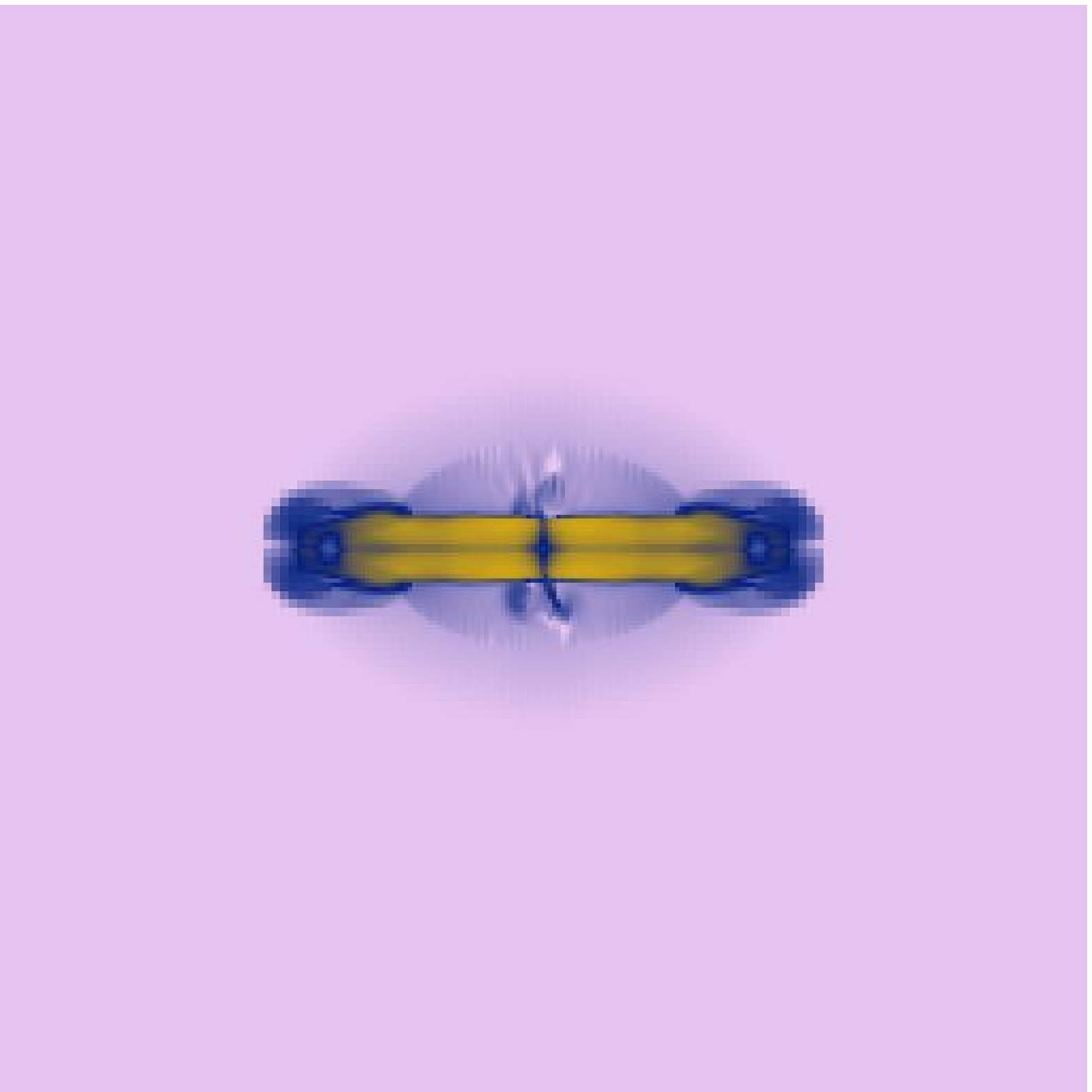}}}
\centering{\resizebox*{!}{5.5cm}{\includegraphics{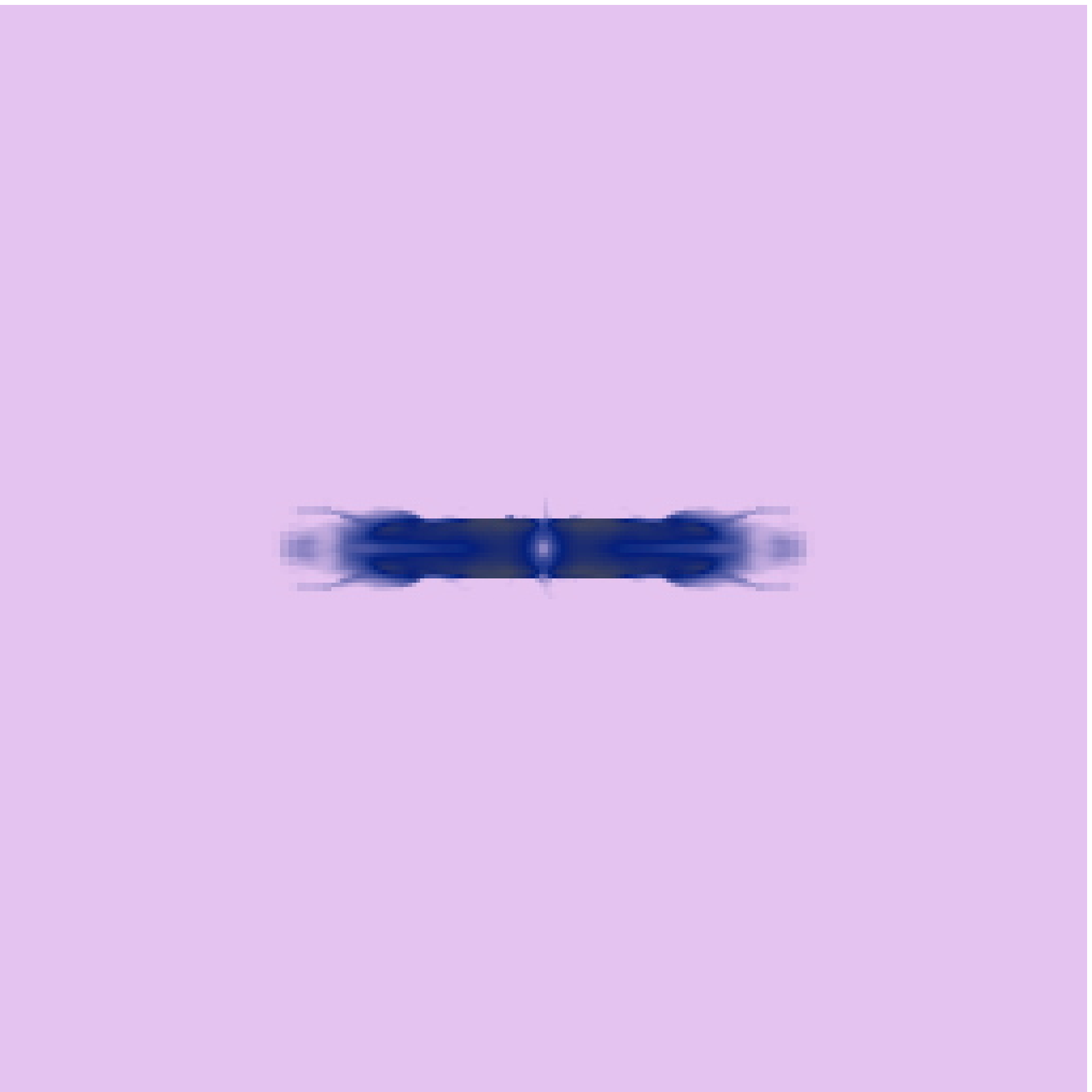}}}
\centering{\resizebox*{!}{5.5cm}{\includegraphics{./IM_COOL/BINI1d-2/magneticX_table.ps}}}\\
\centering{\resizebox*{!}{5.5cm}{\includegraphics{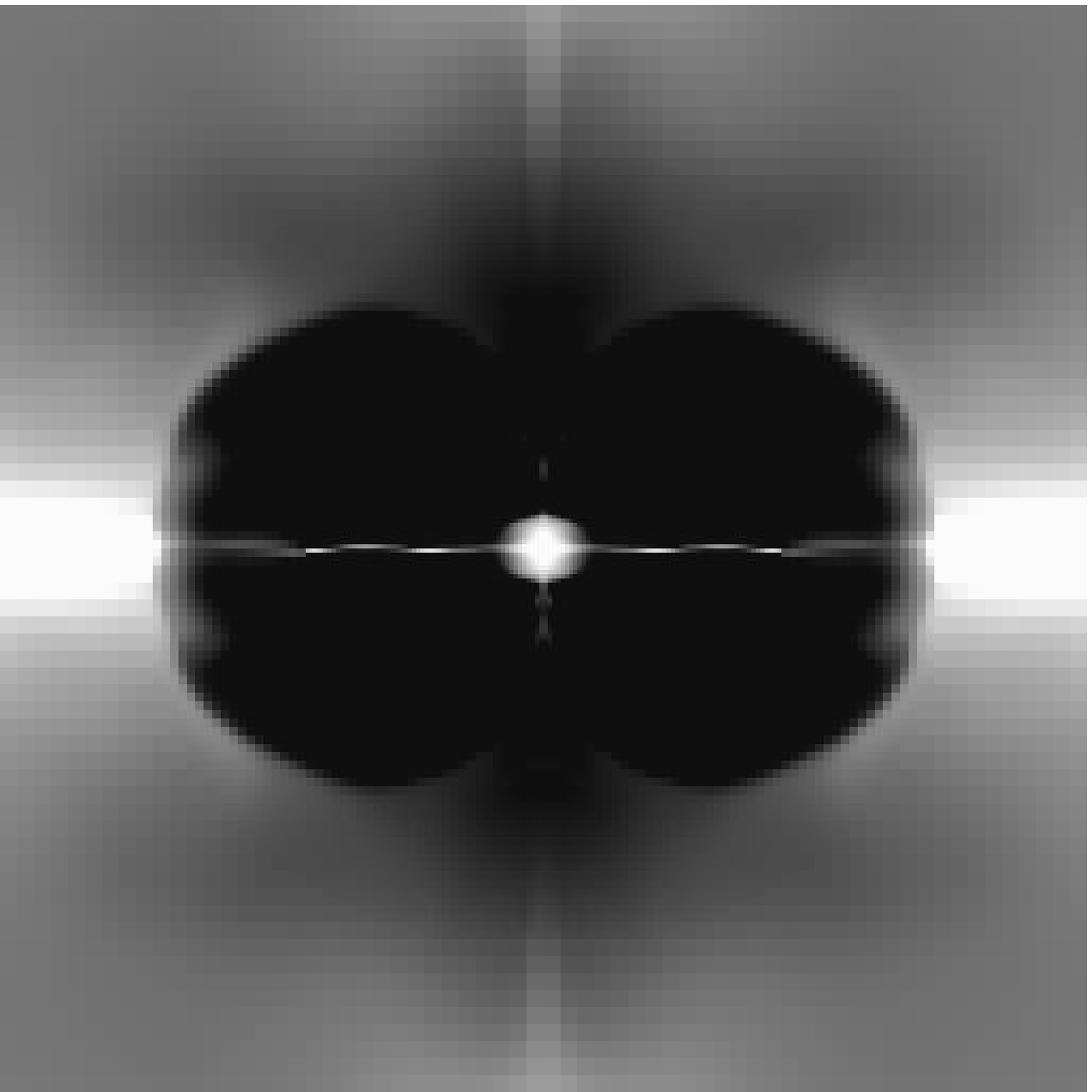}}}
\centering{\resizebox*{!}{5.5cm}{\includegraphics{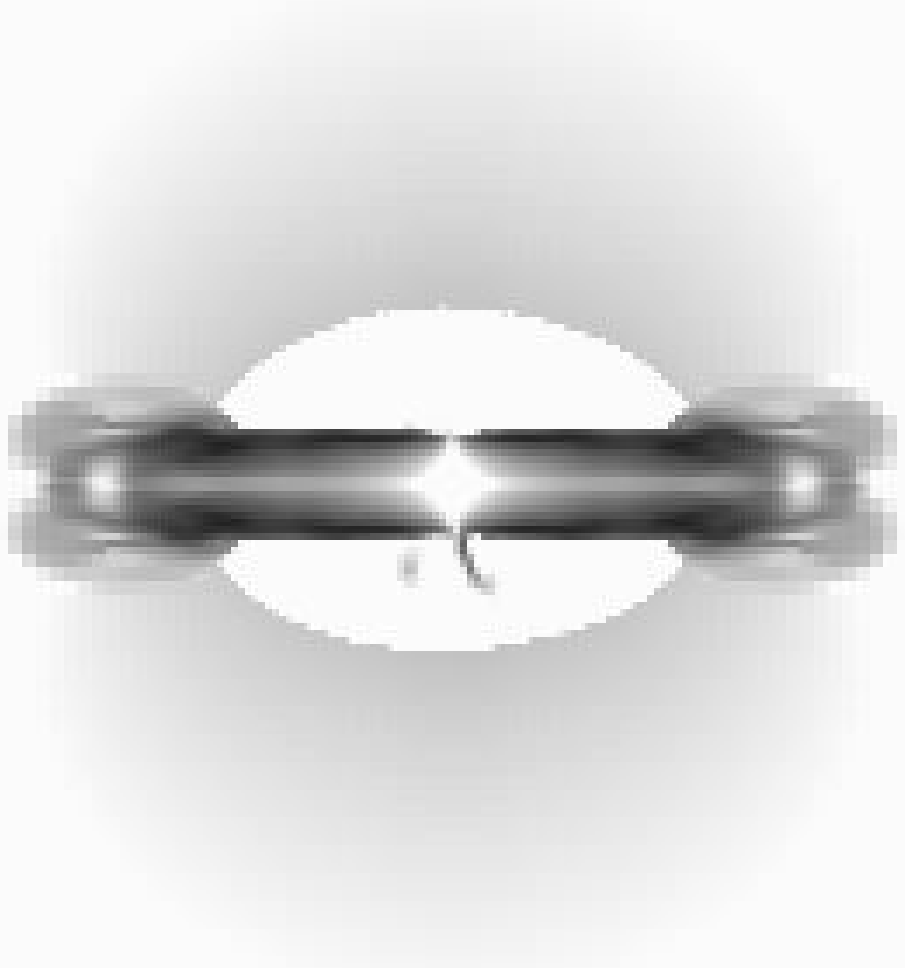}}}
\centering{\resizebox*{!}{5.5cm}{\includegraphics{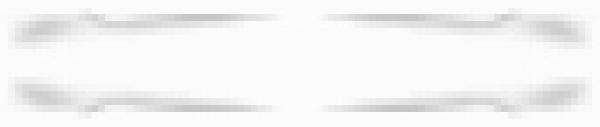}}}
\centering{\resizebox*{!}{5.5cm}{\includegraphics{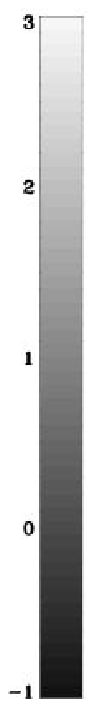}}}
\caption{Slice through the (Oxz) plane of the gas density in $\log \rm
  H/ cm^{-3}$ (first row), of  the temperature in $\log \rm K$ (second
  row), of the magnetic amplitude in  $\log \rm \mu G$ (third row) and
  the $\beta$  parameter in $\log  \beta$ (fourth row) for  an initial
  magnetic  field  of  $B_{\rm  IGM}=10^{-4}\,  \mu$G  (left  column),
  $B_{\rm   IGM}=10^{-5}\,   \mu$G   (middle   column)   and   $B_{\rm
    IGM}=10^{-6}\,  \mu$G (right  column) at  $t\simeq3$  Gyr. Picture
  size is 40 kpc.}
\label{gal_sat_snapshots}
\end{figure*}

After the disc forms, the magnetic field is amplified by rotation.  If
the  resulting   field  is  strong  enough,  the   Lorentz  force  can
significantly affect the dynamics of the flow, and alter the formation
of the  disc itself. The simulation with  $B_{\rm IGM}=10^{-4}\, \mu$G
is quite extreme in this respect: the magnetic field is so strong that
it prevents  the formation of the thin,  centrifugally supported disc,
leaving instead a low density,  highly magnetised torus, in which star
formation  is  very weak.   The  magnetic  field  lines become harder to deform by gas motions, except along  the z axis, where field lines
remain  vertical and  do not  prevent  gas from  collapsing. This  is
clearly shown in  Figure~\ref{gal_sat_snapshots}, where we can compare
the  disc morphology  in our  3 scenarios.The  3 maps  of  the $\beta$
parameter are usefull to  discriminate between the high magnetic field
case $B_{\rm  IGM}=10^{-4}~\mu$G, for which  $\beta$ is very  small in
most of the torus region, demonstrating that the torus is magnetically
supported  against gravity, and  the low  magnetic field  case $B_{\rm
  IGM}=10^{-6}~\mu$G,    for   which   the    disc   we    obtain   is
indistinguishable  from the pure  hydro case,  and the  magnetic field
inside  the disc  is one  order of  magnitude below  the equipartition
value.  Only in the intermediate case $B_{\rm IGM}=10^{-5}~\mu$G do we
get  a thermally  supported disc,  with  a magnetic  field roughly  in
equipartition only inside the disc ($\beta \simeq 1$).

It is quite interesting to  see thanks to these simulations that dwarf
galaxies  formation  can  be  almost  suppressed  (or  at  least  star
formation within them)  by a strong enough magnetic  field in the IGM.
It is even  more interesting to notice that the  value of the magnetic
field that  we require from cosmological simulation  of galaxy cluster
($B_{\rm IGM}  \simeq   10^{-5}~-~10^{-4}~\mu$G)  is  precisely   the  same
critical value above which dwarf galaxy formation can be regulated, if
not  suppressed,  by  magnetic   fields.   This  point  has  important
consequences that we will address in the conclusion.

\section{A Galactic Dynamo~?}
\label{resultsdynamo}

We  now discuss  more  quantitatively the  amplification mechanism  of
magnetic energy  in the  galactic disc.  We  use for that  purpose the
so--called     ``Galactic     Dynamo''     theory     \citep{parker71,
  wielbinski&krause93,  shukurov04}, for  which the  mean  large scale
magnetic field evolution is governed by a modified induction equation
\begin{equation}
\frac{\partial \vec{B}}{\partial{t}}=
\nabla\times (\vec{u} \times \vec{B}) + \nabla \times (\alpha \vec{B})
+ \nabla \times (\eta \nabla \times \vec{B})\, ,
\end{equation}
where  parameter $\alpha$  represents the  field amplification  due to
small scale turbulent motions  driven by various small scale phenomena
(supernova  explosion,  cloud-cloud  collision or  collapsing  vortex
modes  as explained  in  \citealp{ferriere92a}, \citealp{efstathiou00}
and  \citealp{kulsrud&zweibel08})   and  parameter  $\eta$  represents
magnetic diffusivity induced  by turbulent transport and reconnection.
The standard  $\alpha$-$\Omega$ dynamo  theory relies on  the $\alpha$
term to amplify  the radial magnetic field $B_r$  and on the large-scale induction
term  to amplify  the toroidal  field $B_\theta$.   In the
present simulation,  however, we do not include  any explicit $\alpha$
and $\eta$ terms in our  induction equation.  The original goal was to
rely  on our implementation  of supernova  feedback to  induce helical
motions  self-consistently.  Probably  because of our  limited resolution
(around 150~pc) and  the fact that we use a subgrid model based on an
effective EoS, we didn't  get any $\alpha$ effect: field amplification
in our case is due only to differential rotation. \cite{gresseletal08}, simulating a local patch a galaxy with much more resolution elements, obtained a galactic dynamo driven by supernovae explosion and differential rotation. They outlined that the rotation is the critical driver for the dynamo to operate an exponential growth of the field.

\begin{figure}
\centering{\resizebox*{!}{8cm}{\includegraphics{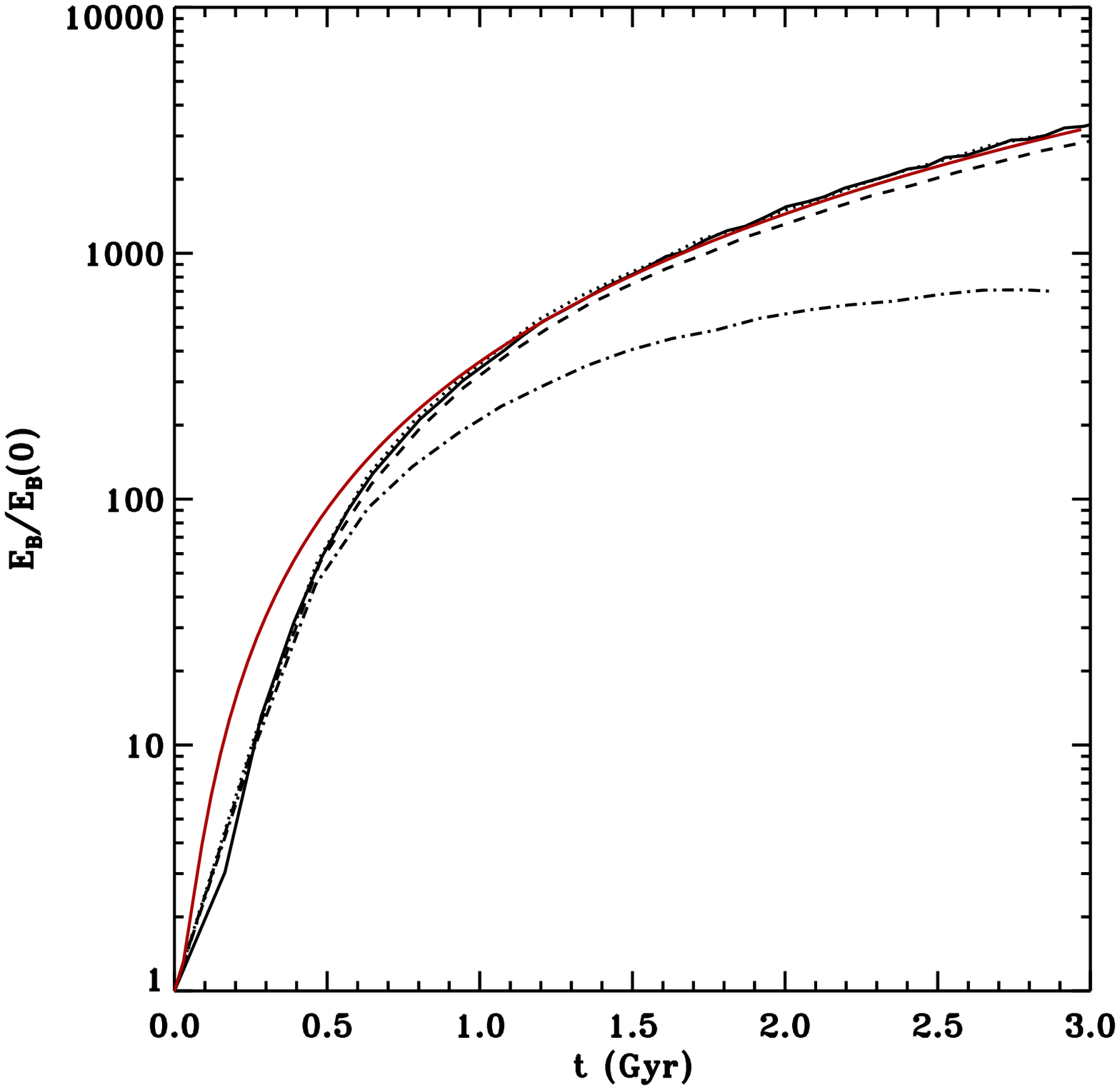}}}
\centering{\resizebox*{!}{8cm}{\includegraphics{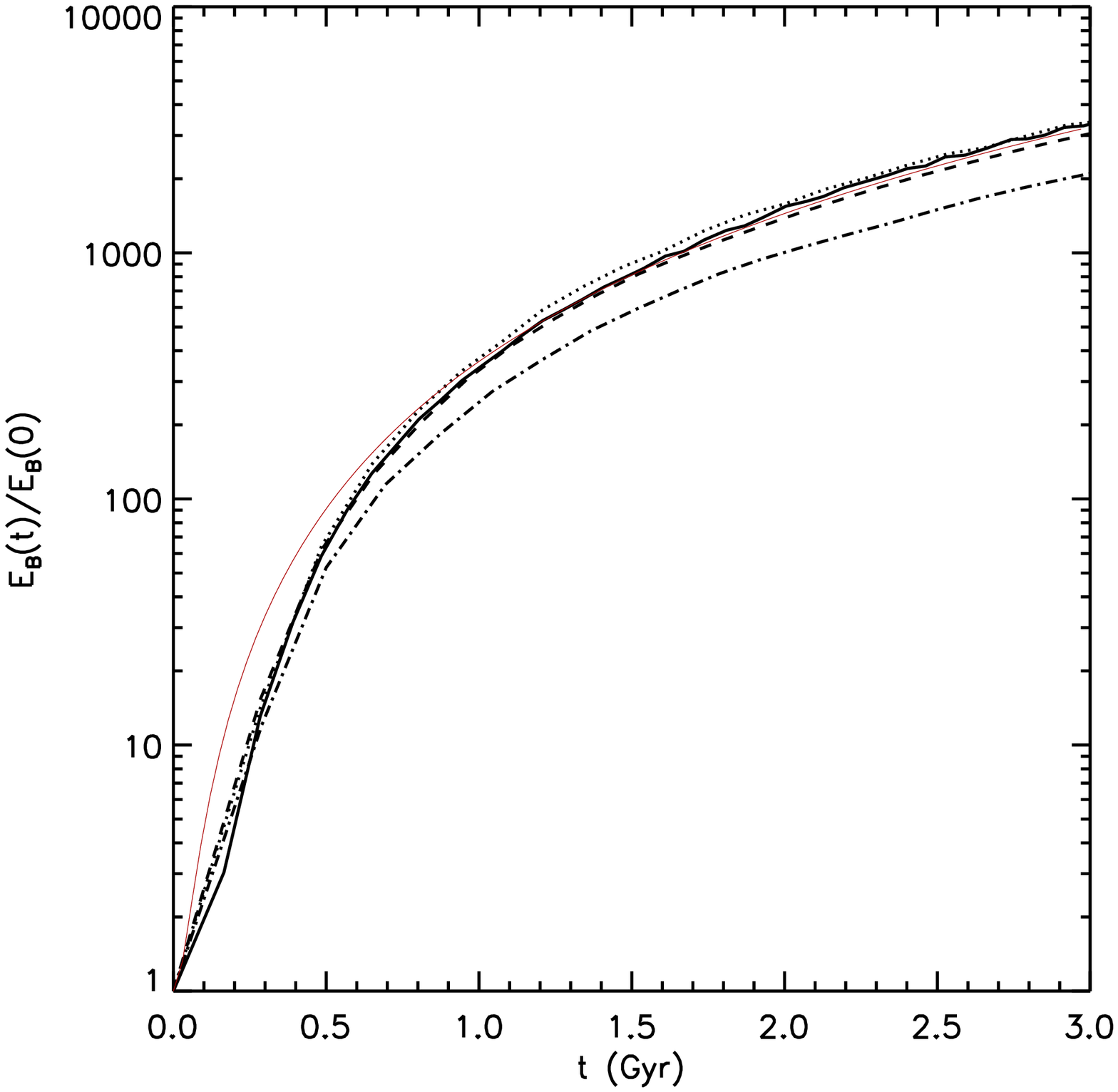}}}
\caption{Total magnetic energy amplification as a function of time (a)
  for  different values of  the initial  magnetic field  (upper panel)
  $B_{\rm IGM}=10^{-7}\,  \mu$G (solid), $B_{\rm  IGM}=10^{-6}\, \mu$G
  (dotted),   $B_{\rm  IGM}=10^{-5}\,   \mu$G  (dashed)   and  $B_{\rm
    IGM}=10^{-4}\, \mu$G (dash--dotted),  and (b) for different values
  of the  Lagrangian refinement criterion  with $B_{\rm IGM}=10^{-7}\,
  \mu$G  (bottom  panel): $m_{\rm  res}=m_{0}$  is  the reference  run
  (solid line),  $m_{\rm res}=5  \times m_{0}$ (dotted  line), $m_{\rm
    res}=25 \times  m_{0}$ (dashed  line) and $m_{\rm  res}=100 \times
  m_{0}$ (dash--dotted line). In each plot, the simple $\Omega$ amplification
  model  is overplotted  in red  for $\Omega_{\rm G}=19\,  \rm  Gyr^{-1}$ as
  average angular velocity.}
\label{Bini_comp}
\end{figure}

An important issue is that we attempt to model a quiescent star-forming galaxy with an EoS mimicking the multiphase nature of the ISM gas, i.e. the gas pressure is artificially boosted to take into account the effect of supernovae heating of the ISM. The direct consequence is that the gas distribution is very smooth in the disc (no small scale clumping can develop), and as a consequence, supernova explosions have a very low impact on the ISM. A next step would be to simulate the evolution of starburst galaxy, within which large clumps ($\sim100$ pc size) would be resolved and within which we could hope that supernovae are able to produce a small-scale dynamo effect. In order to get a global dynamo effect,  additional physics would also be needed: non-ideal effects like Ohmic or turbulent dissipation and cosmic rays have been already identified as key ingredients to get a strong dynamo (see \citealp{hanaszetal09} for instance).

\begin{figure}
\centering{\resizebox*{!}{8cm}{\includegraphics{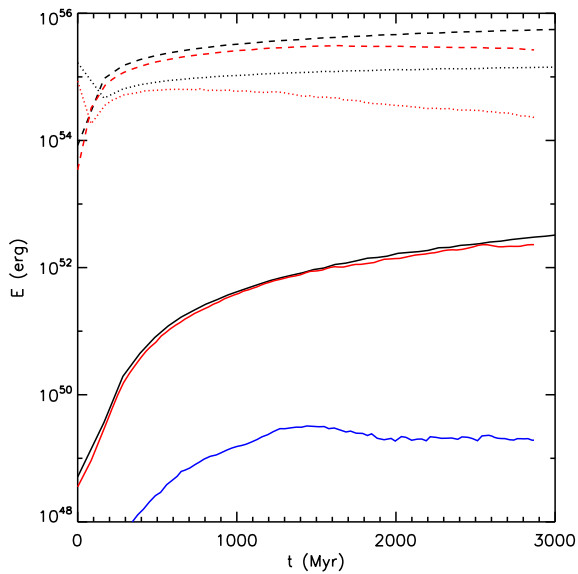}}}
\caption{Evolution of the magnetic energy (solid), thermal energy (dotted), and kinetic energy (dashed) for the simulation without star formation (black) and with SN feedback (red). The initial magnetic field is in both case  $B_{\rm  IGM}=10^{-5}\,   \mu$G, and energies are computed in a small cube of 15 kpc size around the galaxy. The blue solid line is the magnetic energy for the magnetic seeding simulation (see Appendix).}
\label{Ene_evol}
\end{figure}

An investigation of the magnetic energy growth within the disc reveals that there is no substantial increase of the magnetic field when SN explosions are allowed in the galaxy (figure~\ref{Ene_evol}).
There is 3 effects: first, due to supernovae heating (both turbulent and thermal), the early magnetic field amplification appears weaker than in the no-SN case. 
As a result, the magnetic energy is slightly lower than without SNII. 
Second, the longer term magnetic field amplification appears slightly more efficient in presence of SN. Finally, the large-scale galactic wind carry some of the magnetic energy available in the disc, producing a slight decrease in comparison of the simulation without feedback.
As a result, the final magnetic energy value is almost equal to the final energy without SN explosions.
This is a very weak and subtle effect.
We notice that we reproduce roughly the same energy evolutions (kinetic, thermal and magnetic) than~\cite{wang&abel09}.
Interestingly, both kinetic and internal energies are lower when feedback proceeds: this is partly due to the gas consumption by the star forming process, and to the ejection of some material outside from the galaxy.

\begin{figure}
\centering{\resizebox*{!}{8cm}{\includegraphics{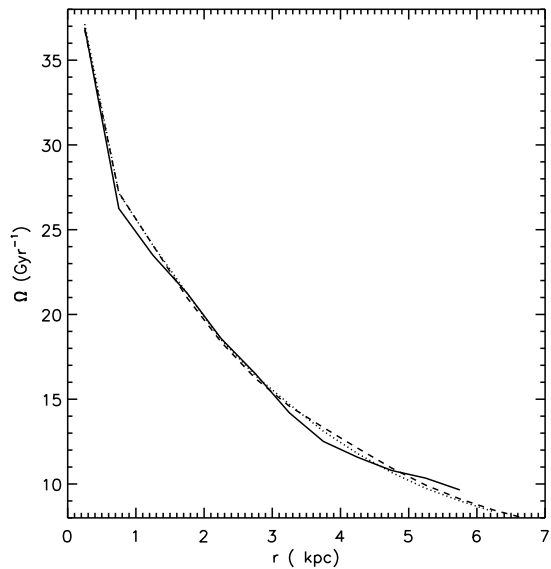}}}
\caption{Mass-weighted average angular velocity in the galactic disc as a function of radius at $t=1$ Gyr (solid line), $t=2$ Gyr (dotted line) and $t=3$ Gyr (dashed line) for the simulation without star formation.}
\label{omegavsr}
\end{figure}

The evolution of the magnetic field is rapidly dominated by its tangential component defined  in the cylindrical coordinates ($r$, $\theta$ and $z$, where $z$ is the spin axis of the galactic disc). Thus
\begin{equation}
\partial_t B_r \simeq 0 \:\:\: {\rm and}\:\:\:  \partial_t B_z \simeq 0
\end{equation}
when compared to the evolution of $B_\theta$ that rigorously writes 
\begin{eqnarray}
\partial_t B_\theta = &-&B_\theta \partial_r v_r - B_\theta \partial_z v_z + B_r \partial_r v_\theta +B_z \partial_z v_\theta \nonumber \\
&-& v_r \partial_r B_\theta - {v_\theta \over r}\partial_\theta B_\theta - v_z \partial_z B_\theta - {v_\theta \over r} B_r \, .
\label{induction_full}
\end{eqnarray}
For an axisymmetric field the amplification is done by the shearing terms, i.e. combining the third and last terms of this equation it leads to
\begin{equation}
\partial_t B_\theta = r B_r \partial_r \Omega \, ,
\label{induction}
\end{equation} 
where  we introduce  the  angular velocity  $\Omega(r)$. Then the radial shear governs the evolution of the magnetic field. We have verified that the vertical shear component ($r B_z \partial_z \Omega$) is negligible here. 
If the right-hand side terms of equation~(\ref{induction}) are not time-dependent, this equation resumes to
\begin{equation}
B_\theta = r B_r \partial_r \Omega \times t\, .
\label{equdynamo}
\end{equation}
If we  assume that  initially the magnetic  field is dominated  by the
radial  component, we  can derive  a  very simple  expression for  the
evolution  of the  total magnetic  energy in  the disc  by integrating
$(B_r^2+B_\theta^2)/8\pi$
\begin{equation}
E_{\rm B}(t)/E_{\rm B}(0) = 1 + \left( \Omega_{\rm G} t\right)^2\, ,
\label{OmegaG}
\end{equation}
where  $\Omega_{\rm G}^2= < (r \partial_r \Omega)^2>$ is the  average ($B_r^2$ weighted) angular velocity squared of  the galaxy. 
This simple analytical  formula compares very well with the actual magnetic energy amplification we have measured in our simulations,  as can be  seen in Figure~\ref{Bini_comp}.
Then one can infer that most of the energy comes from winding up the  radial field within the disk.
The best fit to our pure $\Omega$  amplification model on the magnetic energy evolution (equation~(\ref{OmegaG})) from our simulations gives $\Omega_{\rm G} \simeq  19\, \rm Gyr^{-1}$,  which corresponds, using a simple dimensional analysis, to  a disc  {\it magnetic  scale length}
$r_d \simeq  1.9$~kpc for a  circular velocity of $V_{200}  = 35$~km/s (with $\Omega_{\rm G} \simeq V_{200}/r_d$).
This value is consistent with the angular velocity profiles measured at different times through the course of the simulation (see figure~\ref{omegavsr}). 
Only in our high magnetic field case, $B_{\rm IGM}=10^{-4}~\mu$G,  do we  see a significant  deviation from
this simple model.  This is  due to magnetic braking: angular momentum
is removed  from the disc by  Alfven waves propagating  along the more
rigid field lines  \citep{hennebelle&fromang08}.  The angular velocity
of the disc is therefore  slowly decreasing.  For all the other cases,
magnetic braking is negligible  and the magnetic amplification is very
similar to the pure $\Omega$ amplification model.

We  have estimated  the  effect  of mass  resolution  on the  magnetic
amplification.     Figure~\ref{Bini_comp}     shows    the    measured
amplification in  4 different cases: our  fiducial with $m_{\rm res}=m_0$,
and 3  lower mass  resolution with $m_{\rm res}=5$,  $25$ and  $100 \times
m_0$. The  collapsing halo is  therefore described with fewer and fewer
AMR cells  ($3.10^6$, $8.10^5$, $4.10^5$  and $3.10^5$ total  cells in
decreasing order of $m_{\rm res}$). The disc is however always refined
at  the maximum  level of  refinement,  with a  spatial resolution  of
$150$~pc.  As a  consequence,  only our  lowest resolution  simulation
shows a  significant deviation  from the fiducial  case, which  can be
considered as  fully converged, within  the framework of  our sub-grid
effective EoS.

A more detailed inspection of the magnetic field amplification shows that the contribution of  pure compressional amplification of the magnetic energy ($\rho^{4/3}$) is dominant in the first 100 Myr of galaxy formation due to the low velocity amplitude of the gas in the centre of the NFW profile.
But after ~200 Myr, this contribution to the magnetic field evolution, is negligible in comparison to the measured magnetic energy in the disc.
So it is reasonable to not consider the amplification by compression over the whole galaxy evolution (which runs over Gyrs).

Our simulation  results compare very  well with the work  presented in
\cite{kotarbaetal09}, where these  authors report a  similar magnetic
amplification  using 2  different Smooth  Particle  Hydrodyamics (SPH)
codes.   They used  various force  softening length,  down  to 100~pc,
quite similar to  our grid spacing.  They also  used a similar physical
model,   since   they   didn't   allow   the   gas   to   cool   below
$10^4$~K. Although  their SPH simulations  were all initialized  as an
initial  Milky--Way  disc   configuration,  their  results  are  quite
consistent with a  pure $\Omega$ amplification. In a few  cases only did they
obtain a  faster, exponential growth of magnetic  energy.  These cases
all corresponds to numerical schemes where the magnetic divergence was
allowed  to  deviate  significantly  from  zero.   They  are  probably
associated with  numerical instabilities.  In the majority  of the SPH
results,  for which  the  magnetic divergence  was well--behaved,  the
magnetic energy was growing like $t^2$.

\begin{figure*}
\centering{\resizebox*{!}{5.5cm}{\includegraphics{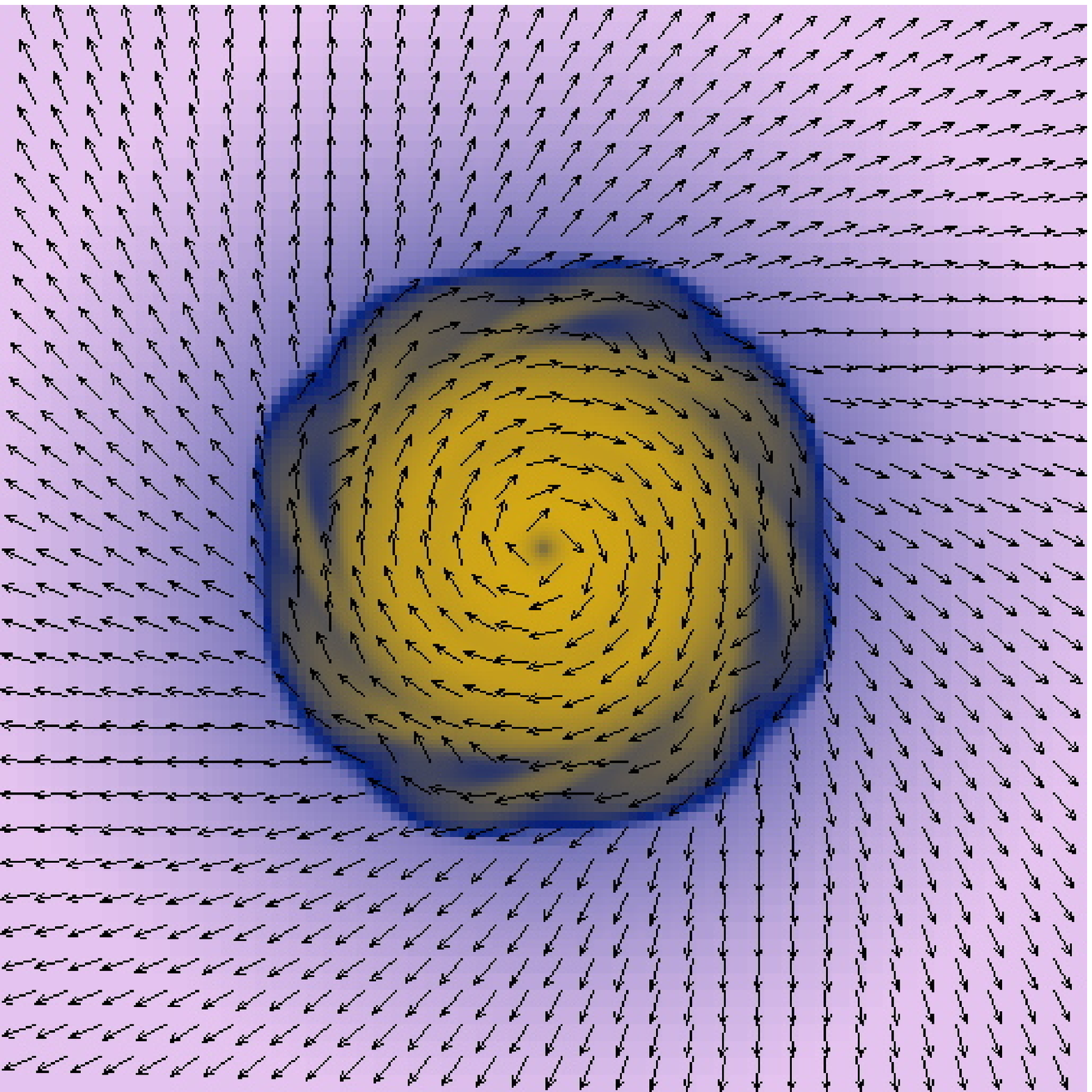}}}
\centering{\resizebox*{!}{5.5cm}{\includegraphics{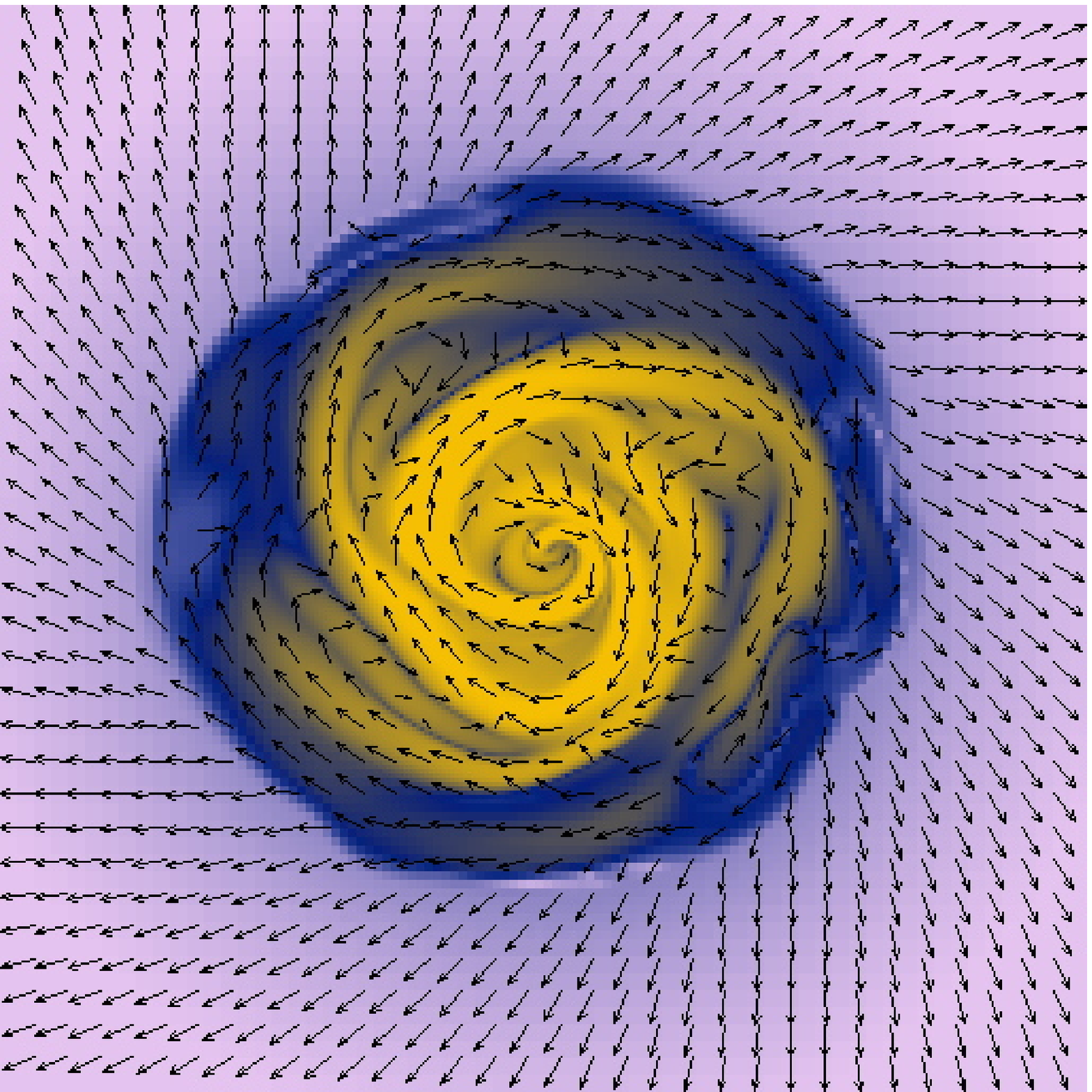}}}
\centering{\resizebox*{!}{5.5cm}{\includegraphics{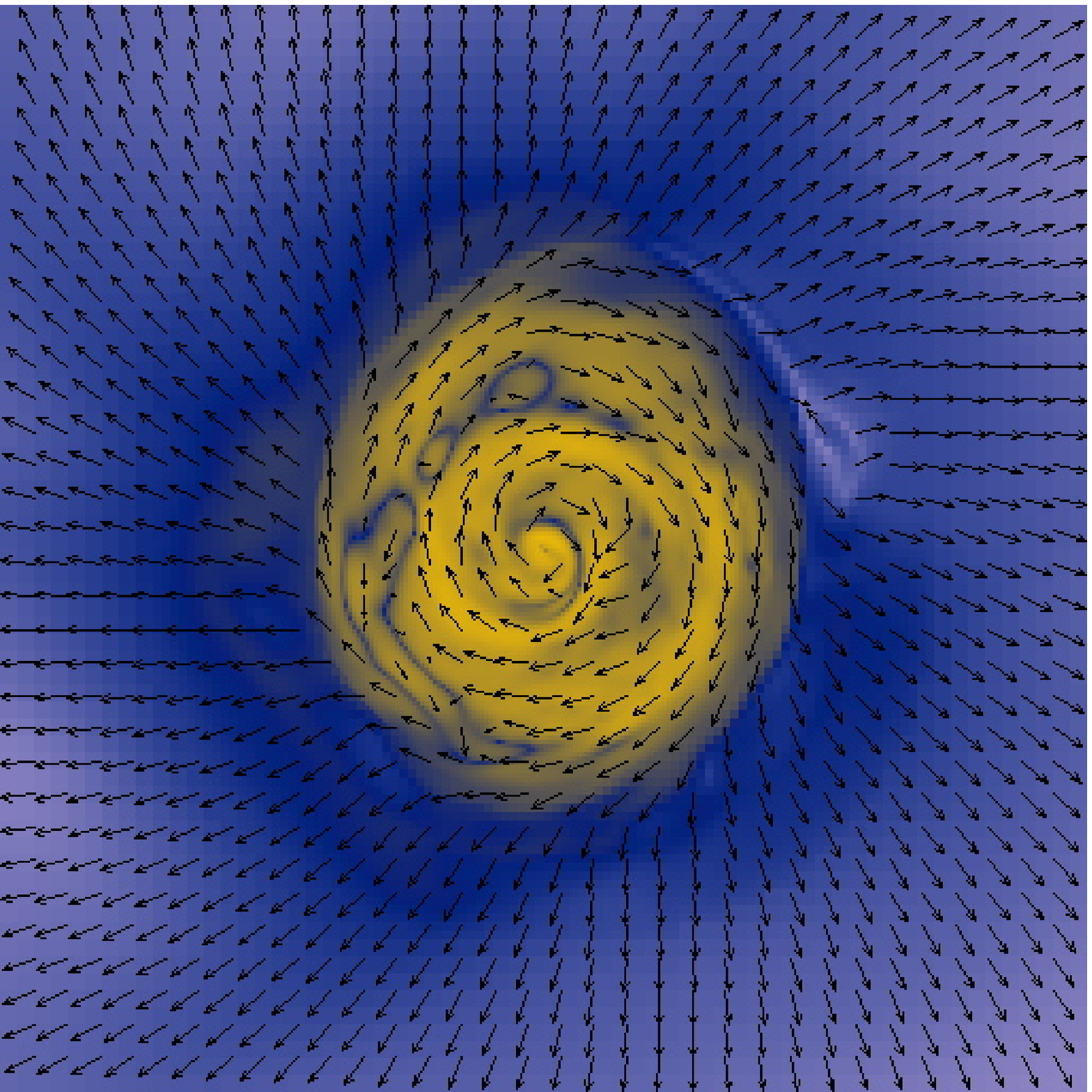}}}
\centering{\resizebox*{!}{5.5cm}{\includegraphics{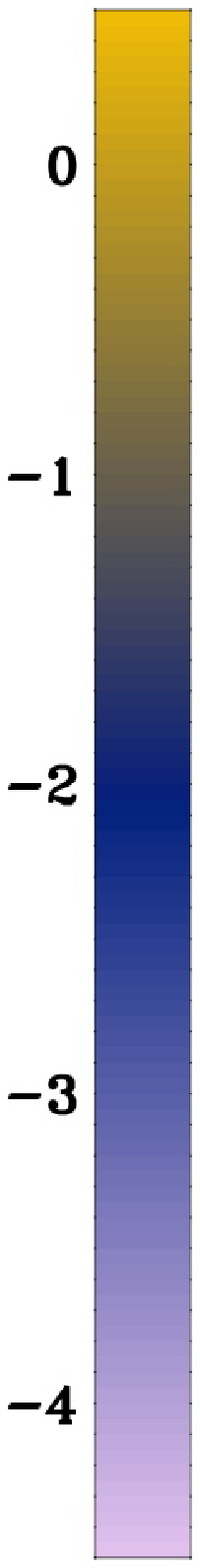}}}
\centering{\resizebox*{!}{5.5cm}{\includegraphics{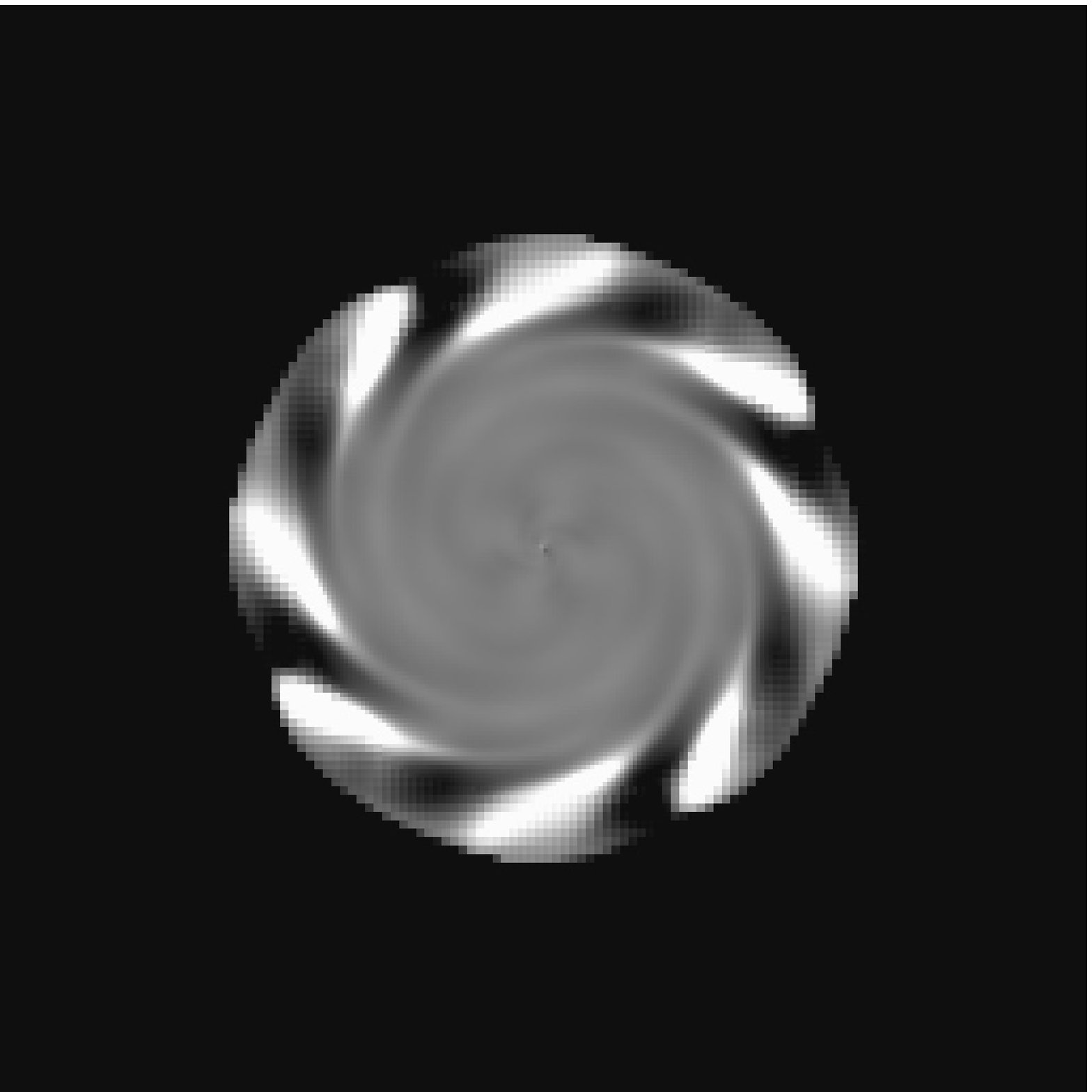}}}
\centering{\resizebox*{!}{5.5cm}{\includegraphics{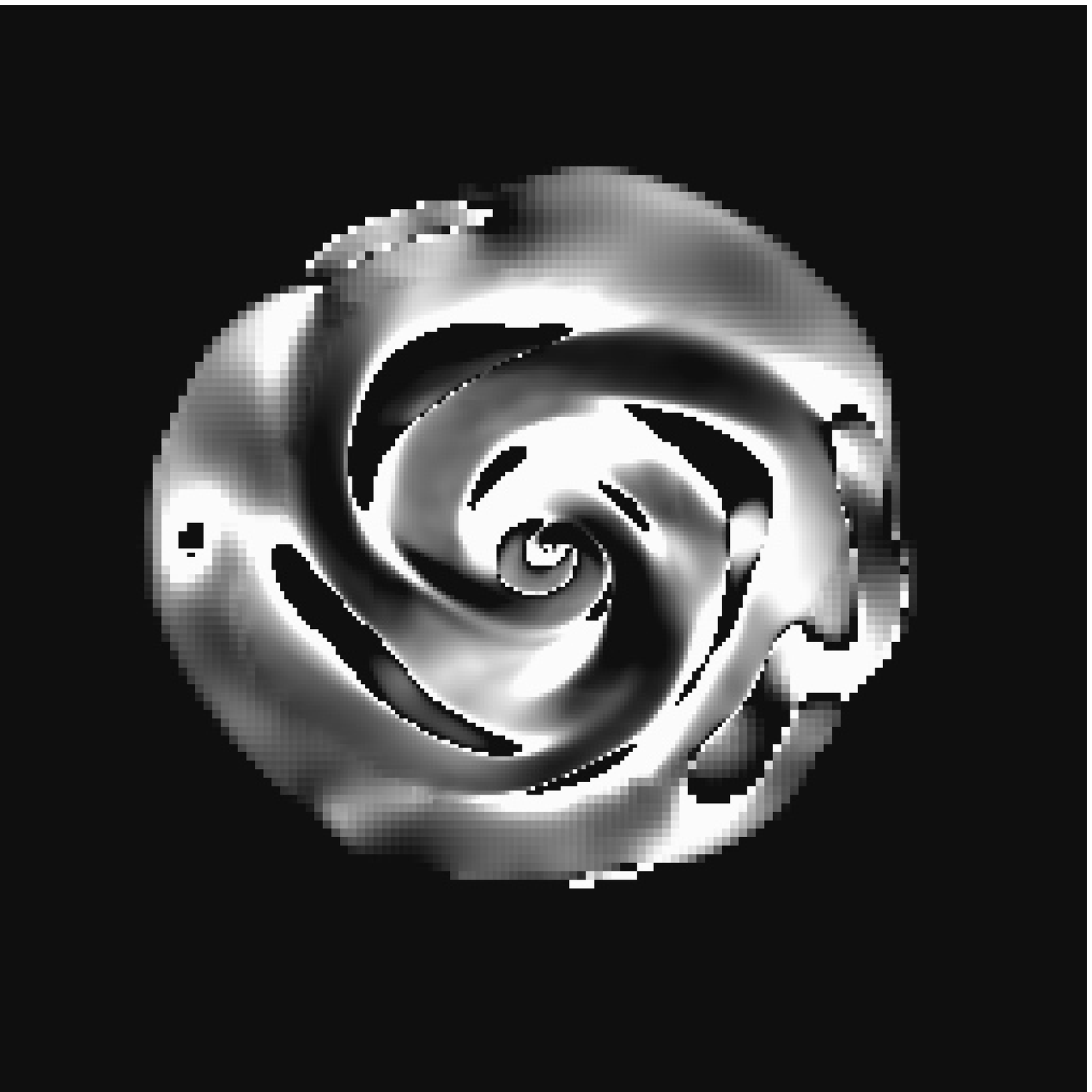}}}
\centering{\resizebox*{!}{5.5cm}{\includegraphics{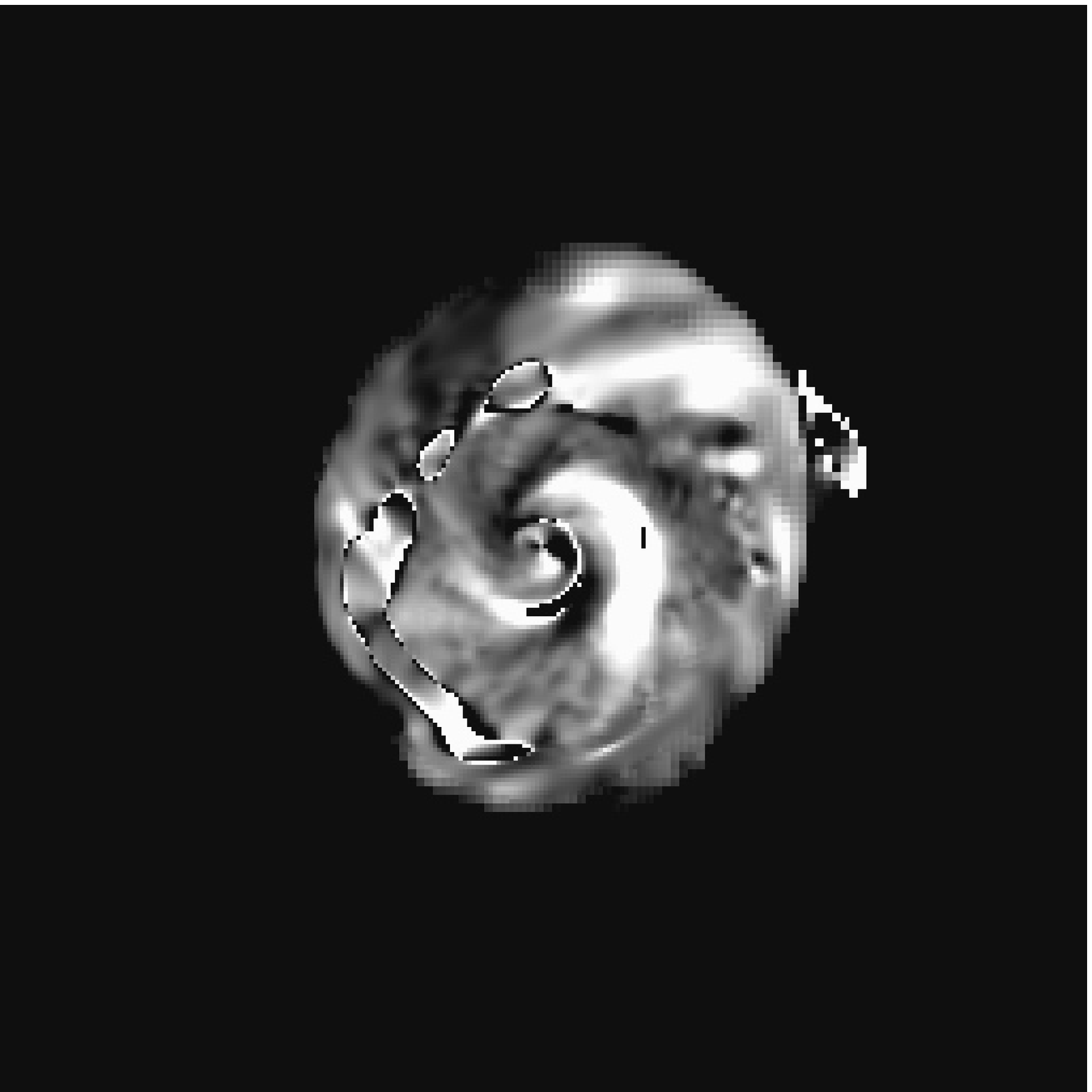}}}
\centering{\resizebox*{!}{5.5cm}{\includegraphics{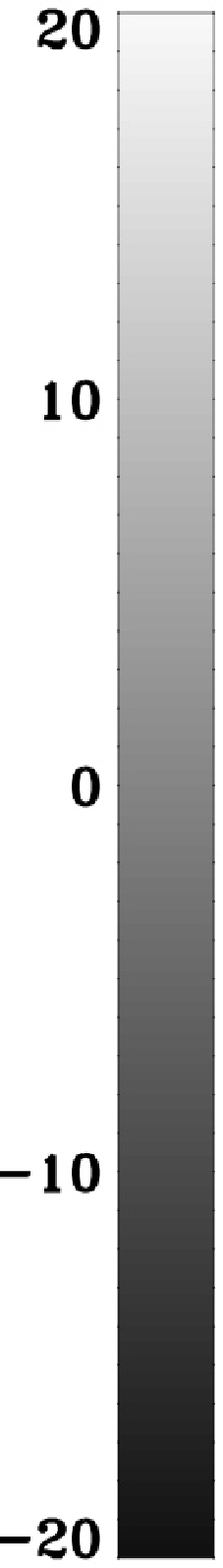}}}
\caption{Magnetic  amplitude in  $\log\, \mu$G  with  magnetic vectors
  overplotted  (upper  panels) and  the  pitch  angle $\theta_p
  $  in  degrees (bottom  panels)  for  the  galaxy without  star
  formation at  $t \simeq  3$ Gyr (left  panels), for the  galaxy with
  star formation  and without feedback  at $t=4$ Gyr  (middle panels),
  and for  the galaxy with  star formation and supernova  feedback at
  $t=3$  Gyr  (right panels)  in  the  upper  part of  the  equatorial
  plan. Initial magnetic field is $B_{\rm IGM}=10^{-5}\mu$G. Picture
  size is 40 kpc. }
\label{Bcool_sf_Z_vec}
\end{figure*}

\cite{wang&abel09}, on  the other hand, have  used an AMR  code with a
much  better spatial  resolution  (down to  25~pc)  and have  included
cooling processes down to 300~K. They were able to resolve, in a dwarf
galaxy smilar to the one studied here, the fragmentation of gas clumps
and the  associated vortex modes.   They however didn't  consider star
formation  and  supernova  feedback.   They concluded  that  the  main
amplification   mechanism  in   their  simulations   was  differential
rotation: even  at the 25 pc  resolution they have  reached, they have
not captured any turbulent dynamo and their associated $\alpha$ effect:
magnetic energy was also growing like $t^2$ in their case.

To support  even further the  simple $\Omega$ amplification model  to explain
our simulation  results, we now  analyze in more details  the magnetic
field morphology in  our disc galaxy.  We consider  for that purpose 3
different  cases: 1-  formation  of  the disc  with  only cooling,  2-
formation of the  disc with cooling and star  formation and finally 3-
formation  of the  disc with  cooling, star  formation  and supernova
feedback.     The   magnetic    field   morphology    is    shown   in
Figure~\ref{Bcool_sf_Z_vec}. We also represent  in the same figure the
magnetic pitch angle defined as
\begin{equation}
\tan \theta_p = {B_r \over B_{\theta}} \, .
\label{pitcheq}
\end{equation}
$\theta_p$  gives the  angle  between the  magnetic  field within  the
horizontal  plane and  its azimuthal  component.  $\theta_p=0^{\circ}$
corresponds to  pure azimuthal field lines  and $\theta_p=90^{\circ}$ to
pure radial field lines. We must underline that pitch angles inferred from observations are obtained with synthetic polarisation maps from radio emission. In the following, we extract a slice of the magnetic field through the galactic plane to get the pitch angle as defined by equation~(\ref{pitcheq}), assuming that it provides a good comparison with observational synthetic polarisation if the pitch is independent of the vertical height (it is roughly the case in these simulations).

Without any star formation process, the magnetic field in the galactic
disc is  almost purely  azimuthal: we  can see from  the left  plot in
Figure~\ref{Bcool_sf_Z_vec}  that  in this  case  the  pitch angle  is
reaching  a maximum  value of  $\theta_p=-2^{\circ}$. The  gas  disc is
stabilized by the  effective EoS, so that no  density perturbation can
break  the almost  perfect azimuthal  symmetry (only  modes  with $m=0$
emerge). After  collapse, the radial  component of the  magnetic field
points away  from the centre of  symmetry above the  galactic plane and
towards the centre below the plane. Using Equation~(\ref{induction}), we
see that the toroidal magnetic field also changes sign above and below
the  plane: the  field  is said  to  be anti-symmetric.   This kind  of
particular symmetry  is called the  A0 mode (anti-symmetric  and $m=0$)
and emerges  naturally in  halo without strong  disc--halo interations
(like  galactic   winds  or  galactic   fountains)  \citep{becketal96,
  beck09}.   The Milky-Way,  for  example,  is believe  to  be a  good
example  of such  a A0  magnetic field  in the  halo \citep{hanetal97,
  sunetal08}.

The situation  appears to  be quite different  when star  formation in
taken into account.  Because mass is removed from the gas disc to form
stars, our effective EoS reaches lower density and the gas sound speed
is smaller.  The disc Toomre parameter get smaller, closer to 1, and a
strong  spiral mode  can  develop. We  see  from the  central plot  of
Figure~\ref{Bcool_sf_Z_vec} that the magnetic field is not axisymmetric
anymore. The  strong density perturbation triggers  fluctuation in the
angular  velocity, so that  $\Omega(r)$ is  not monotone  anymore, but
fluctuates    around   $V_{200}/r$.    We    can   see    again   from
Equation~(\ref{induction}) that, when  $B_r$ changes sign,
the  corresponding toroidal  component  can decrease,  or even  change
sign. We  see indeed in Figure~\ref{Bcool_sf_Z_vec}  that the toroidal
field almost reverts its main direction between the spiral arm and the
inter-arm regions.   As a consequence,  the pitch angle amplitude  $\vert \theta_p\vert$ is
rather  high in  the  low density,  inter-arm regions  $\vert\theta_p\vert\simeq
 40^\circ $ (with both negative and positive signs) and is lower  within the  spiral arms  with $\vert\theta_p\vert\simeq 20^\circ$ (always with negative values). These values are in  good agreement with the observed pitch
angles  in nearby galaxies  (\citealp{becketal96}, \citealp{rohdeetal99}, \citealp{patrikeevetal06}).  The
same  relation between the  pitch angle  and the  spiral arms  is also
observed within  spiral galaxies: magnetic  field is aligned  with the
optical     spiral    arms~(\citealp{krauseetal89,    neiningeretal91,
  berkhuijsen97, beck07}), and it also appears sometimes that stronger
ordered magnetic     fields     are     seen     within     the     inter--arm
regions~(\citealp{ehleetal96, fricketal00}).

The figure~\ref{Bcool_sf_Z_vec} clearly shows positive and negative signs of the pitch angle. Therefore a simple spiral density flow can not explain the observations, which show only negative pitch angles. 
But spirals may contribute to the increase of the pitch angle compared to a pure axisymmetric rotation, at least at the inner edge of the spirals.
Thus, the always negative pitch angle seen in the observations is one of the arguments favouring the dynamo. 

The Magneto Rotational Instability (MRI) could also be a growth mode of the magnetic field in galaxies \citep{kitchatinov&rudiger04} or to sustain turbulence in the disc \citep{piontek&ostriker05}. 
The MRI can give a coherent picture for the pitch angle of polarised emission if the magnetic field has several reversals along the vertical direction \citep{elstneretal09}. 
The MRI needs a strong toro\"idal field to be efficient, and it is clearly shown that the field in our galaxy simulations is preferentially polo\"idal at late times (see figure~\ref{Bcool_sf_Z_vec}).
However there is a possible MRI channel for the magnetic field growth in the early evolution of the galaxy ($<300$ Myr) when the circular magnetic component is still lower or comparable to the vertical field, then, later on, the amplification by differential rotation proceeds.
But it is beyond the scope of this paper to disentangle this early MRI from the pure differential rotation suggested before and would require simulations with higher resolution to properly describe the MRI driven-turbulence.

When  supernova explosions  are  finally included  in  the model,  the
previous  magnetic field  topology is  conserved.  The  spiral  arm is
somewhat  weaker, because  the  increased velocity  dispersion due  to
supernova  blast waves  stabilizes the  disc. The  disc  appears less
coherent and  more perturbed by small scales  perturbations. Note that
this is precisely the small  scale turbulence we need for the $\alpha$
effect,  but  despite  this  small  scale  perturbations,  the  radial
component did  not grow significantly in our  simulation. The magnetic energy evolution conserves the same properties than in the run without star formation: the differential rotation of the disc is the main driver of the galactic magnetic field in these simulations. We probably
need  sub--parsec  scale resolution  in  order  to resolve  supernova
induced cyclonic motions that can  drive a turbulent dynamo, like in
\cite{balsaraetal04}.  Besides  the lack of  small scale amplification
of the radial component of  the field, our simulation recovers most of
the  qualitative  features of  Galactic  dynamo  models, that  predict
similar    pitch    angle     distribution    and    magnetic    field
topologies~(\citealp{krasheninnikova89,           donner&brandenburg90,
  elstneretal92}).  The existence of  this small scale $\alpha$ effect
in galaxies was recently shown by direct simulations of \cite{gresseletal08} but the problem of its saturation is still under debate (\citealp{brandenburg&subramanian05} for
example). The value of the  parameter $\alpha$ needs to be rather high
in order to obtain a  fast enough galactic dynamo. The conservation of
the  total magnetic  helicity  in  the galaxy  makes  large value  for
$\alpha$ rather difficult  to obtain if there is no exchange of magnetic helicity with the halo.  In this paper,  we rather focus
on the other consequence  of supernova feedback, namely the formation
of  a strong  galactic wind,  and  how this  wind transports  magnetic
energy.

\section{Magnetic field in the galactic wind}
\label{resultswind}

\begin{figure*}
\centering{\resizebox*{!}{8cm}{\includegraphics{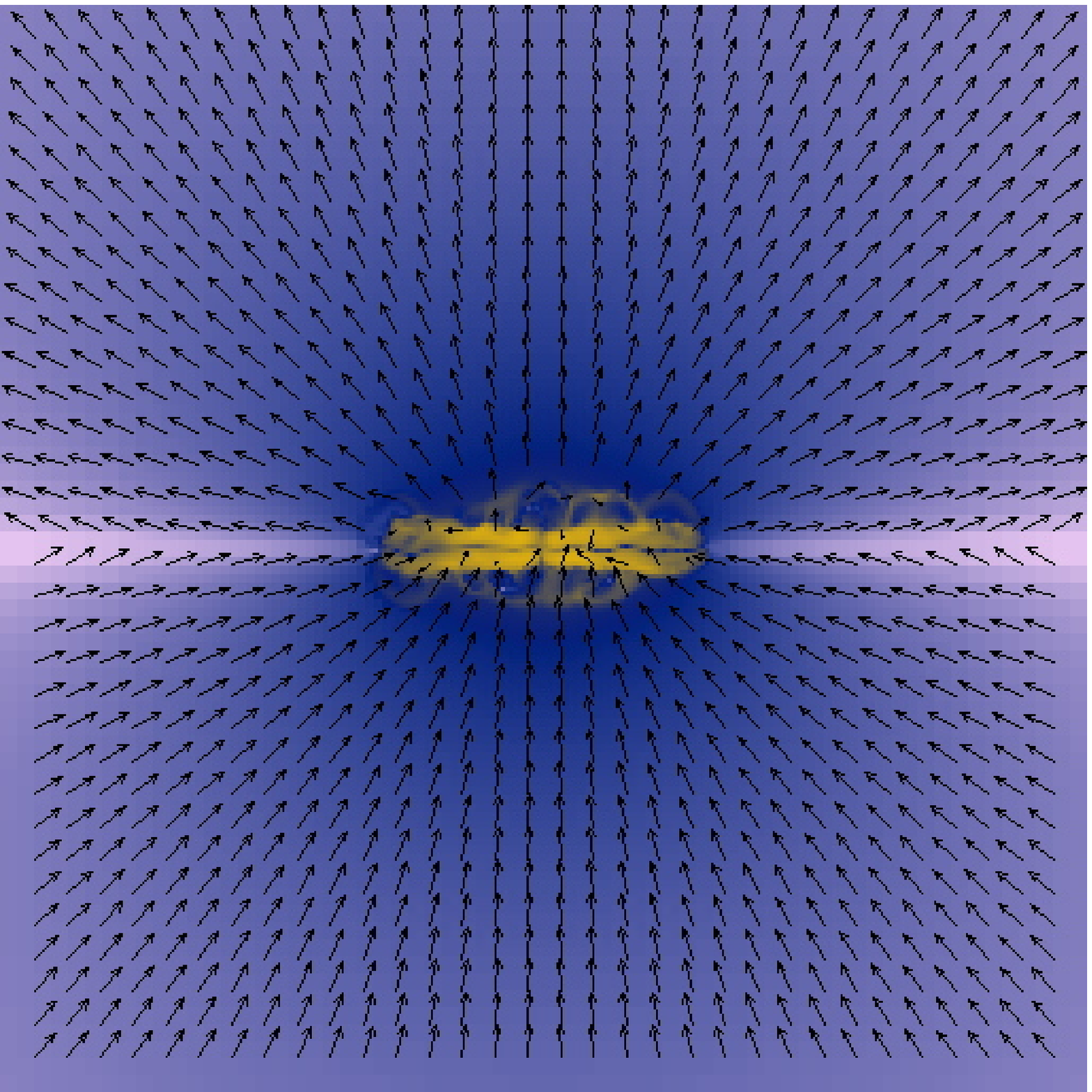}}}
\centering{\resizebox*{!}{8cm}{\includegraphics{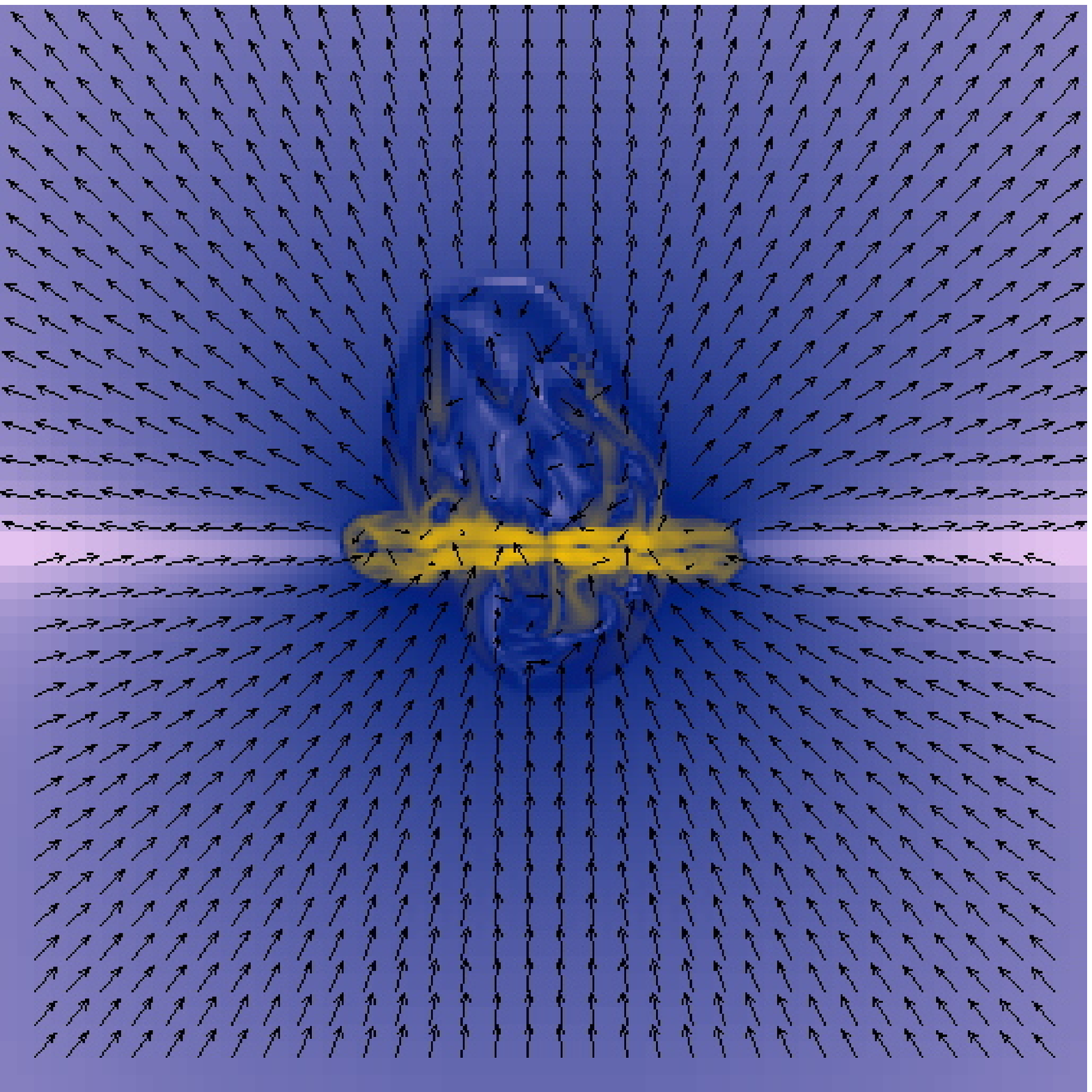}}}
\centering{\resizebox*{!}{8cm}{\includegraphics{./colortable_Bamp_1muG.ps}}}
\centering{\resizebox*{!}{8cm}{\includegraphics{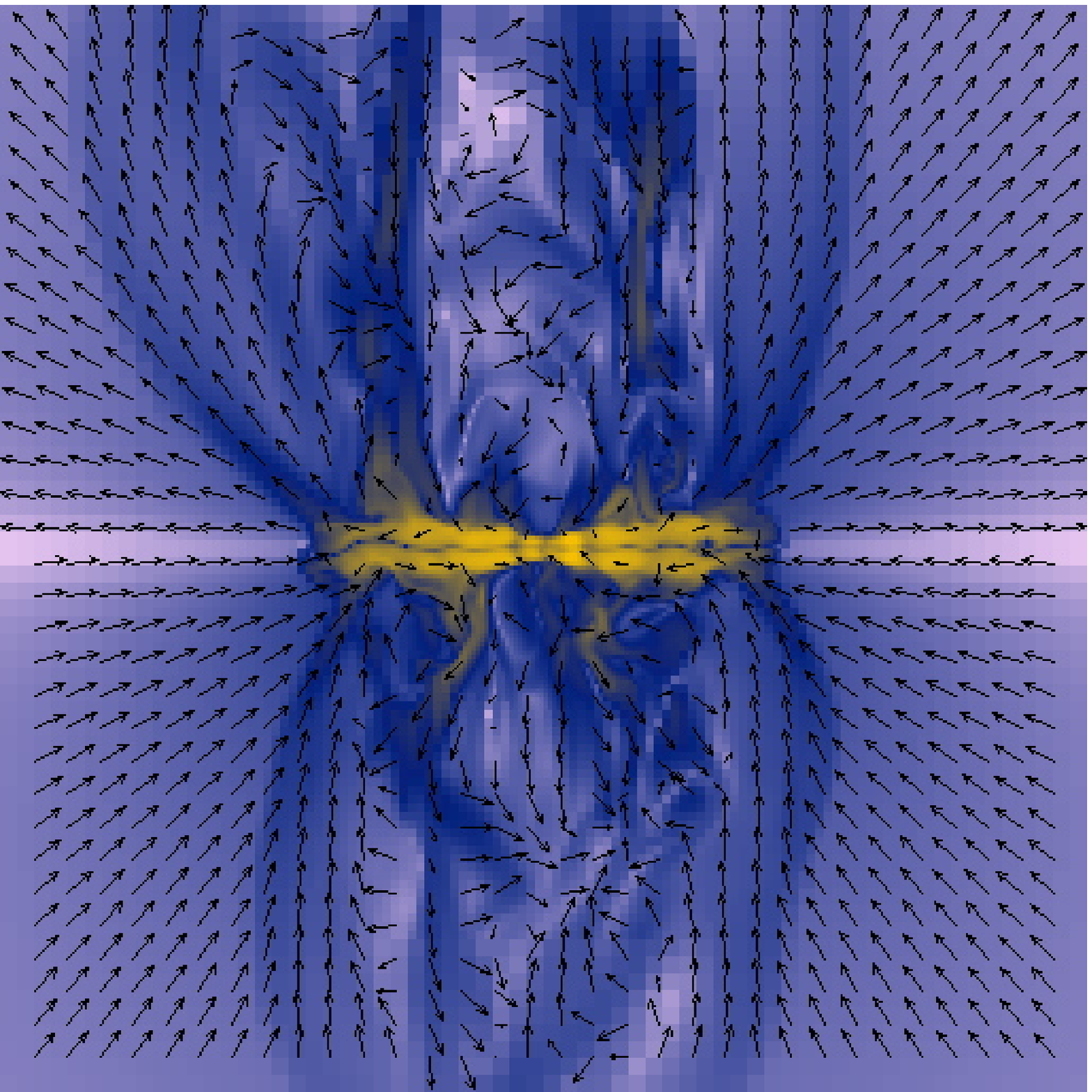}}}
\centering{\resizebox*{!}{8cm}{\includegraphics{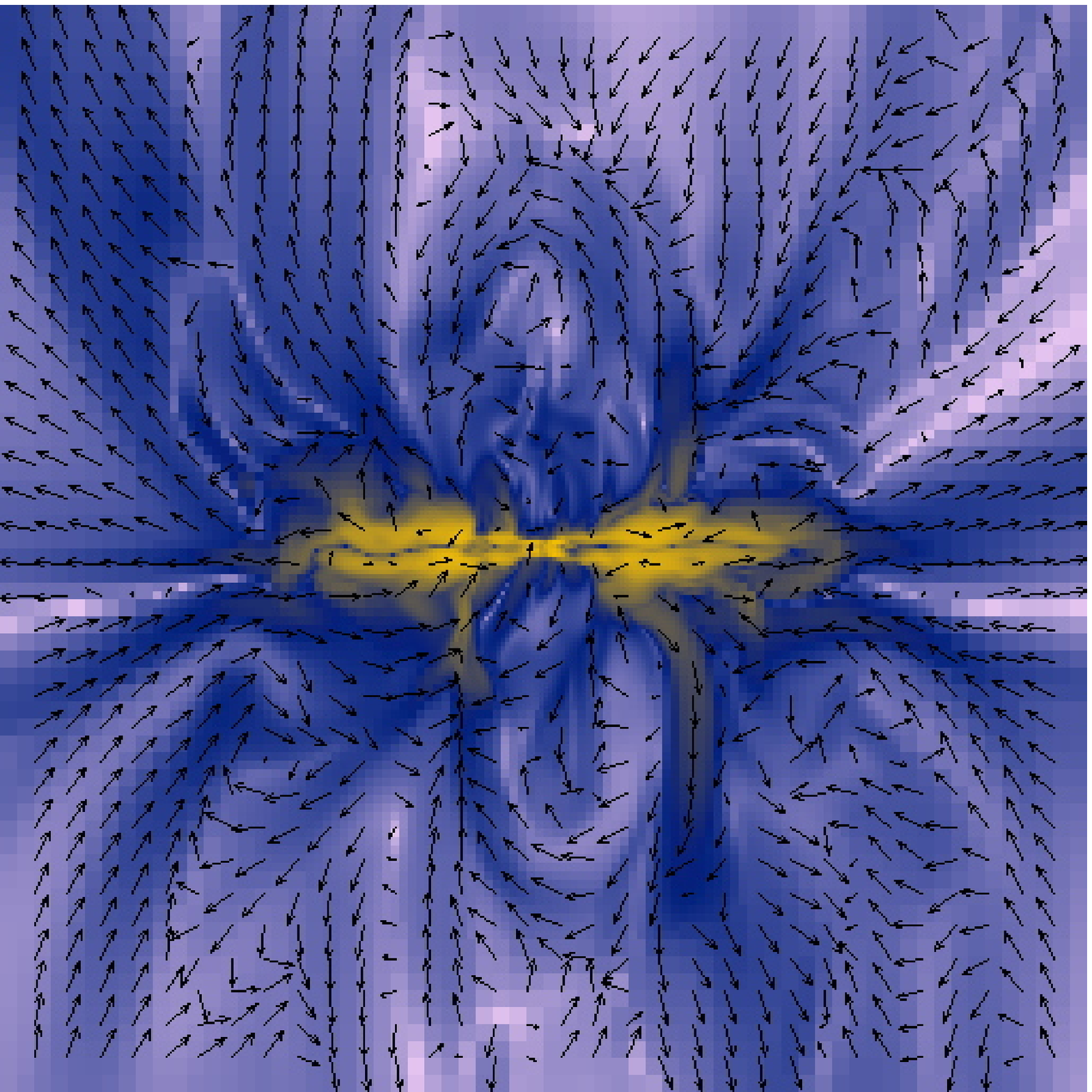}}}
\centering{\resizebox*{!}{8cm}{\includegraphics{./colortable_Bamp_1muG.ps}}}
\caption{Magnetic field  in $\log \,  \mu$G units for the  galaxy with
  star formation and supernova feedback with $B_{\rm IGM}=10^{-5}\,
  \mu$G at different epochs $t=1$ Gyr (upper left), $t=1.5$ Gyr (upper
  right),   $t=2$   Gyr   (bottom   left)  and   $t=3$   Gyr   (bottom
  right). Picture size is 40 kpc.}
\label{Bwind_nodipole}
\end{figure*}

We now study in more  details the properties of the supernova--driven
galactic wind that forms during our simulations. We restrict ourselves
to the analysis of the case $B_{\rm IGM}=10^{-5}\, \mu$G: we obtained in this
case a  normal star--forming  disc, and in  the same time  a magnetic
field close  to equipartition in the  gaseous disc. A  larger value of
$B_{\rm IGM}$ would result in the formation of a non--star--forming torus,
while a smaller value would  give rise to a weakly magnetised galactic
disc,  significantly  below   equipartition.

\begin{figure}
\centering{\resizebox*{!}{8cm}{\includegraphics{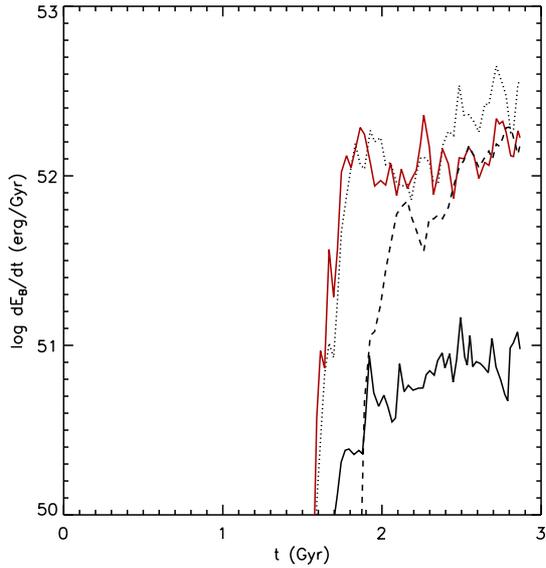}}}
\caption{Magnetic   energy   flux   as   a  function   of   time   for
  $B_{\rm IGM}=10^{-5}\rm  \mu$G measured  at a  radius of  $4r_s$ (dotted
  line) and $9r_s$ (dashed line)  from the halo centre. The analytical
  prediction of  \cite{bertoneetal06} is overplotted as  the red solid
  line. We have overploted the result for the direct magnetic energy injection with zero initial magnetic field (solid line) as described in Appendix~\ref{magnetic_dipole_annexe}.  }
\label{fluxBvsr}
\end{figure}

In      agreement     with      the      simulations     shown      in
\cite{dubois&teyssier08winds}, the galactic  wind appears at $1.5$~Gyr
and  fully develops during  the next  Gyr. Figure~\ref{Bwind_nodipole}
shows 4 different snapshots (from 1 to 3~Gyr) of the magnetic field in
a plane  perpendicular to the  disc. These snapshots illustrate  the 3
different phases  of the galactic  wind, described in more  details in
\cite{dubois&teyssier08winds}: 1-  {\it the bubbling  phase}, when the
supernova  luminosity is  too weak  to break  the ram-pressure  of the
infalling  gas ($t=1$~Gyr in  Fig.~\ref{Bwind_nodipole}), 2-  {\it the
  snow-plow phase}, when the supernova luminosity is greater than the
ram-pressure of  the infalling gas,  and the wind starts  to propagate
through    the    halo    ($t=1.5$~Gyr    and    $t=2$~Gyr    in
Fig.~\ref{Bwind_nodipole}) and 3- {\it  the blow away phase}, when the
wind has  reached its final,  stationnary shape with  a characteristic
noozle-like structure ($t=3$~Gyr in Fig.~\ref{Bwind_nodipole}).

We can see from Figure~\ref{Bwind_nodipole} that the magnetic field in
the  wind  is  highly  turbulent,  with an  average  amplitude  around
$10^{-3}\, \mu$G, but showing  strong variations within the wind, with
highly magnetised filament (as  high as $10^{-1}\, \mu$G) entrained by
low  density bubbles  (as  low as  $10^{-4}\,  \mu$G).  The  turbulent
structure of the  magnetic field is typical of  convective flows, with
buyoantly rising  vortices and their associated  shearing motions. The
magnetic field  lines appear  however mainly colinear  to the  flow, a
magnetic  configuration that  is  actually observed  in galaxies  with
outflows  (\citealp{brandenburgetal93,   chyzyetal06, heesenetal09}).   Whether  the
turbulent flow we observe in  the galactic wind amplifies the magnetic
field is  unclear in  our simulation.  As  we discuss below,  the wind
magnetic energy  is clearly decaying because of  the overall expanding
flow. Moreover, in the wind region, our spatial resolution is slightly
degraded (around 300 pc), because  the grid has been derefined in this
low density region.  This suggest  that we are not resolving the small
scale magnetic  energy amplification  due to the  turbulent convective
flow, leading to a potential underestimation of the magnetic energy in
the wind.

In order  to estimate the amount  of magnetic energy that  the wind can
extract  from  the magnetised  galactic  disc,  we  have measured  the
magnetic energy flux  in 2 thin shells of  radius $4r_s \simeq 20$~kpc
and  $9r_s \simeq  45$~kpc. Results  are  shown in
Figure~\ref{fluxBvsr}. After the wind starts (around $t=1.5$~Gyr), the
magnetic energy  flux quickly jumps  to a value  of $10^{52}$~erg/Gyr,
and stays  roughly constant during  the next Gyr. The  magnetic energy
flux is  decreasing when going  to larger radii (roughly  as $r^{-1}$),
demonstrating  that  the  effect  of  expansion  is  overcoming  field
amplification in the wind turbulence. It seems that the flux still grows after 2 Gyr. It is possible that when the wind reaches a stationary regime, this growth is correlated with the increase of magnetic energy within the disc (see figure~\ref{Ene_evol}).

It is enlightening to interpret our numerical results in the framework
of the analytical theory of \cite{bertoneetal06}. This will also guide
our discussion on magnetic  field amplification in general.  According
to \cite{bertoneetal06}, the magnetic energy flux which is injected at
the base of the galactic wind can be written as
\begin{equation} 
\dot E_{\rm B,in}=\frac{B_{\rm in}^2}{8\pi} 4\pi R_{\rm G}^2 u_{\rm w}\, ,
\end{equation} 
where $B_{\rm in}$ is  the injected magnetic field, $R_{\rm G}  \simeq 7$~kpc is
the  galactic disc radius  and $u_w$  is the  wind velocity.  In their
analytical model, \cite{bertoneetal06} assumed first that the magnetic
energy could  be considered as  some isotropic pressure term.   We see
from  the previous discussion  that the  convective turbulence  in the
wind justifies  this assumption.  They  assume also that  the injected
magnetic field scales with the injected wind gas density as
\begin{equation} 
B_{\rm in} = B_{\rm G} \left(\frac{\rho_{\rm in}}{\rho_{\rm G}}\right)^{2/3} \, ,
\end{equation} 
where $B_{\rm G}$  and $\rho_{\rm G}$ are respectively the  typical magnetic field
and gas  density inside the  galactic disc. Finally, they  related the
wind gas density to the wind mass flux by
\begin{equation} 
\dot M_{\rm w}=\rho_{\rm in} 4\pi R_{\rm G}^2 u_{\rm w} \, .
\end{equation} 
Using the  3 previous  equations, we can  predict the  magnetic energy
flux in the wind to be
\begin{equation} 
\dot E_{\rm B,in}=E_{\rm B,G} \frac{4 u_{\rm w}}{H_{\rm G}}
\left(\frac{\dot M_{\rm w}R_{\rm G}}{3u_{\rm w}M_{\rm ISM}} \right)^{4/3}  \, ,
\label{edotmagana}
\end{equation} 
where $H_{\rm G} \simeq  300$~pc is the disc thickness  and $E_{\rm B,G}$ is the
disc magnetic  energy.  We have plotted the  predicted magnetic energy
flux      corresponding      to      Equation~(\ref{edotmagana})      in
Figure~\ref{fluxBvsr},  using  the  measured  mass flux  ($\dot  M_{\rm w}
\simeq 0.035\, \rm M_{\odot}$/yr), disc magnetic energy ($E_{\rm B,G} \simeq 5
\times   10^{53}$~erg)   and  disc   mass   ($M_{\rm ISM}  \simeq   10^9\,\rm
M_{\odot}$).  We  see that this  formula is quite accurate,  given the
number  of   simplifying  assumptions  we  have   made.   It  slightly
underestimates the  measured flux, by less  that a factor of  2.  If we
integrate this  magnetic energy flux  over $t_{\rm w} \simeq 1$~Gyr  of wind
activity, we find  that roughly $6$~\% of the  disc magnetic energy can
be extracted  by the  galactic wind and  funneled into the  IGM.  We
note  $\epsilon_{\rm w}$ this  global  wind efficiency.   In the  analytical
model of \cite{bertoneetal06}, this efficiency writes
\begin{equation} 
\epsilon_{\rm w} = \frac{4 u_{\rm w} t_{\rm w}}{H_{\rm G}}
\left(\frac{\dot M_{\rm w}R_{\rm G}}{3u_{\rm w}M_{\rm ISM}} \right)^{4/3} \, .
\end{equation}
A wind  mass injection  of 0.03~$\rm M_{\odot}$/yr is  a rather  low value
compared to  starburst galaxies where  wind are usually  observed. Our
simulation  can  be considered  as  a  quiescent  wind case.   Typical
starburst--driven  winds   show  strong  mass  outflow   as  large  as
10~$\rm M_{\odot}$/yr \citep{martin98, martin99}.   On the other hand, the
burst  duration is probably  much smaller,  with $t_{\rm w}  \simeq 100$~Myr
\citep{degrijsetal01}.   Overall, the  efficiency  of magnetic  energy
extraction probably lies above $\epsilon_{\rm w} \simeq$~10~\% for a typical
star  forming dwarf  galaxy. The  other  extreme case  one could  also
consider is when the starburst is  so violent that the whole galaxy is
destroyed. This  scenario is  likely to happen  in the  early universe
inside small  primordial galaxies, as shown by  the recent cosmological
simulations of  \cite{maschenkoetal08}. In this  case, the efficiency
obviously reaches $\epsilon_{\rm w}=$~100~\%.

\begin{figure}
\centering{\resizebox*{!}{7cm}{\includegraphics{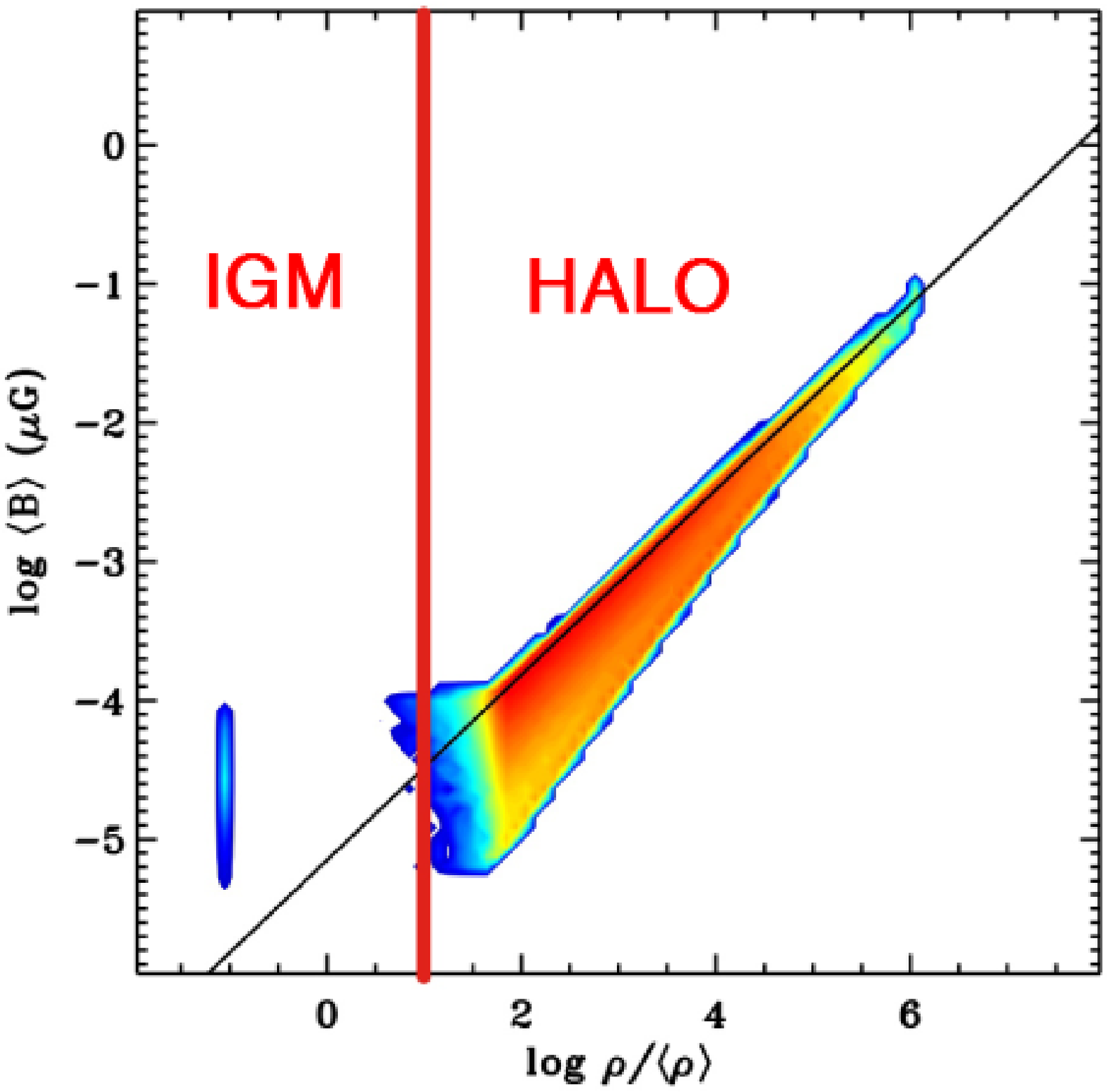}}}
\centering{\resizebox*{!}{7cm}{\includegraphics{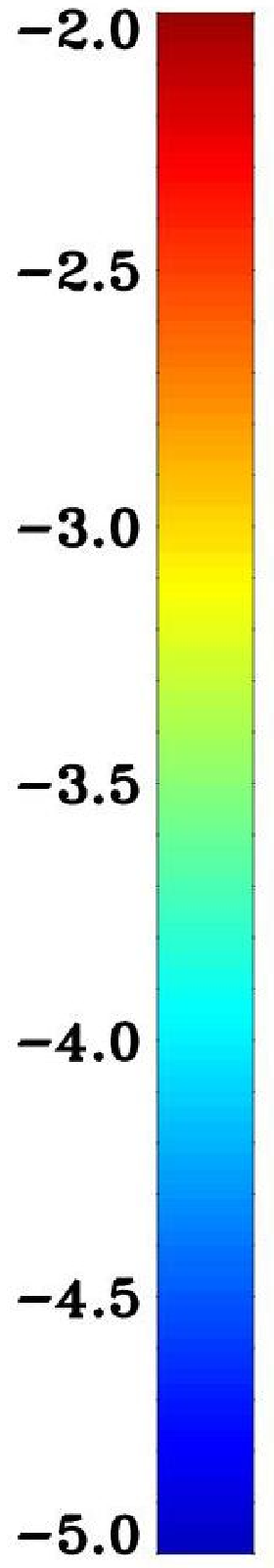}}}
\centering{\resizebox*{!}{7cm}{\includegraphics{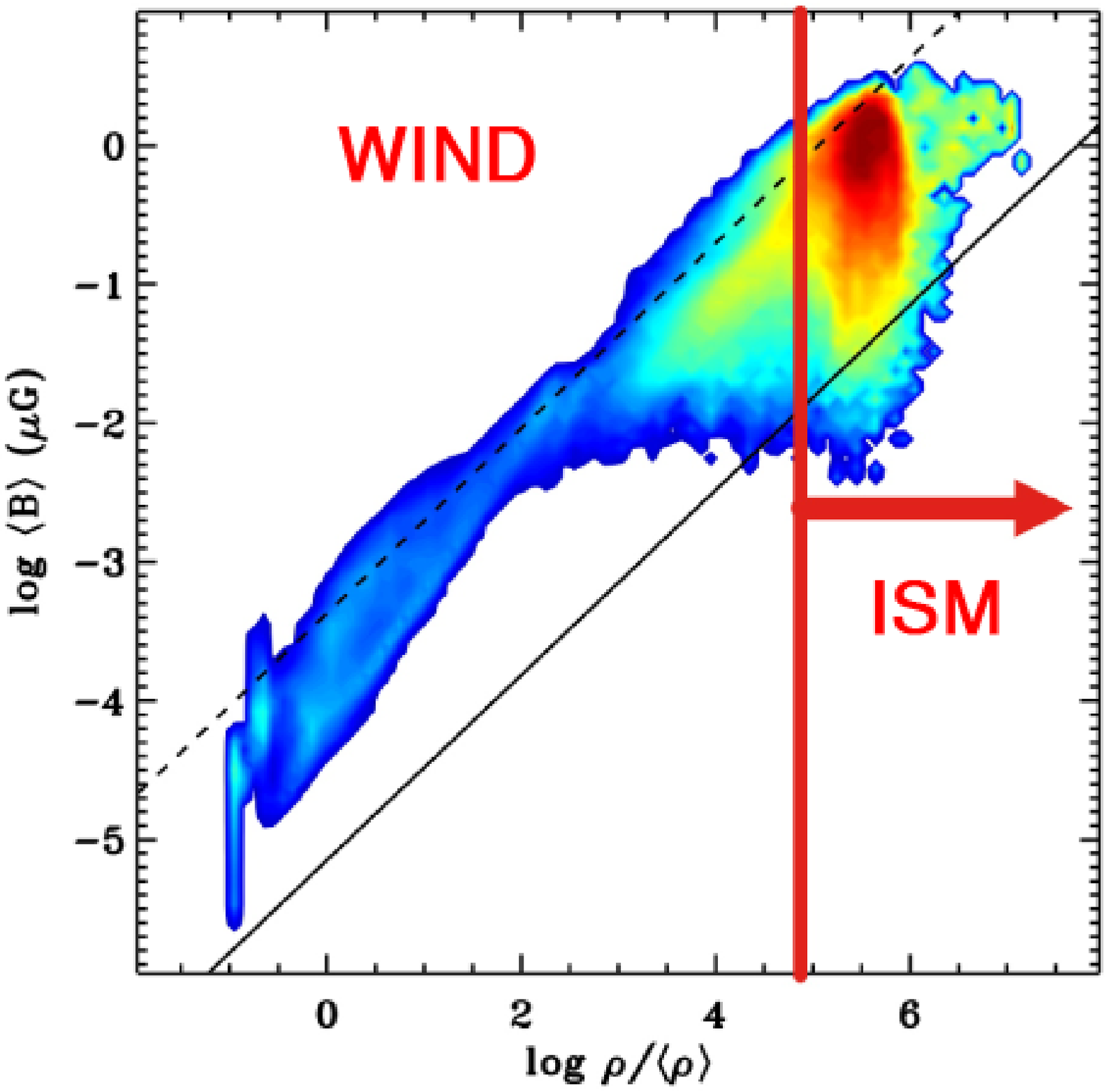}}}
\centering{\resizebox*{!}{7cm}{\includegraphics{./colortable_diagram.ps}}}
\caption{Magnetic amplitude  versus gas over-density  diagram at $t=0$
  Gyr (upper panel) and $t=3$  Gyr (bottom pannel) for the galaxy with
  star  formation, supernova  feedback  and $B_{\rm  IGM}=10^{-5}\,
  \mu$G. Black solid line is for the initial $\rho^{2/3}$ law and black dashed line is for the initial $\rho^{2/3}$ law amplified by a factor 60.}
\label{diag}
\end{figure}

We now want to estimate the magnetic field in the IGM that will result
from the enrichment of the galactic wind. For this purpose, we use the
analytical model developed  by \cite{bertoneetal06} and used recently
in a cosmological  simulation by \cite{donnertetal09}.  We write
for the evolution of the magnetic  energy in the wind blown bubble the
following equation
\begin{equation}
{{\rm d} E_{\rm B} \over {\rm d}t}= \dot E_{\rm B,in} - {\dot R_{\rm w} \over R_{\rm w}} E_{\rm B} \, ,
\label{ebtotbubble}
\end{equation}
$\dot E_{\rm B,in}$ is the magnetic energy injection rate, measured at the
base of the expanding wind and  $R_{\rm w}$ is the bubble radius ($\dot R_{\rm w}$
its time  derivative). The second term  on the right-hand  side of the
previous  equation  stands  for  magnetic  energy losses  due  to  the
expansion  of the  bubble.  This  equation  also assumes  that we  can
neglect  the  turbulent amplification  of  the  magnetic field,  which
turned  out to  be  true in  the  simulations we  have peformed  here.
Following \cite{donnertetal09}, we consider the rather pessimistic case
of a fast bubble expansion and writes $\dot R_{\rm w} / R_{\rm w} = 1/t$.  We also
consider  the  solution of  Equation~(\ref{ebtotbubble})  for which  the
initial bubble energy vanishes
\begin{equation}
E_{\rm B} \simeq \dot E_{\rm B,in} \frac{t_{\rm w}}{2} = \frac{\epsilon_{\rm w}}{2}E_{\rm B,G}\, .
\end{equation}
We note that,  in this scenario, half of the  injected energy has been
lost due to  the expansion of the field  lines. Using the cosmological
wind model of \cite{bertoneetal06}, we consider a final wind radius of
$R_{\rm w}  = 800$~kpc  and  a  wind filling  factor  of $f_{\rm w}=100$~\%  (both
typical values for  the universe at redshift zero),  so that the final
magnetic field in the IGM writes
\begin{equation}
B_{\rm IGM} = \sqrt{\frac{3\epsilon_{\rm w} E_{\rm B,G}}{R_{\rm w}^3}} 
\simeq 1.1 \times 10^{-4}\, \mu \rm{G}\, .
\end{equation}
In conclusion, the galactic wind  has enriched the IGM with a magnetic
field {\it more  than one order of magnitude  larger} than its initial
value.   Interestingly  enough,  the  wind bubble  magnetic  field  is
proportional  to the disc  magnetic field.  Indeed, the  disc magnetic
energy depends on the toroidal magnetic field in the disc as follows
\begin{equation}
E_{\rm B,G}=\frac{B_\theta^2}{8\pi}\pi R_{\rm G}^2 H_{\rm G}\, ,
\end{equation}
so that the new IGM value now writes
\begin{equation}
B_{\rm IGM} = B_\theta \sqrt{\frac{3\epsilon_{\rm w} R_{\rm G}^2 H_{\rm G}}{8R_{\rm w}^3}} \, .
\label{igmfromgal}
\end{equation}
This last formula stands as a nice and concise summary of our work: we
have shown with detailed MHD simulations that the final magnetic field
in the expanding wind bubble  inherits its value from the parent dwarf
galaxy, where an  $\Omega$ effect has amplified the  field by almost 2
orders of magnitude from its initial value.

\section{Discussion}
\label{discussion}

Using a rather  idealized set up to model  dwarf galaxy formation with
MHD, we  have followed the evolution  of the magnetic  field, from the
collapse of an  equilibrium halo, to the formation  of a galactic disc
and finally  to the  ejection of magnetic  field lines by  a supernova
driven  galactic wind. In  order to  illustrate further  the mechanism
powering  up this  cosmic cycle,  for  which dwarf  galaxies play  the
central role, we have  plotted in Figure~\ref{diag} the magnetic phase
space  diagram, showing  a mass  weighted histogram  as a  function of
magnetic field amplitude and gas overdensity. First, when the halo
forms, the  magnetic field  is amplified by  gravitational contraction
from its initial IGM value to its final disc value
\begin{equation}
B_r = B_{\rm IGM}\left( \frac{\rho_{\rm G}}{\rho_{\rm IGM}}\right)^{2/3}\, .
\end{equation}
We  use here  only  the  radial component  $B_r$,  because just  after
collapse, the  magnetic field is  essentially radial in the  disc (see
discussion  above and  Fig.~\ref{Bini_snapshot}).  We  can see  in the
upper part  of Figure~\ref{diag} our initial  configuration, where $B$
and $\rho$ are following this  tight relation.  Note here that we have
neglected the  additional amplification of  the magnetic field  due to
turbulent motions in  the halo.  If we note  $R$ the initial Lagrangian
radius of  the collapsing  halo and $\lambda$  its spin  parameter, we
know  from   the  standard   cosmological  disc  formation   model  of
\cite{moetal98} that the final disc radius writes
\begin{equation}
R_{\rm G} \simeq \frac{\lambda}{200^{1/3}}R \, ,
\end{equation}
so that the initial disc radial component writes
\begin{equation}
B_r \simeq B_{\rm IGM} \frac{200^{2/3}}{\lambda^2} \, . 
\end{equation}
In  the lower  panel of  Figure~\ref{diag}, we  see that  the toroidal
field in  the disc  has grown linearly  with time due  to differential
rotation (the  $\Omega$ amplification), by  a factor of 60,  corresponding to
the growth rate (Eq.~\ref{equdynamo}) integrated over 3~Gyr. 
We finally get  the growth rate of the toroidal  magnetic field in the
disc as a function of the initial magnetic field in the IGM as
\begin{equation}
\partial_t B_\theta \simeq B_r \Omega_{\rm G} 
 \simeq B_{\rm IGM} \frac{200^{2/3}}{\lambda^2} \Omega_{\rm G} \, .
\label{galfromigm}
\end{equation}
In the  lower panel  of Figure~\ref{diag}, we  see that  this amplified
magnetic  field is  funneled  out  of the  galactic  disc, following  a
$\rho^{2/3}$ relation back  to the IGM. This is  the ``galactic wind''
branch that  ends up  the overall  cycle.  We have  seen that  the IGM
magnetic field  has been  amplified by one  order of  magnitude during
this cycle of roughly 3~Gyr.

Although our  simulations, in conjunction  with the work  presented in
\cite{bertoneetal06}   and  \cite{donnertetal09},  provide   a  nice
explanation for  the origin of  an IGM magnetic  field, at a  level as
high as $10^{-4}\mu$G, {\it  assuming equipartition magnetic fields in
  dwarf galaxies}, we still need  to explain the origin of these intense
galactic fields, given  that we start our cosmic  evolution wih a very
small   initial  value  $B_{\rm IGM}   \simeq  10^{-14}\mu$G   just  after
reionization. We  can't rely  on the $\Omega$  amplification alone,  since in
this case,  the magnetic field grows  only linearly with  time. In the
context   of  isolated   galaxies,  the   only  source   of  additional
amplification comes  from small scales turbulent  and cyclonic motions
(the $\alpha$ effect). This turbulent dynamo is believed to amplify
the  radial component of  the galactic  field, the  toroidal component
being generated and amplified by the $\Omega$ effect.

We know however  that galaxies are far from  being isolated. They form
by accretion  of diffuse IGM gas,  usually in the form  of dense, cold
filaments  known  as cold  streams  \citep{dekeletal09},  but also  by
accretion of  satellite galaxies,  orbiting around the  central galaxy
before  being  incorporated more  or  less  violently  into the  final
disc. These 2 important  processes define the hierarchical scenario of
galaxy  formation  \citep{silk&white78,  white&rees78},  and  probably
explain   most   of   the   properties   of   present   day   galaxies
\citep{governatoetal07, mayeretal08}.

We propose here to look at the galactic dynamo problem from a different
angle.  Instead on  relying only on the small  scale dynamo to amplify
the radial component of the field, we should also consider as a source
of radial  field: 1-the toroidal  magnetic field coming  from accreted
satellite galaxies  and 2- the accreted IGM  magnetic field previously
enriched  by galactic  winds or  3-  the accreted  IGM magnetic  field
previously enriched by entirely destroyed progenitor galaxies.  On one
hand,  we  know   from  Equation~(\ref{igmfromgal})  that  the  enriched
magnetic  field  is proportional to the  toroidal  field  of the  parent
galaxy, and on the  other hand, we know from Equation~(\ref{galfromigm})
that  the growth  rate of  the toroidal  field of  the main  galaxy is
proportional  to the  initial IGM  magnetic field.  

Although  a  correct  description   of  satellites  and  cold  streams
accretion  requires   detailed  cosmological  simulations   of  galaxy
formation with MHD,  we can speculate here that  our proposed cycle of
accretion and  ejection of magnetic field  lines (see Fig.~\ref{diag})
probably provides a dynamo loop,  for which the radial field component
is  amplified   by  accretion  of   satellite  galaxies  and   of  the
magnetically enriched  IGM.  A  very rough and  speculative analytical
description   of   this   effect   can  be   obtained   by   injecting
Equation~(\ref{galfromigm}) into Equation~(\ref{igmfromgal}),
\begin{equation}
\partial_t B_{\rm IGM}
 \simeq B_{\rm IGM} \sqrt{\frac{3\epsilon_{\rm w} R_{\rm G}^2 H_{\rm G}}{8R_{\rm w}^3}} \frac{R^2}{R_{\rm G}^{2}} \Omega_{\rm G}
\simeq B_{\rm IGM} \sqrt{0.1\epsilon_{\rm w}} \Omega_{\rm G} \, ,
\label{cosmicdynamo}
\end{equation}
where  we  have  exploited  the result  of  \cite{bertoneetal05}  that
$R_{\rm w}  \simeq 3R$ (a reasonably good  approximation according to
their Figure~2)  and we  have assumed a  typical disc aspect  ratio of
$H_{\rm  G}/R_{\rm  G}  \simeq  1/20$  and a  typical  spin  parameter
$\lambda \simeq 0.05$. We see  from the previous equation that now the
magnetic field  amplification is exponential, and  not linear anymore,
because  we have  connected  the radial  component  to the  progenitor
galaxies'   toroidal   component.   Altough   we   have  clearly   not
demonstrated that this cosmic dynamo actually works, we want to stress
that the cosmological environment might play an important role, namely
by accretion  of amplified field  lines from diffuse gas  or satellite
galaxies.  The  central engines  of this cosmic  dynamo are  the dwarf
galaxies,  because they are  easily disrupted  by galactic  winds, and
therefore provide a important source of diffuse radial field, and also
because they are the progenitors of our Milky Way Galaxy.

Some caveats to this Cosmic Dynamo scenario must be pointed out. We assumed that $B_{\rm IGM}$ is regenerated in a pure radial mode to get that exponential growth. Even though magnetic field lines in large-scale bipolar outflows are mainly perpendicular to the disc (see \citealp{heesenetal09} for example), they carry turbulence that produces non-radial component of the field. Propagation of the outflow can be affected by the particular geometry of the galactic environment and leads to irregular distributions of the magnetic field components. Moreover magnetic fields can contribute constructively or destructively to the $B_{\rm IGM}$ when large-scale bubbles percolate and can substantially alter the final configuration of the field. A very rough way of taking into account these effects on the radial component is to consider that each component of the field is equally distributed, then equation~(\ref{cosmicdynamo}) is simply affected by a factor $1/\sqrt{3}$.

While the  $\alpha$-$\Omega$ dynamo comes from  the connection between
small  scale  turbulence   and  galactic  differential  rotation,  the
proposed Cosmic Dynamo  originates from the coupling of  a large scale
gas cycle (cosmological accretion  and wind ejection) and differential
rotation. We  see from Equation~(\ref{cosmicdynamo})  that this effect
is  probably very  fast, especially  at high  redshift, for  which the
angular  velocity  of  dwarf   galaxies  is  much  higher  than  today
($\Omega_{\rm G}  \propto (1+z)^{3/2}$).  The actual  growth rate will
however  depend  crucially  on  many  complicated  aspects  of  galaxy
formation that  are beyond the scope  of this paper, such  as the wind
filling factor  as a  function of redshift  \citep{bertoneetal05}, the
field  geometry  during satellite  mergers,  the exact  3-dimensional
propagation of  wind bubbles within  the cosmic web, etc. Here it is a priori assumed, that the galactic wind expansion in conjunction with some mergers happen within the 3 Gyr. Thus it has to be shown, that a similar growth could be achieved with an initial field configuration, which is similar to the evolved wind field.

   We believe
that  this   Cosmic  Dynamo,  in  conjunction  with   a strong galactic dynamo, should  be  able to  account  for the  fast
magnetic  field amplification we  see in  our universe,  especially in
light of  the recent measurements performed on  high redshift galaxies
up to redshift $1$ \citep{bernetetal08}. Because the existence of an efficient $\alpha$-$\Omega$ Dynamo is questionable in dwarf galaxies due to low rotation of the disc and the lack of relativistic particles \citep{klein91}, it is important to explore alternative scenarios. It is possible to rely on
the  ideas discussed  by \cite{rees87},  for which  stars are  able to
generate their own magnetic field  by stellar dynamos.  We have tested
this  scenario   in  the   Appendix,  showing  that   for  reasonable
assumptions, we can  reach magnetic field close to  equipartition in a
few Gyr.

We have  also demonstrated with  our cooling halo simulations  that if
the  IGM magnetic  field is  high enough  (namely $B_{\rm  IGM} \simeq
10^{-4}~\mu$G), the Lorentz  force can prevent the formation  of a new
generation  of  dwarf  galaxies.   The  formation  of  a  magnetically
supported torus  prevent any subsequent star  formation, and therefore
breaks  the  central  engine  of  our Cosmic  Dynamo.   This  has  the
interesting   consequence  of   providing   a  saturation   mechanism,
explaining naturally why  the Cosmic Dynamo leads to  a final value of
$B_{\rm  IGM} \simeq  10^{-5}$ to  $10^{-4}~\mu$G. While  the $\alpha$-$\Omega$
dynamo saturates when the small scale turbulent magnetic field reaches
equipartition  in  the  disc  and  magnetic tension  becomes  able  to
suppress  the small scale dynamo   (see
\citealp{subramanian98},   \citealp{brandenburg&subramanian05}),   the
Cosmic Dynamo  saturates when  the large scale,  cosmological magnetic
field  prevents dwarf  galaxies' disc  formation, or  severely affects
star formation within them.

The evolution of magnetic fields in galaxies can now be addressed in a
cosmological context.   In a more or  less near future,  we propose to
extend the present work  using much higher resolution simulations, the
ultimate goal being to resolve  the small scale $\alpha$ effect in our
isolated, cooling halo  simulations. Another ambitious objective would
be to  address the same  problem in a fully  cosmological environment,
capturing the $\alpha$ effect  with some subgrid model, but accounting
properly for  satellite accretion  and wind bubble  propagation. 


\begin{acknowledgements}
The authors  wish to thanks Rainer Beck, Katia Ferri\`ere,  S\'ebastien Fromang and
Ana Palacios for useful  discussions and their precious comments. This
work  has been supported  by the  Horizon Project.   Computations were
done at CCRT, the CEA Supercomputing Centre.
\end{acknowledgements}

\bibliographystyle{aa}
\bibliography{author}

\begin{appendix}

\section{Stellar magnetic feedback scenario}
\label{magnetic_dipole_annexe}


\cite{rees87} was  the first to suggest that  Biermann battery effects
at  the surface  of stars  could create  magnetic fields  from  a zero
field. We  also know  that convective stars  can amplify  the magnetic
energy   extremely   fast    (\citealp{brunetal04})   close   to   its
equipartition  value by  dynamo  effects  like in  the  Sun.  In  this
scenario, massive stars $>10 \, \rm M_{\odot}$ which ones will release
their magnetic energy  in the ISM in powerful  explosions, are the key
ingredient  for  generating  galactic  magnetic fields.  In  the  most
massive stars, it seems that  there is no convective envelope allowing
for a very  fast dynamo, but differential rotation  could be extremely
strong  (\citealp{maederetal08}) and  can  substantially increase  any
seed field.  In some massive stars  the convective zone can  be in the
central part~(\citealp{zahn&brun07})  and the dynamo  can be effective
(\citealp{charbonneau&macgregor01}).  These stars  have also  a very
short life--time (a  few Myr), any amplification process  must have an
$e$-folding   time   extremely  small.   Some   models  predict   that
instabilities in the external radiative envelope are able to sustain a
sufficiently  fast growth  of the  field (\citealp{macdonald&mullan04,
  mullan&macdonald05}).   Though  it  remains  difficult  to  conclude
whether magnetic field are appreciably amplified, there are increasing
observational  evidence for  strong magnetic  fields in  massive stars
(\citealp{alecianetal08, hubrigetal08}).

When  a supernova is  exploding, its  magnetic field  is diluted  in a
larger volume. We can approximate its final value by considerating the
flux  conservation of  the magnetic  field. The  uniform  and averaged
magnetic field in a spherical transformation must follow
\begin{equation}
B_{2}=\left ( { R_{1} \over R_{2} } \right )^2 B_{1}\, .
\end{equation}
For example  the Crab  nebula with $B_{1}\simeq  100\, \mu$G  within a
$R_{1}\simeq  3.4$  pc radius  \citep{kennel&coroniti84}  will have  a
$B_{2}=5.10^{-2}\, \mu$G  field within the radius $R_{2}=150$  pc of a
superbubble. The case  of the Crab nebula is  a very strong particular
case: a fast rotating pulsar inside the centre of the remnant could be
responsible for this very  high magnetic field. It is particularly difficult
 to explain how strong magnetic fields $B_{1}\sim 10^9\,
\mu$G   in  massive   stars  (\citealp{alecianetal08},   $R_{1}\sim  5
R_{\odot}\sim  3.10^6  \, \rm  km$)  can  produce  so high  values  in
supernova remants ($B_{2}\sim 10^{-5}\,  \mu$G for $R_{2}\sim 1$ pc).
This prediction does not take into account the possible amplifications
stages during  the rapid expansion  of the blast wave,  magnetic field
can grow due to the  cosmic ray pressure \citep{parker92}, some strong
Biermann   battery   effect   \citep{hanayamaetal05},  or   turbulence
\citep{bell04}  that   would  allow  stronger   values  in  supernova
remnants.

The  basic  idea to  implement  the release  of
magnetic  field   by  superbubbles  coming  from
supernova explosions is  to put a fixed total  magnetic energy within
the  bubble  (ensuring  that $\nabla  .  \vec  B  =0$) with  a  random
direction.  We assume  that  within a  $r_{\rm  bubble}\simeq 150$  pc
radius,  each  supernova  contributes  constructively for  a  certain
amount of magnetic  energy to the final release.   We arbitrary choose
that  each superbubble (which  corresponds to  an assembly  of several
individual  supernova explosions) releases  a $10^{-3}\,  \mu$G field
following      a     robust     numerical      implementation. This particular choice of $10^{-3}\,  \mu$G magnetic seeding is entirely heuristic and is set to match $\mu$G field after 1 Gyr of galactic evolution.  To  get a  rough  idea of  the
magnetic release by a single  supernova inflating a larger bubble,
we have  to know  the number of  supernovae per star  cluster particle
$N_{\rm  SN,*}=m_{*} \times  \eta_{\rm SN}  / M_{\rm  SN}  \simeq 100$,
where we  have assumed  that $M_{\rm SN}=10\,  \rm M_{\odot}$  and for
$m_*=n_0 \times \Delta x^3\simeq 10^4  \, \rm M_{\odot}$. To simplify
if we  assume that  each single supernova  produces a  magnetic field
with  the  exact same  intensity  and  direction,  it comes  that  the
associated  magnetic field  per supernova  ($10\, \rm  M_{\odot}$) is
$B_{\rm SN} \simeq 10^{-5} \, \mu$G (in a 150 pc radius bubble). We do
not take into account the destructive contributions (magnetic field in
opposite  directions)  but  we  do  not also  take  into  account  the
substantial increase coming  from shocks between supernova explosions
that generate instabilities, shear flows, cosmic rays, etc. 

\begin{figure}[ht]
\begin{center}
\centering{\resizebox*{!}{7.cm}{\includegraphics{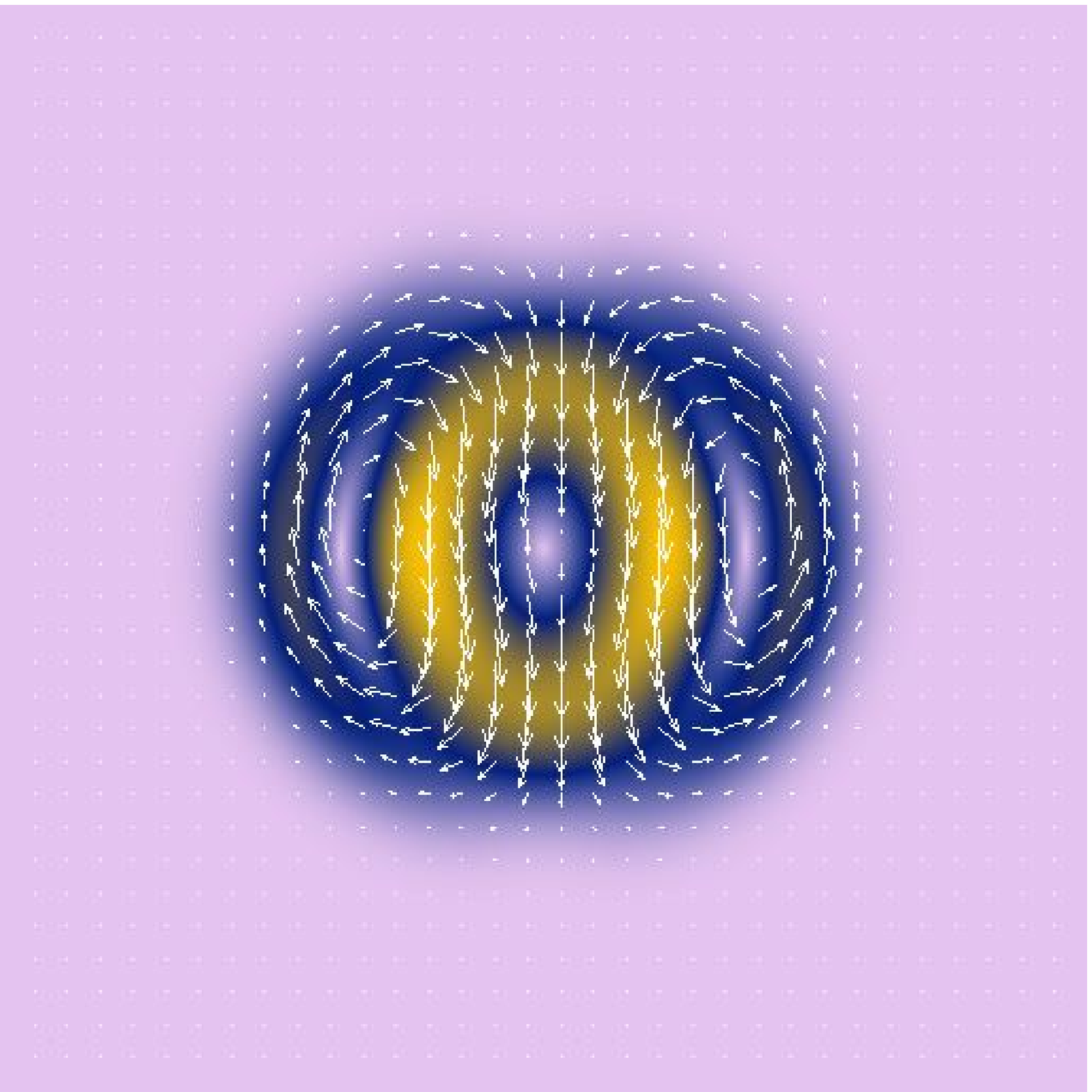}}}
\centering{\resizebox*{!}{7.cm}{\includegraphics{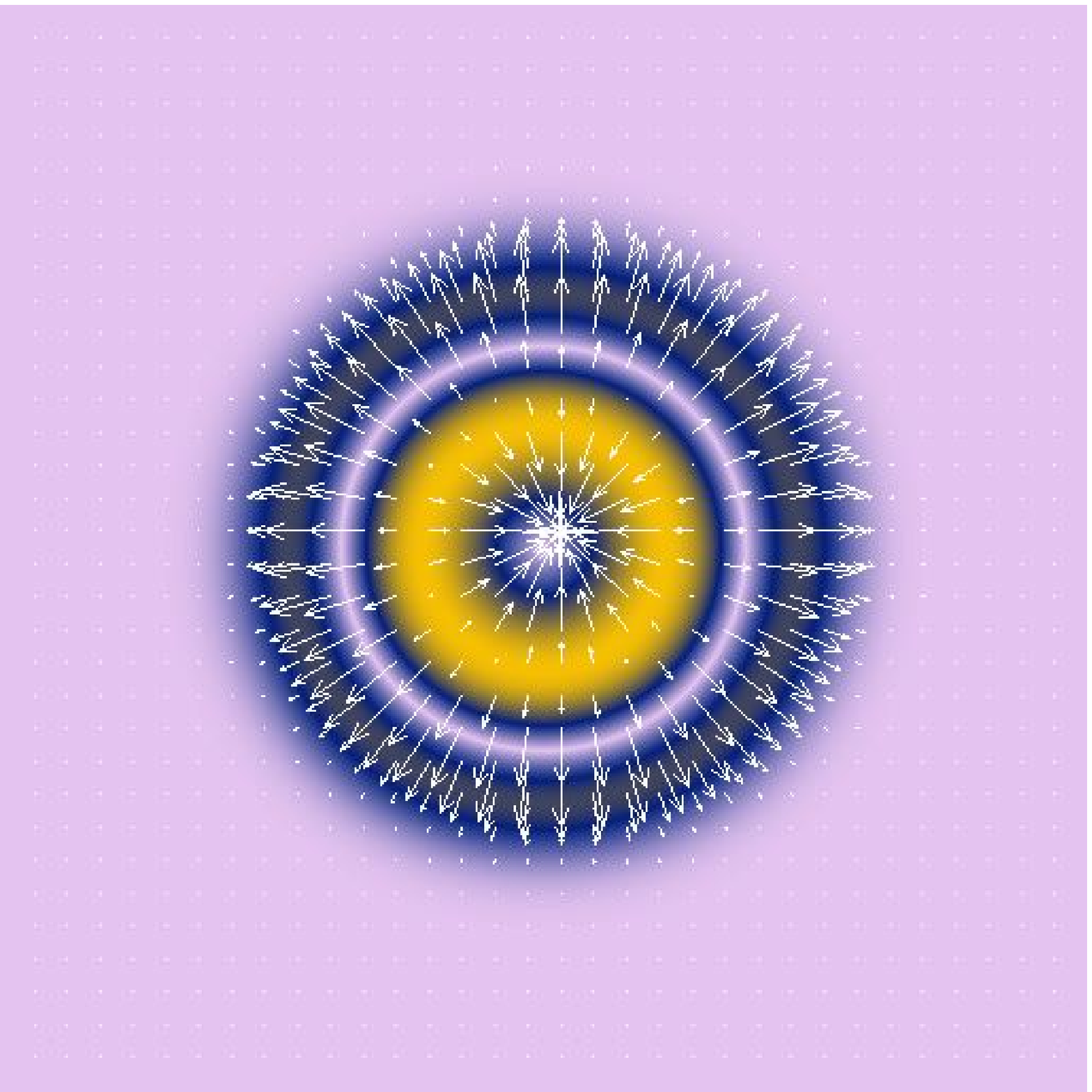}}}
\caption{Magnetic amplitude for the~(\ref{dipole_decroissant}) dipolar
  distribution  where $\vec{m}$  is  aligned with  the zaxis.  Upper
  panel is a slice through the (Oyz) plane and bottom panel is through
  the (Oxy) plane. Magnetic  vectors are overplotted. Picture size
  is $5 r_{\rm SN}$.}
\label{dipole_profile}
\end{center}
\end{figure}

To implement a method of direct injection of the magnetic energy in the code, we must find a form of the magnetic field that ensures $\nabla . \vec{ B}=0$. For
that purpose we  will design a dipolar moment  configuration such that
the  resulting  magnetic  field  is  embedded in  the  radius  of  the
supernova explosion, and that the magnetic energy is not divergent in
the centre  in order to preserve  the stability of the  scheme even at
different  resolutions.  Let us  recall  that  the magnetic  potential
vector created by a unique magnetic dipole follows
\begin{equation}
\vec{A}(r)={4 \pi \over \mu_0 r^3} ( \vec{m} \wedge \vec{r} ) \, ,
\end{equation}
where $\vec m$ is the magnetic  dipolar moment. If we take $\vec m$ as
a  constant value,  one can  easily see  that magnetic  field  will be
infinite in the  centre, then we must choose  a certain distribution of
dipolar  moment  elements  such  that  the  overall  resulting  moment
respects the  compact and convergent criterions.  For  that purpose we
need $\vec m$ to grow as fast as $r^4$ and that sharply decreases down
to zero at a certain radius $r_{\rm SN}$
\begin{equation}
\vec{m} = r^4 {\rm exp} \left( - \left ({r \over r_{\rm SN} }\right )^4 \right ) \vec{m}_0\, .
\label{dipole_decroissant}
\end{equation}
With such a distribution of $\vec m$, we can compute the magnetic field as
\begin{equation}
\vec{B}=\nabla \wedge \vec{A}\, .
\end{equation}
The resulting magnetic field computed on a $512^3$ grid within a $5
r_{\rm SN}$ box size (resolution element is $0.01\, r_{\rm SN}$) is plotted on
Figure~\ref{dipole_profile}.  We  can  renormalize  the  dipolar  moment
distribution  such  that the magnetic  field  gives a  unit
magnetic energy in regards to the measured magnetic energy $E_m$, here
\begin{equation}
m_0=\sqrt {4 \pi \over 6 \mu_0 E_m}=0.85 \, .
\end{equation}
This  value  is not very sensitive  to  the  numerical resolution,  for
example, if $\Delta x =0.3 r_{\rm SN}$, then $m_0=1$. That means that even
if we  poorly resolve the  bubble (it is  the case in  our simulations
$r_{\rm SN}\simeq  2$--$3 \Delta x$),  the final  magnetic energy  is very
close to the converged one.

We  also add  a  saturation  factor   to  avoid  that  the  magnetic  field
excessively  grows  above  the  equipartition  value,  such  that  the
resulting field is multiplied by
\begin{equation}
f_{\rm sat}={1 \over 1+ \left (  0.1{ B \over B_{\rm eq} }\right )^2}\, , 
\end{equation}
where $B$  is the value  of the magnetic  field inside the  bubble and
$B_{\rm eq}$ is the magnetic  amplitude corresponding to its equipartition
state with  internal energy. For  each buble, the $f_{\rm sat}$  factor is
the  maximum of all  individual $f_{\rm sat,i}$  factors computed  on each
cell within  the supernova radius. Such saturation  effects have been
outlined in galactic dynamo simulations (\citealp{cattaneo&hughes96}).

\begin{figure}
\centering{\resizebox*{!}{7.5cm}{\includegraphics{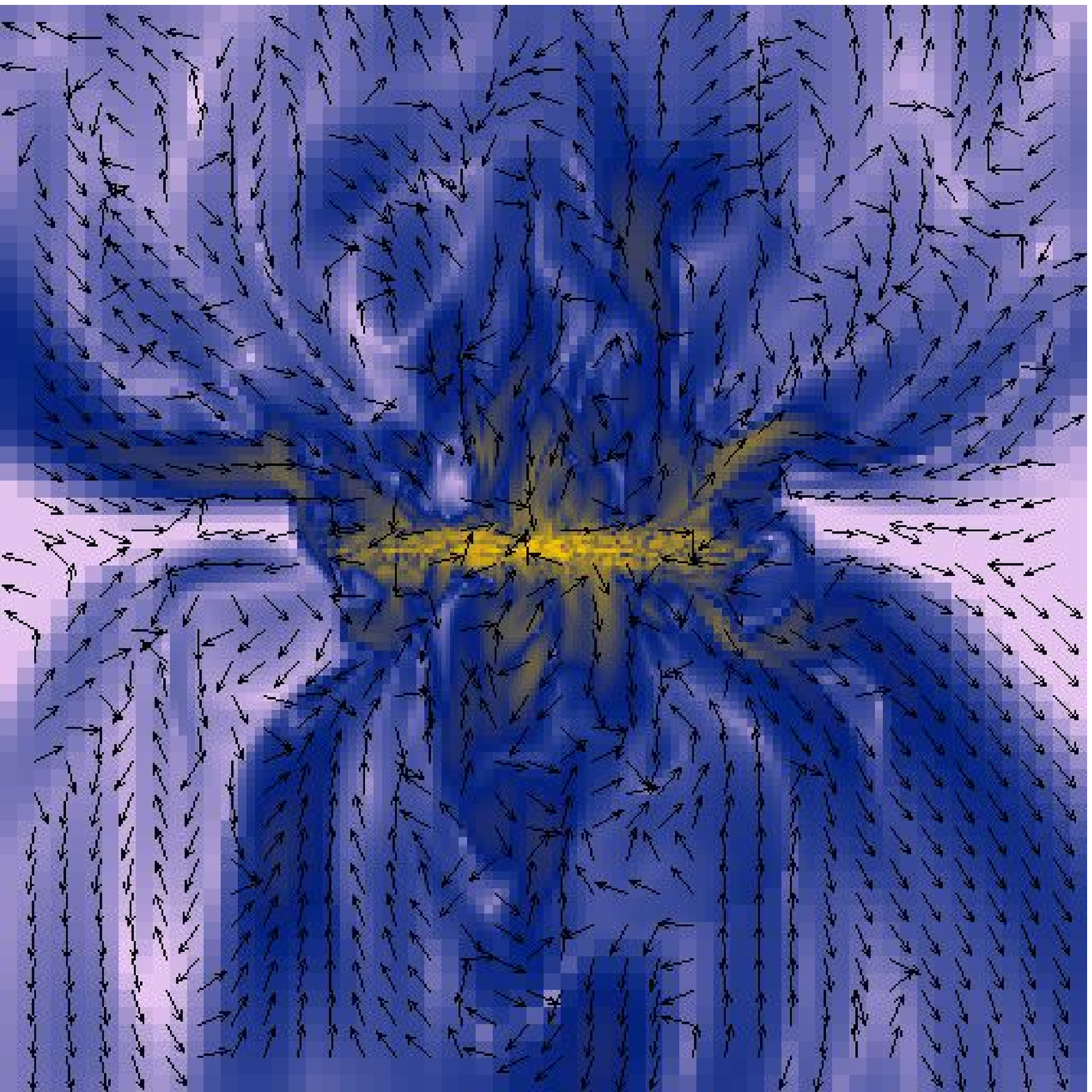}}}
\centering{\resizebox*{!}{7.5cm}{\includegraphics{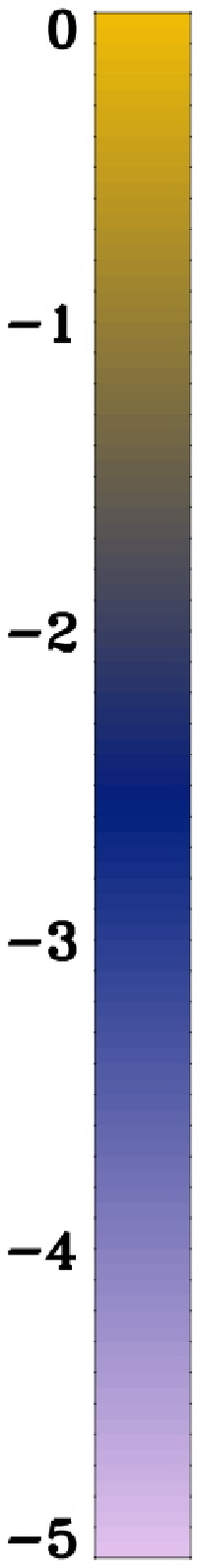}}}
\centering{\resizebox*{!}{7.5cm}{\includegraphics{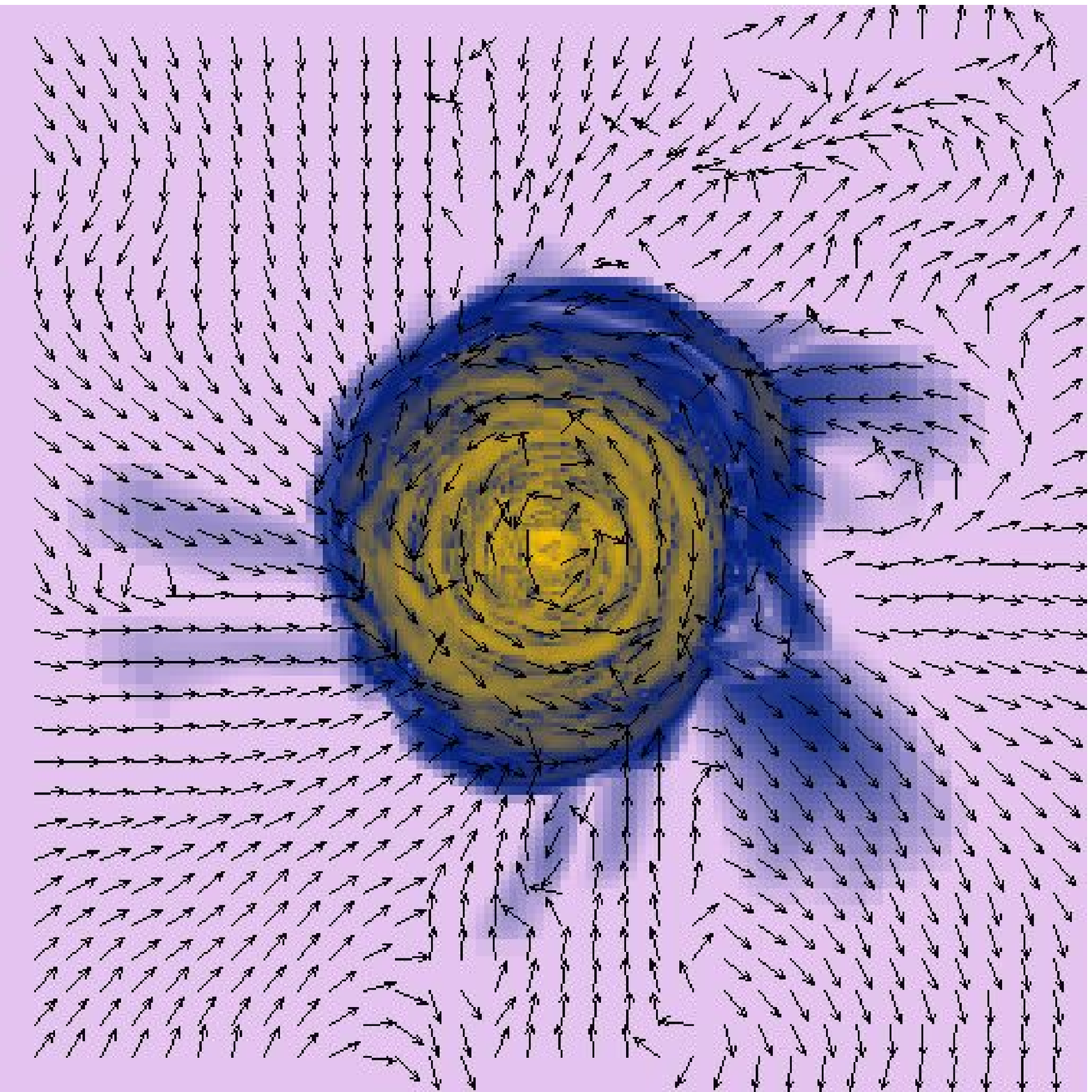}}}
\centering{\resizebox*{!}{7.5cm}{\includegraphics{./colortable_Bamp_dipole.ps}}}
\centering{\resizebox*{!}{7.5cm}{\includegraphics{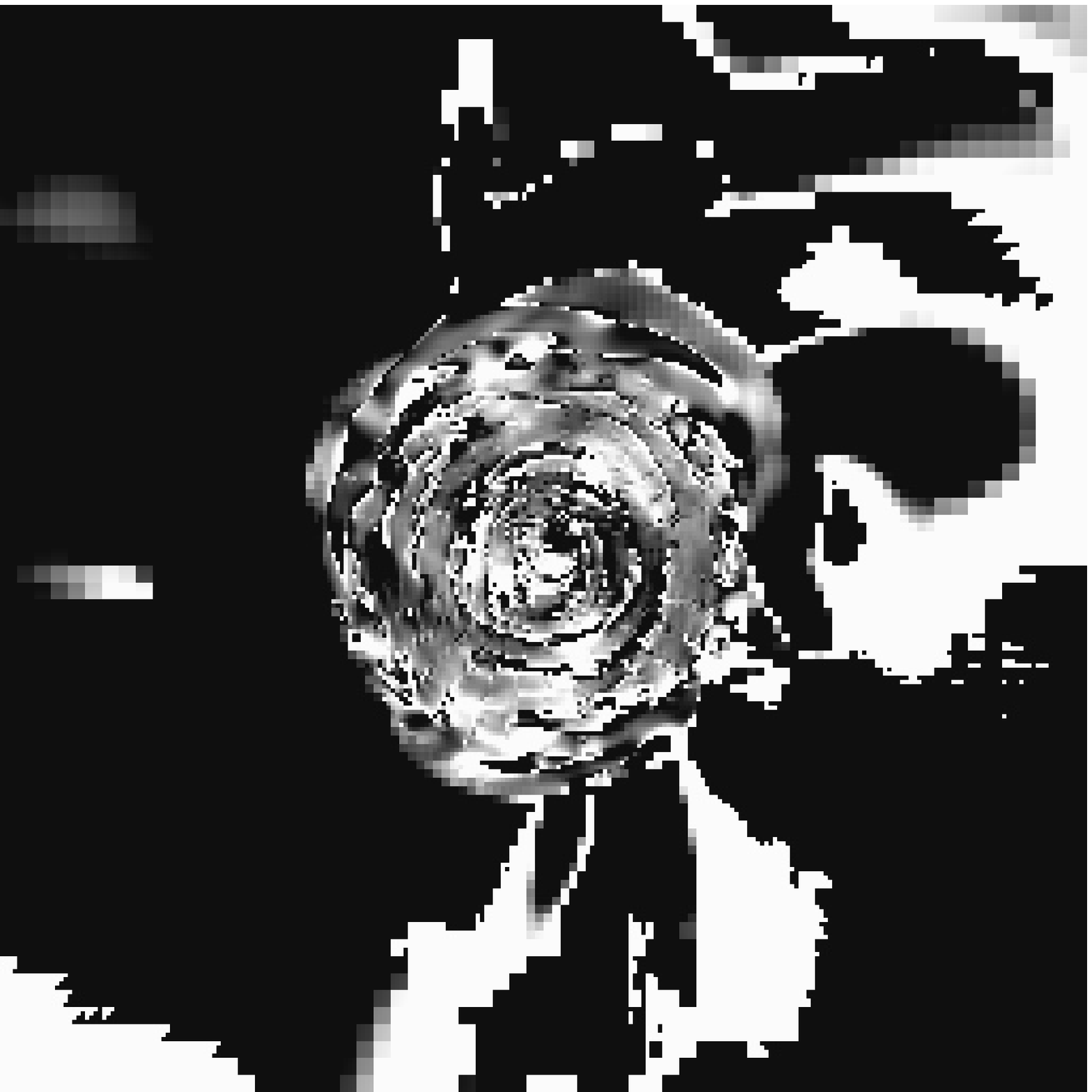}}}
\centering{\resizebox*{!}{7.5cm}{\includegraphics{./colortable_pitch_20_20.ps}}}
\caption{Magnetic  amplitude in  $\log  \, \mu$G  units with  magnetic
  vectors  for the  galaxy with  star formation,  supernova feedback,
  $B_{\rm IGM}=0$  and a magnetic seeding  $B_{\rm SN}=10^{-5}\, \mu$G
  at $t \simeq 3$ Gyr in the (Oyz) plane (upper panel), in the (Oxy) plane
  (middle panel) and the pitch angle  (bottom panel) . Picture size is
  40 kpc.}
\label{Bsn_Z_vec_dipolegen}
\end{figure}

We     produced    the     exact    same     simulation     than    in
Section~\ref{resultswind}  with no  initial  magnetic field.  Magnetic
fields are randomly generated in the  disc but they tend to align with
the   global  rotation   within   the   disc  as   one   can  see   in
Figure~\ref{Bsn_Z_vec_dipolegen}.  After  3  Gyr,  the  amplitude  has
reached a value  of a few $\sim 10^{-2} \, \mu$G  in the central parts
and its configuration is  no more like a A0 type. As  there is no more
initial  magnetic  field,  the   A0  configuration  does  not  appear.
Moreover the  randomisation of  the magnetic field  injection destroys
any  particular   symmetry  and  reconnects  the   field  through  the
equatorial  plane.  This effect  explains  the  smaller  scale of  the
magnetic field  fluctuations but  the global differential  rotation of
the  galaxy maintains  a preferred  component of  the field  along the
rotating vector. A direct consequence of the differential rotation is the amplification applied on magnetic seeds. A computation of the injected magnetic energy from supernovae with the star formation rate, gives $E_{\rm inj}\simeq 10^{48} \, \rm erg/Gyr$. Though the magnetic energy in the disc should reach $10^{48}$ erg after 1 Gyr, figure~\ref{Ene_evol} shows that the energy measured in the galaxy for this simulation (blue line) is $\sim 3.10^{49}$ erg after 1 Gyr. This is an evidence that the differential rotation is still amplifying the seed field generated by supernovae.

Radial  contribution  is   non-negligeable,  the  pitch  angle  $\vert
\theta_p \vert$ is everywhere greater than $10^{\circ}$ and goes up to
$90^{\circ}$ in some regions of  the galaxy. The $10^{\circ} $ minimum
value  is predictable:  if the  mean  field amplitude  is about  $\sim
10^{-2}\,  \mu$G preferentially toroidal,  putting a  $10^{-3}\, \mu$G
field in the supernova remnant  along the radial component produces a
$\vert  \theta_p  \vert  \simeq  5^{\circ}$  pitch  angle.   Then  the
successive explosions and the differential rotation rise this value up
to $90^{\circ}$.

As shown  in figure~\ref{fluxBvsr}, the magnetic energy  flux is lower
than when  the simulation starts with  an initial magnetic  field by a
factor  $\sim 40$,  but this  value depends  a lot  on  the particular
choice of the magnetic seeding $B_{\rm SN}=10^{-5}\, \mu$G. Thus, with
a $E_{\rm B,in}\simeq 10^{51}\, \rm erg.Gyr^{-1}$ at the bottom of the
wind,  we can predict  that the  final magnetic  field within  the hot
tenuous bubble will reach $\sim 10^{-5}\, \mu$G. It corresponds to the
initial  cosmological  seeds  required  in  large  scale  cosmological
simulations \citep{dolagetal05, dubois&teyssier08cluster}.  Relying on
the  assumption  that  some  supernova remnants  (Crab  nebula,  Vela
nebula)     are    permeated     with    strong     magnetic    fields
\citep{kennel&coroniti84, helfandetal01} we verify that this mechanism
can magnetise a  dwarf galaxy and therefore produce  a magnetised wind
to enrich the  IGM.  Using a very low  magnetic seed within supernova
bubbles  (5000 weaker  than the  field  within the  Crab nebulae),  we
showed that the galaxy rapidly reaches 1 $\mu$G in $\sim 1$ Gyr and is
able  to fill  the  IGM with  a  large-scale magnetic  field of  $\sim
10^{-5}\,  \mu$G.  It  is  therefore sufficient  to  explain an  early
magnetisation of the Universe.

\end{appendix}

\end{document}